\documentclass{article}[12pt]

\usepackage[english]{babel}

\usepackage[letterpaper,top=2cm,bottom=2cm,left=3cm,right=3cm,marginparwidth=1.75cm]{geometry}

\usepackage{amsmath,amsfonts,amssymb,bm}
\usepackage{graphicx}
\usepackage[colorlinks=true, allcolors=blue]{hyperref}
\usepackage{hyzsyntax}
\usepackage{setspace}
\usepackage{epigraph}
\usepackage{longtable}
\usepackage{cite}
\usepackage{setspace}
\usepackage{float}
\usepackage{longtable}
\LTcapwidth=\textwidth
\usepackage[most]{tcolorbox}
\usepackage{enumitem}

\counterwithin{equation}{section}
\counterwithin{figure}{section}
\counterwithin{table}{section}

\newcommand{\U}{\textup{U}}
\newcommand{\SU}{\textup{SU}}
\newcommand{\SO}{\textup{SO}}
\newcommand{\Sp}{\textup{Sp}}
\newcommand{\adx}{\mathrm{ad}_X}
\newcommand{\kl}{\mathfrak{l}}
\newcommand{\kn}{\mathfrak{n}}
\newcommand{\kp}{\mathfrak{p}}
\newcommand{\Ind}{\mathrm{Ind}}
\newcommand{\Hom}{\mathrm{Hom}}

\title{\Huge Atomic Higgsings of 6D SCFTs II: \\
Induced Flows \vspace{1cm}}

\author{\Large Jiakang Bao and Hao Y. Zhang \vspace{0.5cm} \\
\href{jiakang.bao@ipmu.jp}{jiakang.bao@ipmu.jp}, \href{hao.zhang@ipmu.jp}{hao.zhang@ipmu.jp}\vspace{1cm}}

\date{}

\begin{document}
\maketitle

\onehalfspacing

\begin{tabular}{ll}
    & Kavli Institute for the Physics and Mathematics of the Universe (WPI), \\
    & University of Tokyo, Kashiwa, Chiba 277-8583, Japan \\
\end{tabular}

\vspace{2cm}

\begin{abstract}
We study a specific type of atomic Higgsings of the 6d $\mathcal{N}=(1,0)$ theories, which we call the induced flows. For the conformal matter theory associated with a pair of nilpotent orbits, the induced flows are given by the inductions of the orbits. We also consider the induced flows for the orbi-instanton theories (as well as some little string theories) that are associated with the homomorphisms from the discrete subgroups of $\SU(2)$ to $E_8$. This gives a physical definition of the inductions among these discrete homomorphisms, analogous to the inductions of the nilpotent orbits. We analyze the Higgs branch dimensions, the monotonicity of the Weyl anomalies (or the 2-group structure constants for LSTs) and the brane pictures under the induced flows.
\end{abstract}

\newpage

\tableofcontents

\section{Introduction and Summary}\label{intro}
The concept of the renormalization group (RG) flows is always an intriguing topic when studying the quantum field theories. Physically, it is crucial for understanding how physical quantities would evolve across different energy scales, revealing fixed points that could possibly correspond to critical phenomena and phase transitions. In higher dimensions, when the theories have sufficiently many supersymmetries, the moduli spaces parametrized by the vacuum expectation values (VEVs) would acquire rich structures. Mathematically, this boils down to the study of geometric singularities (with a certain structure imposed by the amount of supersymmetry) and their stratifications.

Let us first recall the existing literature that studies the Higgs branch of 6d SCFTs using M-/F-theory that precedes our previous paper \cite{Bao:2024eoq}. The approach of using the nilpotent VEVs and the flat $E_8$ connections to systematically study the 6d RG flows were initiated in \cite{Heckman:2015bfa} and systematically pursued in \cite{Heckman:2016ssk,Hassler:2019eso,Baume:2021qho} for conformal matter theories (whose understanding from both F-theory and M-theory \cite{DelZotto:2014hpa} proves to be crucial in this analysis). Moreover, the orbi-instanton theories were investigated in \cite{Mekareeya:2017jgc,Frey:2018vpw,Fazzi:2022hal,Fazzi:2022yca}. Using this approach, the change of the anomalies along the $(1,0)$ RG flows was studied in \cite{Heckman:2015ola,Mekareeya:2016yal,Baume:2023onr,Fazzi:2023ulb}.  On top of these algebraically encoded RG flows, \cite{Heckman:2018pqx} provided many ingredients towards a more complete classification, with an eye of incorporating tensor branch flows.

In our previous paper \cite{Bao:2024eoq}, we studied the Higgs branches for any given 6d superconformal field theories (SCFTs) with at least $\calN = (1,0)$. Our study is based on the classification of 6d SCFTs \cite{Heckman:2013pva,Heckman:2015bfa}, using F-theory \cite{Vafa:1996xn,Morrison:1996na,Morrison:1996pp}. The step-by-step minimal Higgsings, which we call \emph{atomic Higgsings}, can be performed on the generalized quivers of the 6d theories. With our algorithm, we can produce the full Hasse diagrams that encoding the partial orderings of the 6d theories under the RG flows, which also coincide with the partial orderings of the leaf structures of the corresponding symplectic singularities.

More concretely, our atomic Higgsings include the following (non-exclusive) types:
\begin{enumerate}[label=(\roman*)]
    \item Atomic nilpotent VEV Higgsings: For any non-abelian flavour symmetry, we can turn on its minimal nilpotent orbit. The residual unbroken flavour symmetry is the commutant of $\text{Im}(\rho)$ in $\mathfrak{g}_{\mathbb{C}}$, where $\rho$ is the homomorphism from $\mathfrak{sl}(2,\mathbb{C})$ to $\mathfrak{g}_{\mathbb{C}}$ corresponding to the minimal nilpotent orbit in light of the Jacobson-Morozov theorem. The transverse slice would be the closure of this minimal nilpotent orbit. If we examine the concrete change on the tensor branch, we could either only get reduction of gauge ranks, or a combination of gauge rank reduction and blowing down $-1$ curves (the latter would also be counted as an ``atomic combo Higgsing" as indicated below).
    \item Atomic plateau Higgsings: Such a Higgsing is triggered by the semisimple part of the flavour symmetry. This happens when there is a chain of $-n$ curves (or alternating $-1$ and $-4$ curves) all with non-trivial fibre decorations.
    \item Atomic combo Higgsings: This is the case where neither of the above two cases can be performed individually. However, a simultaneous step of reducing the gauge algebras and blowing down the $-1$ curves is possible. It is conjectured that the transverse slice is always of quaternionic dimension 1.
    \item Atomic endpoint-changing Higgsings: This type of the Higgs branch RG flows would change the Dirac pairings of the BPS strings. For $(2,0)$ theories, it could be triggered by a VEV of the $(1,0)$ hypermultiplet in the $(2,0)$ tensor multiplet. This amounts to the complex deformation of the geometric singularity and the separation of the NS5 branes in the Type II setting. For theories with T-brane VEVs, the generalized quiver would split into multiple pieces. If there is a plateau, it would become two (or more) plateaux. If the theory is associated with a pair of nilpotent orbits, the resulting IR theory would be determined by their induced orbits.
\end{enumerate}
Readers are referred to \cite{Bao:2024eoq} for more details on how to perform these atomic Higgsings.

In particular, the Higgsings of types (i) and (ii) can be systematically understood via their F-theory descriptions \cite{Heckman:2016ssk,Heckman:2018pqx}. The Higgsings of type (iii) can be understood on a case-by-case basis by the F-theory descriptions as classified in \cite{Bao:2024eoq}. However, the Higgsings of type (iv) are more involved. Except for a few families of theories, a more systematic and efficient approach to determine all the RG flows is still required. So far, the most general way is to perform a brute-force search for all the possible IR theories based on the compatibility of the 6d tensor branch descriptions and the monotonicity of certain quantities along RG flows. In this paper, we shall study the endpoint-changing flows for theories associated with nilpotent orbits and/or discrete homomorphisms in a more systematic manner. These two families of theories admit M-theory constructions, which will prove crucial in our analysis: M5-branes probing the transverse Kleinian singularity for conformal matter theories, and the whole system placed in the worldvolume of an extra M9-brane for orbi-instanton theories.

We should also mention another perspective in the study of Higgsing the 6d supersymmetric theories, namely the magnetic quivers \cite{Cabrera:2018ann,Bourget:2019aer,Bourget:2022tmw,Hanany:2022itc}\footnote{The magnetic quivers have been extensively studied recently with a vast range of literature. Therefore, it is impossible to cite all them, and we only mention a few relevant ones here. Interested readers are referred to the references in these papers as well.}. The statement is that the Higgs branch of a 6d theory, as an equality of the moduli spaces, is the same as the Coulomb branch (or more precisely, moduli space of dressed monopole operators) \cite{Gaiotto:2008sa,Cremonesi:2013lqa}
of a 3d $\mathcal{N}=4$ magnetic quiver\footnote{Mathematically, the Coulomb branches have been rigorously defined following the Braverman-Finkelberg-Nakajima construction \cite{Nakajima:2015txa,Braverman:2016wma,Braverman:2017ofm,Nakajima:2019olw}.}. The decay and fission algorithm of the unitary magnetic quivers provides a systematic way to Higgs the theories with 8 supercharges in higher dimensions \cite{Bourget:2023dkj,Bourget:2024mgn}. More recently, this was extended to orthosymplectic magnetic quivers in \cite{Bao:2024eoq,Lawrie:2024wan}. When the magnetic quivers are known, the decay and fission algorithm gives a straightforward method that allows us to cross check the algorithm using the generalized quivers with the F-theory descriptions. More importantly, this tells us what the elementary transverse slices are in most cases. Nevertheless, there are many 6d theories whose corresponding magnetic quivers are not known. In these cases, we can still apply the algorithm via the F-theory descriptions as in \cite{Bao:2024eoq}, either for those associated with classical Lie algebras/groups or for those that are exceptional.

As aforementioned, we continue our previous study on the atomic Higgsings in this paper, with a focus on the theories associated with nilpotent orbits and/or discrete homomorphisms. In particular, we shall mainly discuss the flows that are closely related to the concept of inductions. Hence, we shall refer to them as the \textbf{induced flows}. For the induced flows, many of them belong to the endpoint-changing Higgsings, but there are also cases that are either plateau Higgsings or combo Higgsings. For a given theory, there are two boundary conditions with certain induction data that we are interested in. Therefore, we shall focus on the \emph{long quivers} here so that the two sides are not correlated. Such a correlation would prevent us from analyzing the induction of one nilpotent VEV at a time, and we could get around such a complication while maintaining full generality. Specifically, we demand the number of tensor multiplets to be sufficiently large, so that at least one tensor multiplet in the UV theory is completely unaffected by the nilpotent VEV or discrete holonomy both from the left and from the right. Whenever a short quiver theory is found to admit an induced flow, it turns out that we can always ``redefine" it as a long quiver theory\footnote{Let us expound more on this point. Given a short quiver, say associated with a pair of boundary conditions $(\mathcal{B}_1,\mathcal{B}_2)$, it could be possible that there exists another pair of boundary conditions $(\mathcal{B}'_1,\mathcal{B}'_2)$ that give the same generalized quiver, and the quiver is not short with respect to $(\mathcal{B}'_1,\mathcal{B}'_2)$. The induced flows would follow from the inductions for $\left(\mathcal{B}'_1,\mathcal{B}'_2\right)$. If the generalized quiver is always short regardless of such degenerations, then we have not found any induced flows for these ``too short'' quivers. We consider induced flows for short quiver theories as the last missing piece towards completely understanding the Higgsings of all 6d conformal matter theories and orbi-instanton theories.}.

Induced flows can be systematically understood via their M-theory description. Recall that conformal matter theories are engineered via the M-theory picture with M5-branes on Kleinian singularities. An induced flow is then given by the following two simultaneous operations in the transverse space $\mathbb{C}^2/\Gamma_{\mathfrak{g}}\times\mathbb{R}_{\perp}$. The M5-branes are separated into two stacks along a direction perpendicular to $\mathbb{R}_{\perp}$, and a complex structure deformation is performed on the singularity so that it would split into two pieces, whose new position should overlap with each of the newly-created M5 brane stack for atomic-ness. 

% Let us first consider the cases associated with the nilpotent orbits. Examples include the conformal matter theories and the orbi-instanton theories. Then the induced flows are determined by the inductions of the orbits. 

However, a fine print of nilpotent VEV of the flavour symmetry \cite{Cecotti:2010bp}\footnote{Here the nilpotent VEV is sometimes called the ``T-brane data", where ``T'' stands for \textit{triangular}, referring to the fact that a nilpotent VEV of unitary-type flavour symmetry should be thought of as non-Abelian generalization of triangular-shaped D-brane pattern. In IIA/M-theory on a Calabi-Yau 3-fold $X$, the effect of T-brane data is identified by \cite{Anderson:2013rka} as ``periods of 3-form potential valued in the intermediate Jacobian of $X$".} has to be added to the M-theory picture. Specifically, tracking how the VEV changes along this separating move precisely requires us to consider induction of nilpotent orbits (specifying the ``boundary condition" in the F-theory description). In other words, when there is an induced flow from the UV theory $\mathcal{T}\left(\mathcal{O}^{\text{UV}}\right)$ to the IR theory $\mathcal{T}\left(\mathcal{O}^{\text{IR}}\right)$, the orbit $\mathcal{O}^{\text{UV}}$ is induced from the orbit $\mathcal{O}^{\text{IR}}$. Many details of the nilpotent orbits, as well as their inductions, are reviewed in Appendix \ref{preliminaries}.

Having the above M-theoretic understanding of the induced flows of the conformal matter theories, it is natural to attempt a generalization by incorporating an M9-brane into the picture to turn it into an orbi-instanton theory, or add two M9-branes to turn it into an $E_8 \times E_8$ heterotic little string theory (LST) \cite{Aspinwall:1997ye}. Now, the boundary condition (of the F-theoretic generalized quiver) on the M9 side is no longer given by a nilpotent orbit, but instead a flat $E_8$ holonomy in the asymptotic infinity $S^3/\Gamma$ that is topologically non-trivial. Such holonomies are known to be classified by the topological class of homomorphisms from $\Gamma$ to $E_8$ \cite{DelZotto:2014hpa}.

For the discrete homomorphisms $\text{Hom}(\Gamma,E_8)$, where $\Gamma$ is a finite $\SU(2)$ subgroup following the McKay correspondence, the full classifications are only known for the A-type cases, namely the cyclic groups. For the other cases, some partial results can be found in \cite{frey2001conjugacy}. In \cite{Frey:2018vpw}, it was proposed that each discrete homomorphism corresponds to a distinct orbi-instanton theory (if we fix the nilpotent orbit side). This would then give physical classifications of the discrete homomorphisms.

One of the results of our paper is to introduce the inductions of the discrete homomorphisms in this physical sense. When there is an atomic Higgsing from a UV orbi-instanton theory to an IR orbi-instanton theory (which could have either one or multiple components) following our algorithm introduced in \cite{Bao:2024eoq}, we say that the associated homomorphism $\alpha^{\text{IR}}$ induces the homomorphism $\alpha^{\text{UV}}$. Of course, the mathematical rigour (e.g. existence and uniqueness of this induction map) and a geometric meaning would still be desired. The physical definition of such inductions can also be phrased using the class of the little string theories that are associated with a pair of discrete homomorphisms, which would then admit induced flows.

% The induced flows have a nice interpretation as the inductions of the orbits mathematically, and we may extend the concept of inductions to the discrete homomorphisms. 

\paragraph{Conventions} Let us clarify some conventions used in this paper. In the literature, the Hasse diagrams could be plotted in two different ways where the processes of the Higgsings could either go from the top to the bottom or from the bottom to the top. The Higgsings in this paper would go from the top to the bottom.

For the classical cases where the orbits can be labelled by certain partitions, we shall sometimes use the word ``rows'' for its entries and the word ``columns'' for the entries in its transpose for brevity. Moreover, the key ingredient in the inductions would be the Levi subalgebras. Often, a Levi subalgebra $\mathfrak{l}$ of a (semisimple) Lie algebra $\mathfrak{g}$ can be decomposed into $\mathfrak{l}=\mathfrak{s}\oplus\mathfrak{a}$, where $\mathfrak{s}$ (resp.~$\mathfrak{a}$) is the semisimple (resp.~abelian) part. In our discussions, we shall always omit the abelian part, i.e., the factors of $\mathfrak{gl}(1)$, since it has a trivial contribution. Physically, they correspond to the free hypers that generate the smooth part of the moduli space.

For the (symplectic) geometry, an important quantity would be the dimension. In the literature, one might use either the complex dimension or the quaternionic dimension. Here, we would keep the subscripts in the expressions, namely $\dim_{\mathbb{C}}$ and $\dim_{\mathbb{H}}$, if it might cause any confusions. When mentioning the dimensions in words, we would also write the word ``quaternionic'' (resp.~``complex'') explicitly for the quaternionic (resp.~complex) dimensions when necessary.

For convenience, we shall use $(\dots)^{\otimes}$ to indicate the same legs adjacent to a curve in the generalized quiver. We shall also use $\langle\dots\rangle^{\otimes}$ to denote the repeated pattern in the curve configuration. For instance,
\begin{align}
    \overset{\kso(8)}{4} \ \ (1 \ \ [\SO(8)])^{\otimes3}\quad=\quad [\SO(8)] \ \ 1 \ \ \underset{\underset{[\SO(8)]}{1}}{\overset{\kso(8)}{4}} \ \ 1 \ \ [\SO(8)]
\end{align}
and
\begin{equation}
    [\SO(8)] \ \ 1 \ \ \left\langle\overset{\kso(8)}{4} \ \ 1\right\rangle^{\otimes2} \ \ [\SO(8)]\quad=\quad[\SO(8)] \ \ 1 \ \ \overset{\kso(8)}{4} \ \ 1 \ \ \overset{\kso(8)}{4} \ \ 1 \ \ [\SO(8)].
\end{equation}
Moreover, we write $\mathfrak{sp}(n)=\mathfrak{c}_n$, and likewise for the Lie groups.

\paragraph{Organization of the paper} The rest of this paper is organized as follows. In \S\ref{flowsindorb}, we analyze the induced flows of the 6d $(1,0)$ theories that flows to $(2,0)$ theories, in which the notion of the induced nilpotent orbits plays a central role. After presenting some unitary theories as a warm-up to illustrate our idea, we proceed to applying our idea to more general cases including the DE-type conformal matters and the $(1,0)$ theories with DE-type Dirac pairings. This requires using the formalism of the induced nilpotent orbits in its full generality. In \S\ref{indE8}, we analyze the induced flows of the orbi-instanton SCFTs, where the physical idea of atomic induced flows motivates us to introduce the mathematical notion of the induced discrete homomorphisms in parallel to the induced nilpotent orbits. This analysis generalizes to a class of little string theories, for which we need to introduce a pair of discrete homomorphisms into $E_8$. In \S\ref{athm}, we shall verify the $a$-monotonicity for all the induced flows between conformal matter theories of A-types and D-types, supporting the validity of these flows.

In addition, we have a few technical appendices to supplement our results. In Appendix \ref{preliminaries}, we give a concise review of nilpotent orbits and their inductions. In Appendix \ref{atomicInduction_e6}, we make an explicit list of the nilpotent orbit inductions from $\mathfrak{g}\times\mathfrak{gl}(1)$ to $\mathfrak{e}_6$, where $\mathfrak{g}$ is any non-abelian Lie algebra of rank 5, for all possible $\mathfrak{g}$-orbits. In Appendix \ref{guessMQ}, we comment on the 3d $\calN = 4$ magnetic quivers for some D-type orbi-instanton theories in our discussions.

%\newpage

\section{Flows via Induced Orbits}\label{flowsindorb}
In this section, we identify a large family of atomic RG flows in 6d SCFTs. Specifically, we consider the induced flows among the conformal matter theories with pairs of $(\calO_L, \calO_R)$ nilpotent deformations.

\begin{figure}[h]
    \centering
    \includegraphics[width=15cm]{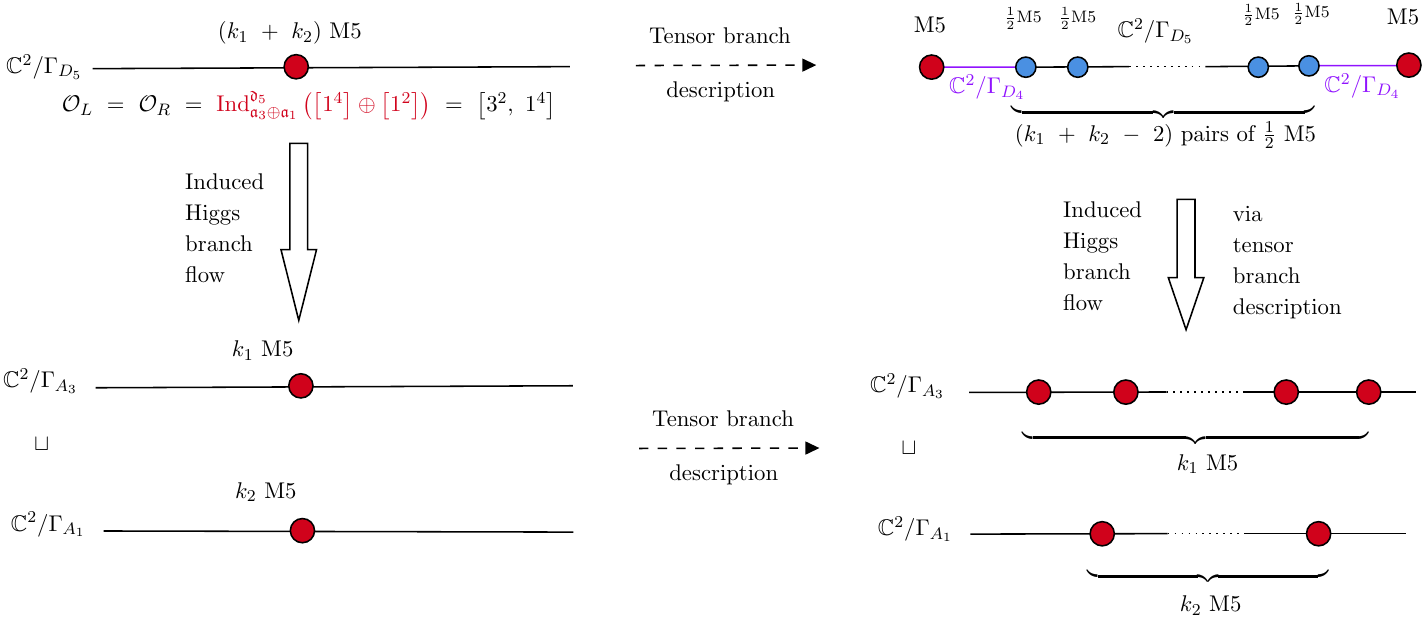}
    \caption{Illustration of the M-theory interpretation of an induced flow. On the left, we have the strong coupling point of an induced flow. The IR theory is a disjoint union of two theories, constructed by M5-branes probing $A_3$ and $A_1$ Kleinian singularities. The UV theory takes all the M5-branes probing a D-type Kleinian singularity, with the nilpotent VEVs induced from the trivial nilpotent orbits in the IR to get an atomic RG flow. On the right, we have the full tensor branch description of this atomic RG flow. We remark that, in the UV theory, each $\bbC^2/\Gamma_{D_4}$ singularity terminates at the full M5-branes on the left and right end, corresponding to the absence of flavour symmetry on both ends of the quiver in the F-theory description. Only in the UV theory, the M5-branes are fractionated into half M5-branes. As explained in \cite{Heckman:2016ssk}, the effect of $\calO = [3^2, 1^4]$ on each side is to shrink the outermost pair of half M5-branes and reduce the singularity type in the next segment from $\bbC^2/\Gamma_{D_5}$ to $\bbC^2/\Gamma_{D_4}$.}\label{fig:M-th}
\end{figure}

We first give an M-theory understanding of the induced flows for the conformal matter theories. For the most generic case, three operations are performed simultaneously in the transverse space $\bbC^2/\Gamma_{\kg} \times \bbR_\perp$:
\begin{enumerate}
\item The stack of M5-branes are separated into two stacks, along a direction that is perpendicular to $\bbR_\perp$ (thus within $\bbC^2/\Gamma_{\kg}$). 
\item A complex structure deformation is performed so that $\bbC^2/\Gamma_{\kg} \rightarrow \bbC^2/\Gamma_{\kl_1} \oplus \bbC^2/\Gamma_{\kl_2}$, where the direction of $\bbR_\perp$ is held fixed.
\begin{itemize}
    \item A special sub-case is that the M5-brane move is trivial (i.e., no M5 separation) and the complex structure deformation only reduces the rank of the McKay dual gauge group by 1. In this case, we always get an induced flow that is also a combo flow.
\end{itemize}
\item When an orbit $\calO_{\kg}$ in the Lie algebra $\kg$ is associated to the UV theory, the resulting orbit for the IR theory is given by $\mathcal{O}_{\mathfrak{l}}$ in the Levi subalgebra $\mathfrak{l}$ of $\mathfrak{g}$ such that
    \begin{equation}
        \calO_{\kg}=\Ind^{\kg}_{\kl}\left(\calO_{\kl}\right). \label{eqn:inducedOrbit}
    \end{equation}
    The orbit $\mathcal{O}_{\mathfrak{l}}$ can be written as $\mathcal{O}_{\mathfrak{l}_1}\oplus\dots\oplus\mathcal{O}_{\mathfrak{l}_n}$ with $\mathfrak{l}=\mathfrak{l}_1\oplus\dots\oplus\mathfrak{l}_n$. For an atomic flow, the Levi subalgebra can be obtained by removing one node in the Dynkin diagram of the Lie algebra $\mathfrak{g}$.
\end{enumerate}
Under this set of moves, an atomic flow would take place whenever the VEV of the UV theory precisely produces the VEV in the IR theory after these operations. This is the case only when the induction condition \eqref{eqn:inducedOrbit} holds for both the left orbit and the right orbit. An illustration can be found in Figure \ref{fig:M-th}. We remark that, to obtain a Higgs branch flow, we \emph{cannot} move the M5-branes apart from each other along the $\bbR_\perp$ direction. Otherwise, we would get a tensor branch flow.

We can also give a description in Type IIB with $k$ NS5-branes probing $\bbC^2/\Gamma$ singularities. Here, an A-type discrete subgroup $\bbZ_k \subset \SU(2)$ would give a theory that is dual to the ``A-type conformal matter theory'' above while a DE-type discrete subgroup $\Gamma$ of $\SU(2)$ would produce a theory with DE-type Dirac pairing to be discussed in \S\ref{DEbase}. Here, on the F-theory tensor branch, there are $k$ 7-branes located at each resolution divisor. Again, the following three steps happen \textit{simultaneously} (where the third step of the induced VEV is completely field-theoretic, so it does not depend on the string frame):
\begin{enumerate}
    \item We split of the NS5-branes that give rise to the gauge symmetry on the tensor branch of the UV theory such that the ``7-brane gauge rank'' is reduced by 1.
    \item We perform a complex structure deformation on the base $\bbC^2/\Gamma$. To get an atomic flow, we need the position of each of the 7-brane gauge rank to match one of the complex structure deformations:
    \begin{itemize}
        \item If the complex structure deformation is trivial, then we get a plateau Higgsing or a combo Higgsing.
        \item If the complex structure deformation is non-trivial, then we get an endpoint-changing flow, which results in a reducible theory in the sense that it has multiple components.
    \end{itemize}
    \item When an orbit $\calO_{\kg}$ in the Lie algebra $\kg$ is associated to the UV theory, the resulting orbit for the IR theory is given by $\mathcal{O}_{\mathfrak{l}}$ in the Levi subalgebra $\mathfrak{l}$ of $\mathfrak{g}$ such that
    \begin{equation}
        \calO_{\kg}=\Ind^{\kg}_{\kl}\left(\calO_{\kl}\right). \label{eqn:inducedOrbit_II}
    \end{equation}
    The orbit $\mathcal{O}_{\mathfrak{l}}$ can be written as $\mathcal{O}_{\mathfrak{l}_1}\oplus\dots\oplus\mathcal{O}_{\mathfrak{l}_n}$ with $\mathfrak{l}=\mathfrak{l}_1\oplus\dots\oplus\mathfrak{l}_n$. For an atomic flow, the Levi subalgebra can be obtained by removing one node in the Dynkin diagram of the Lie algebra $\mathfrak{g}$. 
\end{enumerate}

Based on \cite{Heckman:2016ssk}, the codimension of the Higgs branches and the codimension of the corresponding nilpotent orbits\footnote{Here we slightly abuse the notation, with the understanding that the codimension of a nilpotent orbit is defined as the codimension of its closure.} should satisfy
\begin{equation}
    \text{codim}_{\mathcal{H}^2}\left(\mathcal{H}^1\right)=\text{codim}_{\mathcal{O}^{L,1}}\left(\mathcal{O}^{L,2}\right)+\text{codim}_{\mathcal{O}^{R,1}}\left(\mathcal{O}^{R,2}\right),
\end{equation}
where $\text{codim}_A(B)=\dim_{\mathbb{C}}(A)-\dim_{\mathbb{C}}(B)$ and $\mathcal{H}^i$ denotes the Higgs branch of the theory associated with the orbit pair $\left(\mathcal{O}^{L,i},\mathcal{O}^{R,i}\right)$. Now, take the theory with $\mathcal{H}^1$ to be the one associated with the maximal orbit pair in the Lie algebra $\mathfrak{g}$. In other words, $\mathcal{O}^{L,1}$ and $\mathcal{O}^{R,1}$ are both the principal/regular orbits $\mathcal{O}_{\text{prin}}$. For such a theory, its atomic induced flow would always give the maximal orbit pair in the Levi subalgebra $\mathfrak{l}$. By directly computing the dimensions of the two Higgs branches, we find that
\begin{equation}
    \text{codim}_{\mathcal{H}^1_{\mathfrak{g}}}\left(\mathcal{H}^1_{\mathfrak{l}}\right)=\dim_{\mathbb{C}}\left(\mathcal{H}^1_{\mathfrak{g}}\right)-\dim_{\mathbb{C}}\left(\mathcal{H}^1_{\mathfrak{l}}\right)=2.\label{codim}
\end{equation}
In the language of \cite{Bao:2024eoq}, we found that 
\begin{tcolorbox}[colback=blue!10!white,breakable]
For $\kg' \times \mathfrak{gl}(1)$, a Levi subalgebra of $\kg$ (i.e., $\mathrm{rk}(\kg) - \mathrm{rk}(\kg') = 1$), 
a $(\kg, \kg)$ conformal matter theory with a pair of principal nilpotent orbits $\calO_L = \calO_R = \calO_{\kg, \mathrm{prin}}$ is always related to a $(\kg', \kg')$ conformal matter theory with $\calO_L = \calO_R = \calO_{\kg', \mathrm{prin}}$ by an \textbf{atomic combo flow}.
\end{tcolorbox}
\noindent Moreover, for induced orbits, we have
\begin{equation}
    \text{codim}_{\mathfrak{l}}\left(\mathcal{O}_{\mathfrak{l}}\right)=\text{codim}_{\mathfrak{g}}\left(\mathcal{O}_{\mathfrak{g}}\right),\label{codimprin1}
\end{equation}
where we have still used $\mathcal{O}_{\mathfrak{g}}$ to denote the nilpotent orbit in $\mathfrak{g}$ induced from $\mathcal{O}_{\mathfrak{l}}$ in the Levi subalgebra $\mathfrak{l}$. Since the complex dimension of the principal orbit is equal to the dimension of the Lie algebra minus the rank of the Lie algebra and since $\text{rank}(\mathfrak{g})=\text{rank}(\mathfrak{l})$, we have
\begin{equation}
    \text{codim}_{\mathcal{O}_{\mathfrak{l},\text{prin}}}(\mathcal{O}_{\mathfrak{l}})=\text{codim}_{\mathcal{O}_{\mathfrak{g},\text{prin}}}(\mathcal{O}_{\mathfrak{g}}).
\end{equation}
Therefore, following \eqref{codim}, we get
\begin{align}
    &\dim_{\mathbb{C}}\left(\mathcal{H}^2_{\mathfrak{g}}\right)-\dim_{\mathbb{C}}\left(\mathcal{H}^1_{\mathfrak{g}}\right)-\dim_{\mathbb{C}}\left(\mathcal{H}^2_{\mathfrak{l}}\right)+\dim_{\mathbb{C}}\left(\mathcal{H}^1_{\mathfrak{l}}\right)\nonumber\\
    =&\,\text{codim}_{\mathcal{O}^{L,2}_{\mathfrak{g}}}\left(\mathcal{O}^{L,2}_{\mathfrak{g}}\right)-\text{codim}_{\mathcal{O}^{L,2}_{\mathfrak{l}}}\left(\mathcal{O}^{L,2}_{\mathfrak{l}}\right)+\text{codim}_{\mathcal{O}^{R,1}_{\mathfrak{g}}}\left(\mathcal{O}^{R,2}_{\mathfrak{g}}\right)-\text{codim}_{\mathcal{O}^{R,1}_{\mathfrak{l}}}\left(\mathcal{O}^{R,2}_{\mathfrak{l}}\right)\nonumber\\
    =&\,0.
\end{align}
Recall that we have associated $\mathcal{H}^1$ with the pair of principal orbits. Due to \eqref{codimprin1}, we have
\begin{equation}
    \dim_{\mathbb{C}}\left(\mathcal{H}^2_{\mathfrak{g}}\right)-\dim_{\mathbb{C}}\left(\mathcal{H}^2_{\mathfrak{l}}\right)=2. \label{eqn:qdim_1}
\end{equation}
In other words, the transverse slice of an atomic induced flow is always of \emph{quaternionic} dimension 1.

Here, let us mention some specific nilpotent orbits and their physical implications:
\begin{itemize}
    \item The zero/trivial orbit, as the name suggests, is the smallest orbit 0 in the Lie algebra. It corresponds to the partition $\left[1^n\right]$ for the classical case, and the Bala-Carter (BC) is simply 0 (for both the classical and the exceptional cases). The corresponding theory would always have the trivial VEV on the zero orbit side. Therefore, if both sides have the zero orbits, this is often the starting UV theory.
    \item The principal/regular orbit is the largest nilpotent orbit in the Lie algebra. As a result, its closure is the nilpotent cone. The theory has the largest VEV turned on when the corresponding side is the principal orbit. Therefore, if both sides have the principal orbits, the theory often appears as the bottom IR theory in the nilpotent hierarchy.
    \item An orbit is called a Richardson orbit if it can be induced from some zero orbit. Therefore, a theory corresponds to the Richardson orbits can have an induced flow to a theory without any VEVs in a smaller algebra (namely, the Levi subalgebra).
    \item When an orbit cannot be induced from any other orbits, it is called a rigid orbit. Therefore, a theory associated with a rigid orbit does not admit any induced flows.
\end{itemize}

Before proceeding to more specific discussions, we want to comment on the relation between our physical setup and that in \cite{Chacaltana:2012zy}, an earlier instance where the notion of the induced nilpotent orbits was introduced to study strongly coupled QFTs in the high energy theory literature. The physical setup in \cite{Chacaltana:2012zy} is class $\mathcal{S}$ theories, where a nilpotent orbit labels a puncture on the Riemann surface involved in the 6d-to-4d compactification. The class $\mathcal{S}$ theory is then compactified on an $S^1$ to give a 3d $\calN = 4$ theory $T^\rho[\kg]$, which is not weakly coupled in general. There, the inductions of the nilpotent orbits are used to track the puncture data under mass deformations as a 3d Coulomb branch operation.

\subsection{Warm-Up: A-Type Theories}\label{Atype}
To motivate our discussions, we will start by a special family of cases with an $A_k$-type (2,0) theory base with $\ksu(n_i)$ paired gauge symmetry. The flavour number $N^f_i$ for each tensor multiplet is subject to the condition
\be
2n_{i} = n_{i-1} + n_{i+1} + N^f_i.
\ee
When $n_i$ are the same for all $i$ (say $n$), this theory admits an engineering of $n$ NS5-branes probing a $\bbC^2/\bbZ_{k+1}$ singularity. A general case where $n_i$ are not necessarily the same can also be understood based on the previous construction with an extra pair of T-brane deformation $\left(\calO^L, \calO^R\right)$ \cite{Cecotti:2010bp,Anderson:2013rka,Heckman:2016ssk}. Specifically, the theory (where we have left the flavour symmetries implicit)
\be
    \overset{\ksu(n_1)}{2} \ \ \overset{\ksu(n_2)}{2} \ \ \dots \ \  \overset{\ksu(n_{k-1})}{2} \ \ \overset{\ksu(n_{k})}{2} \ \
\ee
corresponds to the partitions $\bm{p}^{L,R} $ of an integer $N$ such that we have a plateau of maximal gauge symmetries $\mathfrak{su}(N)$ in the middle. The conjugate partitions are given by their transposes:
\begin{equation}
    \left(\bm{p}^L\right)^{\text{T}}=\left[\left(p^L_1\right)^{\text{T}},\dots,\left(p^L_{k_L}\right)^{\text{T}}\right],\quad\left(\bm{p}^R\right)^{\text{T}}=\left[\left(p^R_1\right)^{\text{T}},\dots,\left(p^R_{k_R}\right)^{\text{T}}\right].
\end{equation}
Then to the left of the plateau that contains $n_L$ tensors, we have
\be
    n_i = \sum_{j = 1}^{i} \left(p^L_j\right)^{\text{T}},
\ee
and likewise for the gauge ranks for the tensors to the right of the plateau.

Suppose that $\mathcal{O}^L$ is induced from $\mathcal{O}^L_{\mathfrak{l}}=\mathcal{O}_{L,1}\oplus\mathcal{O}_{L,2}$ (and likewise for $\mathcal{O}^R$). Then we can have the following atomic induced flows:
\begin{equation}\mathcal{T}\left(\mathcal{O}^L,\mathcal{O}^R\right)\rightarrow\mathcal{T}\left(\mathcal{O}_{L,1},\mathcal{O}_{R,1}\right)\sqcup\mathcal{T}\left(\mathcal{O}_{L,2},\mathcal{O}_{R,2}\right).
\end{equation}
The total number of curves/tensor multiplets is reduced by 1 after such an atomic induced flow. If $N$ is even and the Levi subalgebra is $\mathfrak{l}=\mathfrak{sl}(N/2)\oplus\mathfrak{sl}(N/2)$, there could be another possible set of atomic induced flows:
\begin{equation}
    \mathcal{T}\left(\mathcal{O}^L,\mathcal{O}^R\right)\rightarrow\mathcal{T}\left(\mathcal{O}_{L,1},\mathcal{O}_{R,2}\right)\sqcup\mathcal{T}\left(\mathcal{O}_{L,2},\mathcal{O}_{R,1}\right).
\end{equation}
In general, one can take a Levi subalgebra which is a direct sum of more than two simple Lie algebras. However, this would correspond to multiple steps of atomic Higgsings. For an atomic flow, there can be at most two components in the Levi subalgebra. In terms of the Dynkin diagram of $\mathfrak{sl}(N)$, any choice of a possible Levi subalgebra with an atomic induced flow is given by removing a node in the Dynkin diagram. Therefore, one can also remove the node on one of the two ends. The IR theory under this flow would only have one component\footnote{The $\mathfrak{gl}(1)$ part in the Levi subalgebra gives the trivial contribution with a smooth part generated by the free hypers.}. In other words, $\mathcal{O}^L_{\mathfrak{l}}=\mathcal{O}_{L,1}$ and $\mathcal{O}^R_{\mathfrak{l}}=\mathcal{O}_{R,1}$. In this case, the total number of curves/tensor multiplets is not changed.

\paragraph{Example} Let us illustrate this with an example. Consider the theory
\begin{equation}
    \underbrace{\underset{[N_f=1]}{\overset{\ksu(3)}{2}} \ \ \underset{[N_f=1]}{\overset{\ksu(5)}{2}} \ \ \underset{[N_f=1]}{\overset{\ksu(6)}{2}} \ \ \overset{\ksu(6)}{2} \ \ \dots \ \ \overset{\ksu(6)}{2} \ \ \underset{[N_f=1]}{\overset{\ksu(6)}{2}} \ \ \underset{[N_f=1]}{\overset{\ksu(5)}{2}} \ \ \underset{[N_f=1]}{\overset{\ksu(3)}{2}}}_{L}.\label{Atypeexample}
\end{equation}
This corresponds to the pair of nilpotent orbits $([3,2,1],[3,2,1])$ in $\mathfrak{sl}(6)$. The transpose $\bm{r}^{\text{T}}$ of $\bm{r}=[3,2,1]$ is still $[3,2,1]$. There are three possibilities of splitting $\bm{r}^{\text{T}}=\bm{s}^{\text{T}}\sqcup\bm{t}^{\text{T}}$:
\begin{itemize}
    \item When $\bm{s}^{\text{T}}=[3]$ and $\bm{t}^{\text{T}}=[2,1]$, we have $\bm{s}=[1^3]$ and $\bm{t}=[2,1]$. Therefore, we have the following induced flows:
    \begin{align}
        &\eqref{Atypeexample}\rightarrow\underbrace{\underset{[\SU(3)]}{\overset{\ksu(3)}{2}} \ \ \overset{\ksu(3)}{2} \ \ \dots \ \ \overset{\ksu(3)}{2} \ \ \underset{[\SU(3)]}{\overset{\ksu(3)}{2}}}_{K} \ \ \sqcup \ \ \underbrace{\underset{[N_f=1]}{\overset{\ksu(2)}{2}} \ \ \underset{[N_f=1]}{\overset{\ksu(3)}{2}} \ \ \overset{\ksu(3)}{2} \ \ \dots \ \ \overset{\ksu(3)}{2} \ \ \underset{[N_f=1]}{\overset{\ksu(3)}{2}} \ \ \underset{[N_f=1]}{\overset{\ksu(2)}{2}}}_{L-K-1},\\
        &\eqref{Atypeexample}\rightarrow\underbrace{\underset{[\SU(3)]}{\overset{\ksu(3)}{2}} \ \ \overset{\ksu(3)}{2} \ \ \dots \ \ \overset{\ksu(3)}{2} \ \ \underset{[N_f=1]}{\overset{\ksu(3)}{2}} \ \ \underset{[N_f=1]}{\overset{\ksu(2)}{2}}}_{K} \ \ \sqcup \ \ \underbrace{\underset{[N_f=1]}{\overset{\ksu(2)}{2}} \ \ \underset{[N_f=1]}{\overset{\ksu(3)}{2}} \ \ \overset{\ksu(3)}{2} \ \ \dots \ \ \overset{\ksu(3)}{2} \ \ \underset{[\SU(3)]}{\overset{\ksu(3)}{2}}}_{L-K-1}.
    \end{align}
    \item When $\bm{s}^{\text{T}}=[3,1]$ and $\bm{t}^{\text{T}}=[2]$, we have $\bm{s}=[2,1^2]$ and $\bm{t}=[1^2]$. Therefore, we have the following induced flows:
    \begin{equation}
        \eqref{Atypeexample}\rightarrow\underbrace{\underset{[\SU(2)]}{\overset{\ksu(3)}{2}} \ \ \underset{[N_f=1]}{\overset{\ksu(4)}{2}} \ \ \overset{\ksu(4)}{2} \ \ \dots \ \ \overset{\ksu(4)}{2} \ \ \underset{[N_f=1]}{\overset{\ksu(4)}{2}} \ \ \underset{[\SU(2)]}{\overset{\ksu(3)}{2}}}_{K} \ \ \sqcup \ \ \underbrace{\underset{[\SU(2)]}{\overset{\ksu(2)}{2}} \ \ \overset{\ksu(2)}{2} \ \ \dots \ \ \overset{\ksu(2)}{2} \ \ \underset{[\SU(2)]}{\overset{\ksu(2)}{2}}}_{L-K-1}.
    \end{equation}
    \item When $\bm{s}^{\text{T}}=[3,2]$ and $\bm{t}^{\text{T}}=[1]$, we have $\bm{s}=[2^2,1]$ and $\bm{t}=[1]$. This means that the Levi subalgebra is taken to be $\mathfrak{l}=\mathfrak{sl}(5)$. Therefore, we have the following induced flows:
    \begin{equation}
        \eqref{Atypeexample}\rightarrow\underbrace{\underset{[N_f=1]}{\overset{\ksu(3)}{2}} \ \ \underset{[\SU(2)]}{\overset{\ksu(5)}{2}} \ \ \overset{\ksu(5)}{2} \ \ \dots \ \ \overset{\ksu(5)}{2} \ \ \underset{[\SU(2)]}{\overset{\ksu(5)}{2}} \ \ \underset{[N_f=1]}{\overset{\ksu(3)}{2}}}_{L}.
    \end{equation}
\end{itemize}

\subsubsection{A-Type Induced Flows via Unitary Magnetic Quivers}\label{AtypeMQ}
In terms of the magnetic quivers, an atomic induced flow could either correspond to a decay or correspond to a fission. With the help of the magnetic quivers, we can also read off the elementary transverse slices for the induced flows. In most cases, the Levi subalgebra has two components. This would give the fission of the magnetic quiver. The transverse slice would be the Kleinian singularity $A_1$ (resp.~the non-normal slice $m$) when the two magnetic quivers after fission are the same (resp.~different) \cite{Bourget:2023dkj,Bourget:2024mgn}.

If the Levi subalgebra is $\mathfrak{sl}(N-1)$, then this corresponds to the decay of the magnetic quiver. The transverse slice would be the Kleinian singularity $A_{b+1}$. Here, $b$ is the balance of the intersecting node of the three legs in the magnetic quiver. If the two nilpotent orbits are $\bm{p}=[p_1^{n_1},\dots]$ and $\bm{q}=[q_1^{m_1},\dots]$, then the balance $b$ is given by
\begin{equation}
    b=(N-p_1)+(N-q_1)+(L+1)-2N=L+1-p_1-q_1,
\end{equation}
where $L$ is the total number of curves. In other words, the transverse slice is $A_{L+2-p_1-q_1}$.

\paragraph{Example} As an example, the magnetic quiver for the theory in \eqref{Atypeexample} is
\begin{equation}
    \includegraphics[width=3cm]{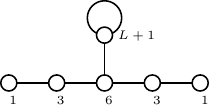}.\label{AtypeexampleMQ}
\end{equation}
The possible atomic induced flows are listed in Table \ref{AtypeexampleMQtable}.
\begin{longtable}{|c|c|c|c|}
\hline
$\mathfrak{s}$ & $\left(\mathcal{O}^L_{\mathfrak{{s}}},\mathcal{O}^R_{\mathfrak{s}}\right)$ & Magnetic quiver & Slice \\ \hline
$2A_2$ & $\left([1^3],[1^3]\right)\&\left([2,1],[2,1]\right)$ & \includegraphics[width=5cm]{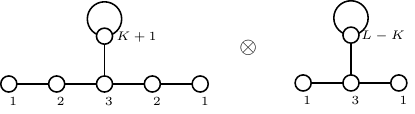} & $m$ \\ \hline
$2A_2$ & $\left([1^3],[2,1]\right)\&\left([2,1],[1^3]\right)$ & \includegraphics[width=5cm]{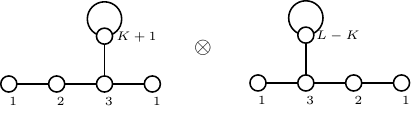} & $\begin{cases}
    A_1,&K=\frac{L-1}{2}\\
    m,&\text{otherwise}
\end{cases}$ \\ \hline
$A_3+A_1$ & $\left([2,1^2],[2,1^2]\right)\&\left([1^2],[1^2]\right)$ & \includegraphics[width=5cm]{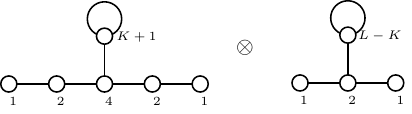} & $m$ \\ \hline
$A_5$ & $\left([2^2,1]\right)$ & \includegraphics[width=3cm]{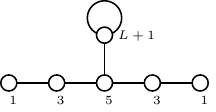} & $A_{2L-3}$ \\ \hline
\caption{The atomic induced flows for the theory in \eqref{Atypeexample} whose magnetic quiver is given in \eqref{AtypeexampleMQ}. We list the magnetic quivers for the IR theories and the transverse slices, along with the associated Levi subalgebras and nilpotent orbit pairs.}\label{AtypeexampleMQtable}
\end{longtable}

\subsubsection{A-Type Induced Flows via D7-Branes}\label{Atype7brane}
Following \cite{DeWolfe:1998zf,Hassler:2019eso}, it could also be helpful to illustrate the procedure of an induced flow using the 7-branes. Instead of Higgsing a UV theory, let us take the reverse. We shall consider a pair of child theories (or a single child theory in certain cases) that would compose the full IR theory and think of the UV theory that could produce the IR theory via an induced flow. Of course, there will be a caveat we need to be careful about when taking this reversed process, which we will comment on shortly.

In general, consider two theories with bases of type $A_{m_i}$ $(2,0)$ theory and fibres $\SU(k_i)$, where $i = 1, 2$. On their tensor branches, we can view the fibre configuration on each curve as $k_i$ 7-branes without any VEV. In other words, we have the zero orbits $\left[1^{k_i}\right]$ in $\mathfrak{sl}(k_i)$.

Starting with some IR theory, the choice of the candidate UV theory is not unique. Here, we pick the combination with base type $A_{m_1 + m_2 + 1}$ (as can be understood from the deformation of the Kleinian singularity or the dual M5-brane counting) and gauge symmetry
$\SU(k_1 + k_2)$ (as from the 7-brane counting). In particular, we take the zero orbit $\left[1^{k_1 + k_2}\right]$ in $\mathfrak{sl}(k_1 + k_2)$. Indeed, the stack of $k_1 + k_2$ 7-branes can be separated into a stack of $k_1$ 7-branes and another stack of $k_2$ 7-branes. Each stack would coincide with one of the two singularities of type $A_{m_i}$ that comes from the complex structure deformation. However, we should be careful as such a flow is \textit{not} atomic.

\begin{figure}[h]
    \centering
    \includegraphics[width=15cm]{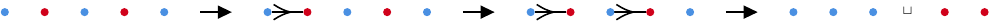}
    \caption{An illustration of the 7-brane picture. We have used different colours to indicate the stacks the 7-branes belong to in the IR theory. The flow from the leftmost one with $[1^5]$ to the rightmost one with $[1^3]\oplus[1^2]$ is not atomic. The atomic Higgsings would go through $[2,1^3]$ and then $[2^2,1]$ before reaching the atomic induced flow. See \cite{Hassler:2019eso} for more explanations on the notation that describes nilpotent VEVs by 7-branes and string junctions.}\label{7branesAtypezeroorbits}
\end{figure}

\begin{figure}[h]
    \centering
    \includegraphics[width=15cm]{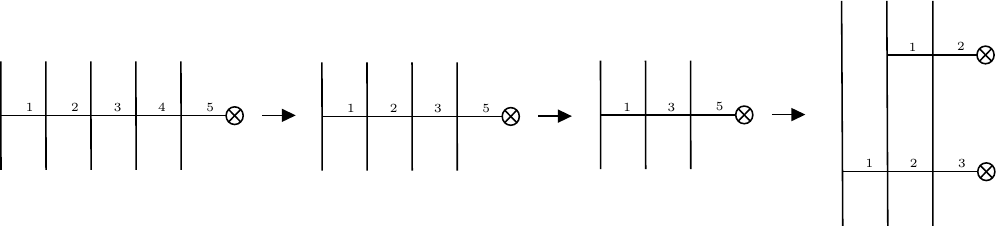}
    \caption{A Type IIA brane picture illustration of the same atomic Higgsing example. The horizontal and vertical lines denote the D6- and D8-branes respectively, and the crosses are the NS5-branes. We have omitted any possible configurations right to the NS5-brane in each brane system. An atomic Higgsing corresponds to separating the whole configuration into two pieces along a direction within the D8s but perpendicular to the D6s.}\label{IIAillustration}
\end{figure}

The reason is as follows. It is possible to find an intermediate theory by turning on some non-trivial nilpotent VEV in the UV theory as a nilpotent orbit inside $\mathfrak{sl}(k_1 + k_2)$ before reducing to the trivial orbits in $\mathfrak{sl}(k_1)\oplus\mathfrak{sl}(k_2)$. There is a natural candidate. We can turn on some VEV parametrized by some $2\times2$ Jordan blocks, each of which can be described by an open string connecting a pair of D7-branes (with no pair of strings sharing an endpoint) following \cite{Hassler:2019eso}. Then we could separate each pair of such D7-branes into the two stacks when performing the flow. Therefore, the VEV is not visible in the IR theory. We can at most have $\min(k_1,k_2)$ of such string junctions, i.e., a VEV of $\left[2^{k_1}, 1^{k_2 - k_1}\right]$. The Higgs branch would indeed change by quaternionic dimension 1. An example is illustrated in Figure \ref{7branesAtypezeroorbits}. The Type IIA picture can be found in Figure \ref{IIAillustration}.

\begin{figure}[h]
    \centering
    \includegraphics[width=10cm]{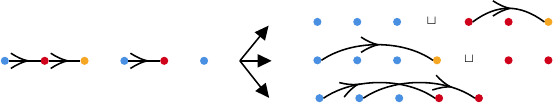}
    \caption{In the starting orbit $[3,2,1]$, we have three buckets with 7-branes of numbers 3, 2 and 1 respectively. The first (second, resp.~third) 7-brane in each bucket is coloured blue (red, resp.~orange). In the top case, the blue ones are in the same component while the red and orange ones are in the other component. As there was a red 7-brane and an orange 7-brane in the same bucket, we connect them with an open string. Therefore, we have the orbit $[1^3]\oplus[2,1]$. In the middle case, for the first component, there was a blue 7-brane and an orange 7-brane in the same bucket, so we connect them with an open string as the red one is now in the other component. This yields the orbit $[2,1^2]\oplus[1^2]$. In the bottom case, the two open strings again come from connecting the 7-branes that were in the same buckets. We have omitted the component with only one single 7-brane in the picture. The resulting orbit is $[2^2,1]$.}\label{7branesAtypeexample}
\end{figure}

More generally, given a 7-brane configuration of an A-type orbit, we can get the induction data, namely the corresponding orbits under the induced flows as follows. For simplicity, we say that the 7-branes connected by open strings live in the same bucket. Then for the $i^{\text{th}}$ 7-brane in each bucket, we paint it with the colour $i$. Now, to get the atomic induced flows, we always separate the 7-branes into at most two components. The 7-branes (in different buckets) with the same colour should always belong to the same component. In each component, the 7-branes which were in the same bucket are connected by open strings one after another, following the order the colours $i$. Notice that if a 7-brane of colour $j$ and a 7-brane of colour $j+k$ that were in the same bucket are now in the same component but all the 7-branes of colours $j+1,\dots,j+k-1$ previously in the same bucket are not in this component, we should also connect the 7-branes of colours $j$ and $j+k$ with an open string. This procedure directly follows from the previous discussions on splitting the partitions via their transposes. The last step in Figure \ref{7branesAtypezeroorbits} provides such as example. We also illustrate this with a further example of the orbits appeared in the induced flows for the theory \eqref{Atypeexample} in Figure \ref{7branesAtypeexample}.

The 7-brane picture gives a nice interpretation of the A-type nilpotent orbits and their induction data. Nevertheless, a complete understanding via the 7-branes for the full theories under the induced flows similar to the analysis in \cite{Hassler:2019eso} would still require future investigations.

\subsubsection{Plateau Higgsings vs Endpoint-Changing Flows }\label{plateauvsendpointchanging}
Before delving into the general cases other than the A-types, let us make the best possible use of the simplest A-type cases by emphasizing the distinction between two types of the atomic flows. Their similarity is that we are decomposing the gauge group on the tensor branch, but the difference is whether we change the Dirac pairing at the same time.

\paragraph{The plateau Higgsing} This is the one where we do not change the Dirac pairing. An atomic plateau Higgsing would look like
\begin{align}
\begin{split}
    &[\SU(k-1)] \ \ \overset{\ksu(k)}{2} \ \ \underset{[N_f = 1]}{\overset{\ksu(k+1)}{2}} \ \ \overset{\ksu(k+1)}{2} \ \ \dots \ \ \overset{\ksu(k+1)}{2} \ \ \underset{[N_f = 1]}{\overset{\ksu(k+1)}{2}} \ \ \overset{\ksu(k)}{2} \ \ [\SU(k-1)]   \\
    \longrightarrow & [\SU(k)] \ \ \overset{\ksu(k)}{2} \ \ \overset{\ksu(k)}{2} \ \ \overset{\ksu(k)}{2} \ \ \dots \ \ \overset{\ksu(k)}{2} \ \ \overset{\ksu(k)}{2} \ \ \overset{\ksu(k)}{2} \ \ [\SU(k)].
\end{split}
\end{align}
In particular, the UV theory is associated with the orbits $\left([2,1^{k-1}],[2,1^{k-1}]\right)$. After the induced flow, we get the zero orbits $\left([1^k,1^k]\right)$. Such Higgsings can be thought of as giving VEVs to the delocalized $\U(1)$ flavour symmetries that span across all the $-2$ curves with $\ksu(k+1)$ gauge symmetries as in \cite{Apruzzi:2020eqi}, which we leave implicit in our tensor branch descriptions.

One can understand this induced flow as getting a reducible IR theory consisting of one strongly coupled SCFT, and one with only a collection of hypermultiplets that is weakly coupled. This corresponds to the $\mathfrak{gl}(1)$ part in the Levi subalgebra. The $2k-2$ weakly coupled (or effectively free) hypers generate the moduli space $\bbH^{2k-2}$ (up to some discrete factors). It is conventional that the dimension counting excludes the smooth part, and the transverse slice is the Kleinian singularity as stated above.

\paragraph{The endpoint-changing flow} In contrary, if we also perform a non-trivial complex structure deformation of the F-theory base (i.e., changing the Dirac pairing) accompanying the change of the gauge symmetries, we would get an endpoint-changing transition. Some simple examples were given in \S\ref{Atype}. As a more complicated example, we have
\be
    \underset{[N_f = 1]}{\overset{\ksu(3)}{2}} \ \ \overset{\ksu(5)}{2} \ \ \underset{[N_f = 1]}{\overset{\ksu(7)}{2}} \ \ \underset{[\SU(2)]}{\overset{\ksu(8)}{2}} \ \ \underset{[N_f = 1]}{\overset{\ksu(7)}{2}} \ \ \overset{\ksu(5)}{2} \ \ \underset{[N_f = 1]}{\overset{\ksu(3)}{2}} \ \  \longrightarrow \ \ \overset{\ksu(2)}{2} \ \ \underset{[SU(4)]}{\overset{\ksu(4)}{2}} \ \ \overset{\ksu(2)}{2} \ \ \sqcup  \underset{[\SU(2)]}{\overset{\ksu(3)}{2}} \ \ \underset{[\SU(2)]}{\overset{\ksu(4)}{2}} \ \ \underset{[\SU(2)]}{\overset{\ksu(3)}{2}}.
\ee
The orbit $[4, 3, 1]$ in $\mathfrak{sl}(8)$ on either of the two sides is induced from $[2^2]\oplus[2,1^2]$ in $\mathfrak{sl}(4)\oplus\mathfrak{sl}(2)$. The transverse slice is $m$ here. In general, the slice would be either $A_1$ or $m$.

% \subsubsection{7-brane Picture for Richardson Orbits}

\subsection{D-Type Conformal Matter Theories}\label{Dtype}
In this section, we shall handle the case where the $D_{m+k}$ conformal matter theory turns into a $D_m$ conformal matter theory and an A-type theory with gauge algebra $\mathfrak{su}(k)$ (tautologically known as the ``$A_{k-1}$ conformal matter theory''). The number of tensor multiplets can be obtained by demanding the conservation of the total number of (half) M5-branes. By examining the long quiver theories and focusing on one end at a time, we have two cases to consider. One is the integer M5 case (described by O$6^-$ in Type IIA) where the unHiggsed flavour symmetry is decomposed from $\SO(2m+2k)$ to $\SO(2m) \times \SU(k)$. The other is the half-integer M5 case (described by O$6^+$ in Type IIA) where the flavour symmetry is decomposed from $\Sp(m+k-4)$ to $\Sp(m-4) \times \SU(k)$.

In both cases, the induction of orbits can be obtained by the following algorithm based on \cite[Theorem 7.3.3]{collingwood1993nilpotent} (which is also reviewed in Appendix \ref{indorb}). Denote by $\calO_{\bm{d}}$ an orbit in $\mathfrak{sl}(l)$ and $\calO_{\bm{f}}$ an orbit in $\mathfrak{g}'$ of BCD-type. Then they would induce an orbit in the Lie algebra in $\mathfrak{g}$ which is of the same type as $\mathfrak{g}'$. Define the partition $\bm{p}=[p_1,\dots,p_N]$ such that
\begin{equation}
    p_i=2d_i+f_i.
\end{equation}
Then the nilpotent orbit in $\mathfrak{g}$ induced from $\mathcal{O}_{\bm{d}}\oplus\mathcal{O}_{\bm{f}}$ is given by the partition $\bm{p}_{\text{X}}$, where the subscript X denotes the X-collapse of $\bm{p}$ (with $\text{X}=\text{B},\text{C},\text{D}$). See Appendix \ref{indorb} for more details.

Now, starting from a UV theory associated with a pair of orbits $\left(\mathcal{O}^L,\mathcal{O}^R\right)$. It can be Higgsed to an IR theory whose components are given by the induction data $\left(\mathcal{O}^L_{\mathfrak{sl}(l)},\mathcal{O}^R_{\mathfrak{sl}(l)}\right)$ and $\left(\mathcal{O}^L_{\mathfrak{g}'},\mathcal{O}^R_{\mathfrak{g}'}\right)$. In terms of the Dynkin diagram of $\mathfrak{g}$, this can again be understood as removing one node from it.

\textbf{A D-type endpoint-changing flow} Here, we recall an example that was brought up in our previous paper \cite{Bao:2024eoq}. This is the $(D_6, D_6)$ conformal matter theory with a pair of nilpotent VEVs $\left([5,3,1^4],[3^2,1^6]\right)_{\mathfrak{so}(12)}$:
\begin{equation}
    [\SU(2)\times\SU(2)] \ \ \overset{\kso(8)}{3} \ \ \overset{\ksp(1)}{1} \ \ \overset{\kso(11)}{4} \ \ \overset{\ksp(2)}{1} \ \ \overset{\kso(12)}{4} \ \ \dots \ \ \overset{\kso(12)}{4} \ \ \overset{\ksp(2)}{1} \ \ \overset{\kso(10)}{4} \ \ 1 \ \ [\SO(6)],
\end{equation}
Then we have an atomic flow to the following IR theory:
\begin{align}
    \begin{split}
    &[\SU(2)\times\SU(2)\times\SU(2)] \ \ \overset{\kso(8)}{3} \ \ 1 \ \ \overset{\kso(8)}{4} \ \ 1 \ \ \overset{\kso(8)}{4} \ \ \dots \ \ \overset{\kso(8)}{4} \ \ 1 \ \ \overset{\kso(8)}{4} \ \ 1 \ \ [\SO(8)]\\
    \sqcup& \ \ 2 \ \ \underset{[N_f=1]}{\overset{\ksu(2)}{2}} \ \ \overset{\ksu(2)}{2} \ \ \overset{\ksu(2)}{2} \ \ \dots \ \ \overset{\ksu(2)}{2} \ \ \overset{\ksu(2)}{2} \ \ [\SU(2)].
    \end{split}
\end{align}
The two components are given by the pairs of nilpotent VEVs $\left([2^2,1^4],[1^8]\right)_{\mathfrak{so}(8)}$ and $\left([2],[1^2]\right)_{\mathfrak{sl}(2)}$ respectively.

\textbf{A D-type combo flow} We now consider another D-type example which does not involve splitting the M5-branes, but we only perform a complex structure deformation on the Kleinian singularity of type $D_4$ to that of type $A_3$. We also pick the trivial nilpotent orbits $\left([1^4], [1^4]\right)_{\ksu(4)}$ for the IR theory, which induces the nilpotent orbits of $\left([2^4], [2^4]\right)_{\kso(8)}$ in the UV theory. Take the number of M5-branes to be 5, we obtain the following RG flow\footnote{Analogous to the delocalized $\U(1)$ symmetries studied in \cite{Apruzzi:2020eqi}, we are tempted to identify this flow as triggered by a VEV of the $\bbZ_2$ delocalized discrete symmetry.}:
\begin{equation}
    [\Sp(1)\times\Sp(1)] \ \ \overset{\kso(7)}{3} \ \ 1 \ \ \overset{\kso(8)}{4} \ \ 1 \ \ \overset{\kso(8)}{4} \ \ 1 \ \ \overset{\kso(7)}{3} \ \ 1 \ \ [\Sp(1)\times\Sp(1)] \quad \longrightarrow [\SU(4)] \ \ \overset{\ksu(4)}{2} \ \ \overset{\ksu(4)}{2} \ \ \overset{\ksu(4)}{2} \ \ [\SU(4)]
\end{equation}
This flow represents a general phenomenon: an atomic flow that does not change the endpoint configuration (that is, neither between two A-type Kleinian singularities nor between two D-type Kleinian singularities) reproduces a combo flow, which was introduced in our first paper \cite{Bao:2024eoq} only by the F-theory consideration.

Here, we have omitted the labels I and II for the very even nilpotent orbits for simplicity. How they can be distinguished under inductions are reviewed in Appendix \ref{indorb}. As pointed out in \cite{Distler:2022yse}, the Higgs branch of the conformal matter theory with the pair $\left(\mathcal{O}^{\text{I}},\mathcal{O}^{\text{I}}\right)$ or $\left(\mathcal{O}^{\text{II}},\mathcal{O}^{\text{II}}\right)$ is different from the one with $\left(\mathcal{O}^{\text{I}},\mathcal{O}^{\text{II}}\right)$. As one of them would have more generators, it is tempting to conjecture that both of the cases would have the induced flows while they would differ by some other (non-induced) flows.

\textbf{A C-type example} We may also consider a C-type example by considering a generalized quiver with an end being a $-4$-curve. The flavour symmetry is thus of the C-type. It is sufficient that we analyze one end at a time, so the analysis applies no matter whether both ends have the C-type flavour symmetries, or only one end has a C-type flavour symmetry while the other end has an A-type flavour symmetry. For instance, consider the following ``$(C_4, C_4)$ conformal matter'' with VEVs $\left([3^2, 1^2],[3^2, 1^2]\right)_{\mathfrak{sp}(4)}$:
\begin{align}
    &[\Sp(1)] \ \ \overset{\kso(12)}{4} \ \ \overset{\ksp(3)}{1} \ \ \overset{\kso(16)}{4} \ \  \overset{\ksp(4)}{1} \ \ \overset{\kso(16)}{4} \dots \ \ \overset{\kso(16)}{4} \ \  \overset{\ksp(4)}{1} \ \  \overset{\kso(16)}{4} \ \ \overset{\ksp(3)}{1} \ \ \overset{\kso(12)}{4} \ \ [\Sp(1)].
\end{align}
Then we have an atomic flow to the following IR theory:
\begin{align}
    \begin{split}
    &[\Sp(2)] \ \ \overset{\kso(12)}{4} \ \ \overset{\ksp(2)}{1} \ \ \overset{\kso(12)}{4} \ \ \dots \ \ \overset{\kso(12)}{4} \ \ \overset{\ksp(2)}{1} \ \ \overset{\kso(12)}{4} \ \ [\Sp(2)]\\
    \sqcup& \ \ [\SU(2)] \ \ \overset{\ksu(2)}{2} \ \ {\overset{\ksu(2)}{2}} \ \ \overset{\ksu(2)}{2} \ \ \overset{\ksu(2)}{2} \ \ \dots \ \ \overset{\ksu(2)}{2} \ \ \overset{\ksu(2)}{2} \ \ [\SU(2)].% \times \SU(2)_{deloc}$
    \end{split}
\end{align}
Both of them are given by the zero orbits, that is, $\left([1^4],[1^4]\right)_{\mathfrak{sp}(2)}$ and $\left([1^2],[1^2]\right)_{\mathfrak{sl}(2)}$.

\subsubsection{D-Type Induced Flows via Orthosymplectic Magnetic Quivers}\label{DtypeMQ}
For D-type conformal matter theories, it is also possible to obtain the magnetic quivers (for those associated with the special orbits) via the Type IIA brane constructions \cite{Mekareeya:2016yal,Hanany:2022itc}. This would possibly involve negatively charged branes. After the Hanany-Witten transitions, one can read off the orthosymplectic magnetic quivers.

The induced flows should then also correspond to the decays and fissions of the orthosymplectic magnetic quivers as in \cite{Bao:2024eoq,Lawrie:2024wan}. This is similar to the A-type cases as discussed in \S\ref{AtypeMQ}. The transverse slice would still be of quaternionic dimension 1 (see the paragraph of general derivation leading to \eqref{eqn:qdim_1}) although what the singularity would be is not clear. For the quiver fission, the resulting IR theory would \emph{generically} have one orthosymplectic quiver and one unitary quiver. For the quiver decay\footnote{In \cite{Lawrie:2024wan}, such Higgsings are classified as a special type of the quiver fission, dubbed unitarizations. Since the same type of the induced Higgsings appears in the unitary quiver decay as discussed above and since there is still one single component in the descendant theory, we shall simply refer to it as the quiver decay to ease our terminology. Nevertheless, the manipulations on the magnetic quivers from the induced Higgsings should be clear regardless of the terminologies.}, we find that given a UV orthosymplectic magnetic quiver $\mathcal{Q}_{\text{osp}}$, the atomic induced flow simply takes it to a unitary quiver $\mathcal{Q}_{\text{u}}$ with all the orthosymplectic nodes changed to the unitary ones, but with the ranks remaining the same.

This induced flow would then require further verification due to the following reason. In terms of quiver subtractions, as the ranks of the nodes in $\mathcal{Q}_{\text{osp}}$ and $\mathcal{Q}_{\text{u}}$ are completely the same, it is tempting to subtract both quivers by the same unitary magnetic quiver slices (which is well-established for $\mathcal{Q}_{\text{u}}$). The magnetic quiver $\mathcal{Q}_{\text{osp}}$ (resp.~$\mathcal{Q}_{\text{u}}$) would eventually be reduced to a single $\mathfrak{c}_1$ (resp.~$\U(1)$) node. Therefore, it could be possible that the Coulomb branches of $\mathcal{Q}_{\text{osp}}$ and $\mathcal{Q}_{\text{u}}$ would only differ by a smooth part. However, we conjecture that this is not the case. In other words, there is indeed a Higgsing from $\mathcal{Q}_{\text{osp}}$ to $\mathcal{Q}_{\text{u}}$. This is supported by the following example. In fact, very recently, the validity of such Higgsings was also argued in \cite{Lawrie:2024wan}, and such processes were called unitarizations therein.

\paragraph{Example} Let us consider the magnetic quiver\footnote{The orbits and the corresponding building legs in the magnetic quivers also appear in \cite{Lawrie:2024wan} in the examples of the class $\mathcal{S}$ theories. This indicates that such legs would satisfy the desired decay process. Nevertheless, for completeness of our argument, we shall further check this in the ``conformal matter'' examples using the Hilbert series.}
\begin{equation}
    \includegraphics[width=3cm]{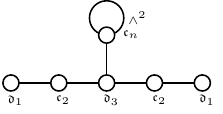}.\label{D32_21_2MQ}
\end{equation}
This is the formal $(D_3,D_3)$ conformal matter theory\footnote{This is formal in the sense that there would be gauge algebras with negative ranks. Nevertheless, we could still study the Coulomb branch of this magnetic quiver, which should be well-defined from the checks here.} with the orbit pair $\left([2^2,1^2],[2^2,1^2]\right)_{\mathfrak{so}(6)}$. In fact, $[2^2,1^2]_{\mathfrak{so}(6)}$ is the D-collapse of the partition $[2^3]$. Our claim is that there is an atomic induced Higgsing to the magnetic quiver
\begin{equation}
    \includegraphics[width=3cm]{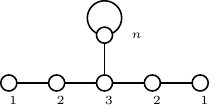},\label{A31_3MQ}
\end{equation}
which is the A-type theory with the orbit pair $\left([1^3],[1^3]\right)_{\mathfrak{sl}(3)}$. As we can see, the orthosymplectic nodes are changed to the unitary ones while the ranks remain the same. Now, we would like to verify that this is indeed a valid Higgsing.

Since $\mathfrak{so}(6)\cong\mathfrak{sl}(4)$, we can see that the orbit $[2^2,1^2]_{\mathfrak{so}(6)}$ is the same as the orbit $[2,1^2]_{\mathfrak{sl}(4)}$. Therefore, it is natural to expect that the magnetic quiver in \eqref{D32_21_2MQ} has the same Coulomb branch as
\begin{equation}
    \includegraphics[width=3cm]{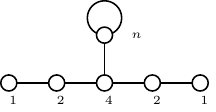}\label{A421_2MQ}
\end{equation}
does. This unitary magnetic quiver is of A-type with the orbit pair $\left([2,1^2],[2,1^2]\right)_{\mathfrak{sl}(4)}$. One way to check that the two Coulomb branches do coincide is to compute their Hilbert series\footnote{There are also various cases studied in \cite{Nawata:2021nse} where the unitary and orthosymplectic magnetic quivers have such an IR duality.}. For instance, for the unitary magnetic quiver, we have
\begin{align}
    n=4:&\quad\text{HS}=1+11t^2+16t^3+82t^4+184t^5+612t^6+1408t^7+3970t^8+9104t^9+22775t^{10}+\dots,\\
    n=5:&\quad\text{HS}=1 + 9 t^2 + 10 t^3 + 57 t^4 + 102 t^5 + 338 t^6 + 680 t^7 + 
 1836 t^8 + 3770 t^9 + 9032 t^{10}+\dots.
\end{align}
However, in the orthosymplectic magnetic quiver, the $\mathfrak{c}_2$ nodes are underbalanced, rendering the Hilbert series more difficult to compute. One way to circumvent this is to compute their parent theories, namely those associated with the pairs of the zero orbits. If these two Coulomb branches coincide, their descendant theories would also have the same symplectic singularities. For $\left([1^4],[1^4]\right)_{\mathfrak{sl}(4)}$, we have
\begin{equation}
    \includegraphics[width=4cm]{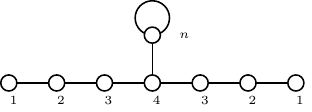}.
\end{equation}
For $\left([1^6],[1^6]\right)_{\mathfrak{so}(6)}$, we have
\begin{equation}
    \includegraphics[width=5cm]{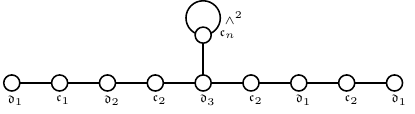}.
\end{equation}
We have checked their Hilbert series perturbatively, and indeed they coincide:
\begin{align}
\begin{split}
    n=2:&\quad\text{HS}=1 + 63 t^2 + 2023 t^4 + 43428 t^6 + 696086 t^8 + 8860325 t^{10} + 
 93249581 t^{12} \\
 &\qquad\quad+ 835269497 t^{14} + 6509509030 t^{16} + 
 44911956047 t^{18} + 278222007609 t^{20}+\dots,
\end{split}\\
\begin{split}
     n=3:&\quad\text{HS}=1 + 31 t^2 + 32 t^3 + 498 t^4 + 992 t^5 + 6089 t^6 + 15936 t^7 + 
 64733 t^8 + 183264 t^9 \\
 &\qquad\quad+ 609515 t^{10} + 1714464 t^{11} + 5082602 t^{12} + 
 13803200 t^{13} + 37821944 t^{14} \\
 &\qquad\quad+ 98383936 t^{15} + 253774372 t^{16} + 631890752 t^{17} + 1550380645 t^{18} \\
 &\qquad\quad+ 3703236224 t^{19} + 8699985237 t^{20}+\dots,
\end{split}\\
\begin{split}
    n=4:&\quad\text{HS}=1 + 31 t^2 + 530 t^4 + 6512 t^6 + 63978 t^8 + 531846 t^{10} + 
 3872310 t^{12} + 25264380 t^{14} \\
 &\qquad\quad+ 150100185 t^{16} + 821730608 t^{18} + 
 4183040471 t^{20}+\dots,
\end{split}\\
\begin{split}
    n=5:&\quad\text{HS}=1 + 31 t^2 + 498 t^4 + 32 t^5 + 5520 t^6 + 992 t^7 + 47437 t^8 + 
 15936 t^9 + 337403 t^{10} \\
 &\qquad\quad+ 176672 t^{11} + 2074757 t^{12} + 
 1519008 t^{13} + 11389466 t^{14} + 10801536 t^{15} \\
 &\qquad\quad+ 57203082 t^{16} + 
 66205280 t^{17} + 267780472 t^{18} + 360084768 t^{19} + 1184061853 t^{20}+\dots.
\end{split}
\end{align}
As a result, the magnetic quivers in \eqref{D32_21_2MQ} and in \eqref{A421_2MQ} have the same Coulomb branch. In this special case, since the unitary magnetic quiver is known, we can even tell the transverse slice when Higgsing to \eqref{A31_3MQ}. It is the Kleinian singularity $A_{n-3}$.

\subsubsection{D-Type Induced Flows via \texorpdfstring{$(p, q)$}{(p,q)} 7-Branes}\label{Dtype7brane}
Let us also make a brief comment on the 7-brane pictures for the D-type cases. In short, the basic idea is that we start with a stack of (half) 7-branes with an $\text{O}7^-$, and the open strings stretched among the branes give the corresponding nilpotent orbit\footnote{We remark that, this setup is more flexible than adjoint Higgsing -- a single junction from the $i^\text{th}$ brane to the $j^\text{th}$ brane in a stack of $k \geq 3$ branes (while keeping the remaining branes intact) cannot be understood as splitting the $k$ branes into 2 stacks. Therefore, it is natural to interpret this junction as $E_{i, j}$, a nilpotent element in the Lie algebra.}. Then we move some of the half 7-branes far away such that they would not be affected by the orientifold and hence recombine into full 7-branes. As a result, there would be two parts of the whole brane system, one with the orientifold giving the nilpotent orbit of D-type and the other giving the nilpotent orbit of A-type.

Again, this would have an easier construction if we take the reverse. For instance, by taking two half 7-branes (and their mirror images) with an orientifold, we would get the orbit $[1^4]_{\mathfrak{so}(4)}$. Now, let us introduce two full 7-branes from infinity, which corresponds to the orbit $[1^2]_{\mathfrak{sl}(2)}$. They would split into half 7-branes due to the orientifold. This would then give rise to the orbit $[1^8]_{\mathfrak{so}(8)}$. However, similar to what we have discussed for the A-type cases in \S\ref{Atype7brane}, this is not an atomic induced flow. One has to have several atomic Higgsings to reach the orbit $[4^2]^{\text{I,II}}_{\mathfrak{so}(8)}$ before getting $[1^4]_{\mathfrak{so}(4)}\oplus[1^2]_{\mathfrak{sl}(2)}$.

In general, the 7-brane pictures for the D-type cases would be more involved. If we would like to mimic the process for the A-type cases, then we should consider the LS duals instead of just taking the transposes, and this method would only work for the special cases. Of course, the way to obtain the induction data that we have mentioned above and also in Appendix \ref{indorb} would work for both the special and the non-special orbits. This would require taking the D-collapses. It could be realized in the 7-brane systems by temporarily ignoring the orientifolds and treating the systems as if they were the A-type cases. Then we can perform the nilpotent Higgsings as in \cite{Hassler:2019eso} to get the partitions whose even parts would have even multiplicities. Nevertheless, a systematic approach using the 7-branes is still yet to be completed, and we plan to investigate this in our future works. In fact, with the 7-brane picture, we could identify several candidate VEVs which flow to the IR configuration. However, the 7-brane picture is not helpful in comparing such VEVs in the UV theory -- such a comparison would instead require an explicit description in terms of the induction of nilpotent orbits, under which the criterion for atomic induced flow then becomes clear.

\subsection{E-type Conformal Matter Theories}\label{exceptional}
Let us first explain how to handle the induced flows for the E-type cases. For such cases, the magnetic quivers are often not known, and the 7-brane picture would be more involved. Nevertheless, we shall see that our method based on the induced orbits would still work.

A technical difference from the AD-type flows is that the nilpotent orbits of the exceptional types are no longer labelled by partitions of integers, but we have to use a more abstract notion of the Bala-Carter labels. In this formalism, the treatment of the induced nilpotent orbits is not as systematic as in the algorithms in the AD-type cases above. Nonetheless, it is possible to give an exhaustive list of the induced orbits as there are finitely of them. The result in \cite{spaltenstein2006classes} is such that, for each induced orbit in $\kg$, we are provided with the rigid induction data.

For our purpose, we need an atomic induction. In other words, given an orbit $\mathcal{O}_{\mathfrak{g}}$ in $\mathfrak{g}$ induced from an orbit $\mathcal{O}_{\mathfrak{l}}$ in the Levi subalgebra $\mathfrak{l}$. There does not exist a distinct orbit $\mathcal{O}_{\mathfrak{l}'}$ in some Levi subalgebra $\mathfrak{l}'$ such that
\begin{equation}
    \mathcal{O}_{\mathfrak{g}}=\text{Ind}^{\mathfrak{g}}_{\mathfrak{l}}(\mathcal{O}_{\mathfrak{l}})=\text{Ind}^{\mathfrak{g}}_{\mathfrak{l}'}\left(\text{Ind}^{\mathfrak{l}'}_{\mathfrak{l}}(\mathcal{O}_{\mathfrak{l}})\right).
\end{equation}
Put differently, this indicates the transitivity of the induced orbits. Any induced flow can be decomposed into multiple steps of atomic induced Higgsings.

Therefore, we can translate the rigid induction data in \cite{spaltenstein2006classes} into the atomic inductions for a given Lie algebra $\mathfrak{g}$ by performing this process reversely. Starting from a rigid orbit, one can step by step find the orbit that it induces in the larger Levi algebra. Eventually, this would lead to the atomic induction data. In Appendix \ref{atomicInduction_e6}, we give the full list of the atomic induction data for the case of $\mathfrak{g}=\mathfrak{e}_6$.

\paragraph{The $E_6$ case} Let us illustrate this with a few examples for the $E_6$ case. Consider the $(E_6,E_6)$ conformal matter of rank 3, where the right orbit is always taken to be $2A_1$. In particular, we have an induction from $\mathfrak{so}(10)$ to $\mathfrak{e}_6$. The atomic induced flow would give $\left[1^{10}\right]_{\mathfrak{so}(10)}$ for the orbit $2A_1$.

If the left orbit is $D_5(a_1)$ (which is a special orbit), then the atomic induced flow gives rise to the orbit $[4^2,1^2]_{\mathfrak{so}(10)}$ (which is a special orbit):
\be
    \overset{\ksu(3)}{3} \ \ 1 \ \ \underset{[\U(1)]}{\overset{\ke_6}{5}} \ \ 1 \ \ \overset{\ksu(3)}{3} \ \ 1 \ \ {\overset{\ke_6}{6}} \ \ 1 \ \ \overset{\ksu(2)}{2} \ \  [\SO(7)] \ \ \longrightarrow \ \ \overset{\ksu(3)}{3} \ \ 1 \ \ \underset{[\Sp(1)]}{\overset{\kso(10)}{4}} \ \ \overset{\ksp(1)}{1} \ \ \overset{\kso(10)}{4} \ \ \overset{\ksp(1)}{1}\ \ [\SO(10)].\label{eqn:E6_induced_D5_special}
\ee
If we take the left orbit to be $A_5$ (which is a non-special orbit), then the atomic induced flow gives rise to the orbit $[5,2^2,1]_{\mathfrak{so}(10)}$ (which is a non-special orbit):
\be
   [\Sp(1)] \ \  \overset{\kg_2}{3} \ \ 1 \ \ {\overset{\kf_4}{5}} \ \ 1 \ \ \overset{\ksu(3)}{3} \ \ 1 \ \ {\overset{\ke_6}{6}} \ \ 1 \ \ \overset{\ksu(2)}{2} \ \ \longrightarrow \ \  [\SO(7)] \ \ \overset{\kg_2}{3} \ \ 1 \ \ \underset{[N_f = 1/2]}{\overset{\kso(9)}{4}} \ \ \overset{\ksp(1)}{1} \ \ \overset{\kso(10)}{4} \ \ \overset{\ksp(1)}{1}\ \ [\SO(10)].
\ee
In the following examples, we will fix the left orbit to be $D_5(a_1)$, and the right orbit always to be some orbit (atomically) induced from the zero orbit in some Levi subalgebra.

Next, let us examine the orbits induced from $\mathfrak{sl}(6)$. We take the right orbit to be $A_2$ whose atomic induced flow gives $[1^6]_{\mathfrak{sl}(6)}$. For the left orbit $D_5(a_1)$, the atomic induced flow yields the orbit $[3,2,1]_{\mathfrak{sl}(6)}$. Therefore, we have an atomic flow
\be
    \overset{\ksu(3)}{3} \ \ 1 \ \ \underset{[\U(1)]}{\overset{\ke_6}{5}} \ \ 1 \ \ \overset{\ksu(3)}{3} \ \ 1 \ \ \underset{\underset{[\SU(3)]}{1}}{\overset{\ke_6}{6}} \ \ 1 \ \ [\SU(3)] \ \ \longrightarrow \ \  [N_f = 1] \ \ \overset{\ksu(3)}{2}  \ \ \underset{[N_f = 1]}{\overset{\ksu(5)}{2}} \ \ \overset{\ksu(6)}{2} \ \ [\SU(7)].\label{eqn:E6_induced_A5}
\ee
We remark that, the IR theories in the flows \eqref{eqn:E6_induced_D5_special} and \eqref{eqn:E6_induced_A5} both come from the same E-type theory, each via a different induced flow of quaternionic dimension 1. The fact that they cannot flow to each other can be seen from the geometry by comparing ``the rightmost $-2$-curve before blowing up the base''. The gauge algebras are $\kso(10)$ over the $-4$-curve and $\ksu(6)$ over the $-2$-curve, which are not contained in one another.

Then we move on to the induced flows of $E_6$ to the Levi subalgebra with multiple non-abelian summands. Here, we increase the length of the UV theory to a $(E_6, E_6)$ conformal matter of rank 5 with 7 M5-branes. We put 3 of the M5-branes on $\mathbb{C}^2/\bbZ_2$ and the 4 remaining M5-branes on $\bbC^2/\bbZ_5$ in the IR theory.

This results in the induction data for $\mathfrak{sl}(2)\oplus\mathfrak{sl}(5)$. The zero orbit $[1^2]\oplus[1^5]$ induces $A_2 + 2A_1$ in $\mathfrak{e}_6$. Here, the orbit $D_5(a_1)$ can be induced from the orbit of $[2] \oplus [2^2, 1]$. Therefore, we get the atomic induced flow
\begin{align}
\begin{split}
    & \overset{\ksu(3)}{3} \ \ 1 \ \ \underset{[\U(1)]}{\overset{\ke_6}{5}} \ \ 1 \ \ \overset{\ksu(3)}{3} \ \ 1 \ \ \overset{\ke_6}{6} \ \ 1 \ \ \overset{\ksu(3)}{3} \ \ 1 \ \ \overset{\ke_6}{6} \ \ 1 \ \ \overset{\ksu(3)}{3} \ \ 1 \ \ \overset{\ke_6}{6} \ \ 1 \ \ \overset{\ksu(3)}{3} \ \ 1 \ \ \overset{\ke_6}{4} \ \ [\U(2)] \\
    \longrightarrow & \ \ 2 \ \ \overset{\ksu(2)}{2} \ \  [G_2] \ \ \sqcup \ \ [N_f = 1] \ \ \overset{\ksu(3)}{2} \ \ \underset{[\SU(2)]}{\overset{\ksu(5)}{2}}  \ \ \overset{\ksu(5)}{2} \ \ [\SU(5)].%\label{eqn:E6_induced_A5}
\end{split}
\end{align}
The same orbit $D_5(a_1)$ can also be induced from the orbit $[1^2] \oplus [3, 1^2]$, giving an alternative atomic induced flow:
\begin{align}
\begin{split}
    & \overset{\ksu(3)}{3} \ \ 1 \ \ \underset{[\U(1)]}{\overset{\ke_6}{5}} \ \ 1 \ \ \overset{\ksu(3)}{3} \ \ 1 \ \ \overset{\ke_6}{6} \ \ 1 \ \ \overset{\ksu(3)}{3} \ \ 1 \ \ \overset{\ke_6}{6} \ \ 1 \ \ \overset{\ksu(3)}{3} \ \ 1 \ \ {\overset{\ke_6}{6}} \ \ 1 \ \ \overset{\ksu(3)}{3} \ \ 1 \ \ {\overset{\ke_6}{4}} \ \ [\U(2)] \\   \longrightarrow & \ \ [\SU(2)] \ \ \overset{\ksu(2)}{2} \ \ \overset{\ksu(2)}{2} \ \  [\SU(2)] \ \ \sqcup \ \ [\SU(2)] \ \ \overset{\ksu(3)}{2} \ \ \overset{\ksu(4)}{2} \ \ \overset{\ksu(5)}{2} \ \ [\SU(6)].%\label{eqn:E6_induced_A5}
\end{split}
\end{align}
In both cases, the quaternionic dimensions of the Higgs branches go down from 28 to 27, indeed producing 1-dimensional flows.

There is one more induction from a direct sum of three non-abelian summands, $\mathfrak{sl}(2)\oplus\mathfrak{sl}(3)\oplus\mathfrak{sl}(3)$. For example, the orbit $D_5(a_1)$ can be induced from the orbit of $[1^2]\oplus[1^3]\oplus[3]$. On the right hand side, we have a triplet of zero orbits inducing $D_4(a_1)$. Physically, this means that we have a theory with 8 M5-branes, which then splits into three components with the numbers of M5-branes equal to 2, 2 and 4 respectively. This would then give
\begin{align}
\begin{split}
    & \overset{\ksu(3)}{3} \ \ 1 \ \ \underset{[\U(1)]}{\overset{\ke_6}{5}} \ \ 1 \ \ \overset{\ksu(3)}{3} \ \ 1 \ \ \overset{\ke_6}{6} \ \ 1 \ \ \overset{\ksu(3)}{3} \ \ 1 \ \ \overset{\ke_6}{6} \ \ 1 \ \ \overset{\ksu(3)}{3} \ \ 1 \ \ \overset{\ke_6}{6} \ \ 1 \ \ \overset{\ksu(3)}{3} \ \ 1 \ \ {\overset{\ke_6}{6}} \ \ 1 \ \ \overset{\ksu(3)}{3} \ \ 1 \ \ \overset{\kso(8)}{4}  \\
    \longrightarrow & \ \ [\SO(7)] \ \ \overset{\ksu(2)}{2} \ \ \sqcup \ \ [\SU(6)] \ \ \overset{\ksu(3)}{2} \ \ \sqcup \ \ 2 \ \ \overset{\ksu(2)}{2} \ \ \overset{\ksu(3)}{2} \ \ [\SU(4)].%\label{eqn:E6_induced_A5}
\end{split}
\end{align}
This flow has quaternionic dimension going down from 25 to 24, which is again reduced by 1. It is worth noting that triple splitting removes 1 quaternionic dimension here, whereas such triple splitting would remove 2 quaternionic dimensions for a $(2,0)$ theory.

Alternatively, we could consider the same splitting pattern for the singularities and the VEVs. Since the M5-branes only split into two pieces, we would have a third stack of M5-branes that is not paired with any singularity:
\begin{align}
    & \overset{\ksu(3)}{3} \ \ 1 \ \ \underset{[\U(1)]}{\overset{\ke_6}{5}} \ \ 1 \ \ \overset{\ksu(3)}{3} \ \ 1 \ \ \overset{\ke_6}{6} \ \ 1 \ \ \overset{\ksu(3)}{3} \ \ 1 \ \ \overset{\ke_6}{6} \ \ 1 \ \ \overset{\ksu(3)}{3} \ \ 1 \ \ \overset{\ke_6}{6} \ \ 1 \ \ \overset{\ksu(3)}{3} \ \ 1 \ \ {\overset{\ke_6}{6}} \ \ 1 \ \ \overset{\ksu(3)}{3} \ \ 1 \ \ \overset{\kso(8)}{4} \\
    \longrightarrow & \ \ [\SU(3)] \ \ \overset{\ksu(3)}{2} \ \ \overset{\ksu(3)}{2} \ \ \overset{\ksu(3)}{2} \ \ [\SU(3)] \sqcup \ \ 2 \ \ \overset{\ksu(2)}{2} \ \ \overset{\ksu(3)}{2} \ \ [\SU(4)] \ \ \sqcup \ \ \text{4 free hypers}.%\label{eqn:E6_induced_A5}
\end{align}
As the 4 hypers would only give the smooth part in the Higgs branch, we shall often omit this, and the induced flow is still an atomic flow.

\paragraph{The $E_7$ and $E_8$ cases} After treating all the induced orbits in $\mathfrak{e}_6$, we now consider some examples of the induced orbits in $\mathfrak{e}_7$ and $\mathfrak{e}_8$. We take the $(E_6,E_6)$ conformal matter theory with orbit pair $(D_5(a_1),A_2)_{\mathfrak{e}_6}$, whose relevant induced orbits in $E_7$ and $E_8$ can be determined as:
\begin{align}
\begin{split}
& \Ind^{\mathfrak{e}_7}_{\mathfrak{e}_6}(D_5(a_1)) = E_6(a_1),\quad\Ind^{\mathfrak{e}_7}_{\mathfrak{e}_6}(A_2) = D_4(a_1) + A_1. \\
& \Ind^{\mathfrak{e}_8}_{\mathfrak{e}_7}(E_6(a_1)) = E_8(a_4),\quad\Ind^{\mathfrak{e}_8}_{\mathfrak{e}_7}(D_4(a_1) + A_1) = D_6(a_1).
\end{split}
\end{align}

For instance, when the left and right orbits are $E_8(a_4)$ and $D_6(a_1)$ in $\mathfrak{e}_8$ respectively, we have the atomic induced flow to the $(E_7,E_7)$ conformal matter theory of rank 3 with orbit pair $(E_6(a_1),D_4(a_1)+A_1)_{\mathfrak{e}_7}$:
\begin{align}
\begin{split}
    &\overset{\ksu(3)}{3} \ \ 1 \ \ {\overset{\ke_6}{6}} \ \ 1 \ \ \overset{\ksu(3)}{3} \ \ 1 \ \   \overset{\kf_4}{5} \ \ 1 \ \  \overset{\kg_2}{3} \ \ \overset{\ksu(2)}{2} \ \ 2 \ \ 1 \ \ \underset{\underset{[\SU(2)]}{1}}{\overset{\ke_7}{8}} \ \ 1 \ \ [\SU(2)] \\
    \longrightarrow & \overset{\ksu(3)}{3} \ \ 1 \ \ {\overset{\ke_6}{6}} \ \ 1 \ \ \overset{\ksu(2)}{2} \ \  \overset{\kso(7)}{3} \ \ \overset{\ksu(2)}{2} \ \ 1 \ \ \underset{\underset{[\SU(2)]}{1}}{\overset{\ke_7}{7}} \ \ 1 \ \ [\SU(2)].
\end{split}
\end{align}

We can further take the atomic induced Higgsing to reach
\begin{align}
\begin{split}
    &\overset{\ksu(3)}{3} \ \ 1 \ \ \underset{[\U(1)]}{\overset{\ke_6}{6}} \ \ 1 \ \ \overset{\ksu(2)}{2} \ \  \overset{\kso(7)}{3} \ \ \overset{\ksu(2)}{2} \ \ 1 \ \ \underset{\underset{[\SU(2)]}{1}}{\overset{\ke_7}{7}} \ \ 1 \ \ [\SU(2)] \\
    \longrightarrow & \overset{\ksu(3)}{3} \ \ 1 \ \ \underset{[\U(1)]}{\overset{\ke_6}{5}} \ \ 1 \ \ \overset{\ksu(3)}{3} \ \ 1 \ \ \underset{\underset{[\SU(3)]}{1}}{\overset{\ke_6}{6}} \ \ 1 \ \ [\SU(3)].
\end{split}
\end{align}

Therefore, we recover the $(E_6,E_6)$ conformal matter theory with orbit pair $(D_5(a_1),A_2)_{\mathfrak{e}_6}$ which was encountered above. In particular, we have seen that this can be further Higgsed to the $A_5$ theory with orbit pair $([3,2,1],[1^6])_{\mathfrak{sl}(6)}$. In particular, $[1^6]_{\mathfrak{sl}(6)}$ is the rigid orbit that induces $D_6(a_1)$ in $\mathfrak{e}_8$ as listed in \cite{spaltenstein2006classes}, which confirms the above sequence of the atomic induced flows.

\paragraph{Non-simply-laced flavours} Now, we move on to the induced flows involving non-simply laced flavour symmetries. From the M-theory picture, all of these induced flows can be seen as the following two steps happening simultaneously:
\begin{itemize}
    \item A DE-type Kleinian singularity splits into a DE-type Kleinian singularity and an A-type Kleinian singularity (and possibly another A-type Kleinian singularity). Here, at least one E-type singularity should be involved, otherwise it is essentially covered in the C-type example in the previous subsection.
    \item A stack of $q + m$ M5-branes splits into $q$ MS5-branes probing the DE-type singularity, and $m$ M5-branes probing the A-type singularity. Here $q \in \mathbb{Q}$ but $q \not \in \bbZ$, and $m \in \bbZ$ in the presence of the A-type Kleinian singularity.
\end{itemize}

As is usually the case for exceptional type Lie algebras, there is a finite list of cases on a single side that we would encounter:
\begin{itemize}
    \item Higgsing from $E_6$ to $D_5$, where an $\SU(3)$ flavour symmetry at 1/2 frozen flux is reduced to $\Sp(1)$;
    \item Higgsing from $E_7$ to $D_6$, where an $\SO(7)$ flavour symmetry at 1/2 frozen flux is reduced to $\Sp(2)$;
    \item Higgsing from $E_8$ to $D_7$, where an $F_4$ flavour symmetry at 1/2 frozen flux is reduced to $\Sp(3)$;
    \item Higgsing from $E_7$ to $E_6$, where an $\SO(7)$ flavour symmetry is reduced to $\SU(3)$ at flux 1/2;
    \item Higgsing from $E_8$ to $E_7$, where an $F_4$ flavour symmetry is reduced to $\SO(7)$ at flux 1/2, and a $G_2$ flavour symmetry is reduced to $\SU(2)$ at flux $\pm 1/3$.
\end{itemize}
There are two more cases, where we have a trivial flavour symmetry in the IR, both from an $\SU(2)$ flavour symmetry in the UV. They come from $E_7$ to $E_6$ at flux $\pm 1/3$, and $E_8$ to $E_7$ at flux $\pm 1/4$. 

In general, the conformal matter theory can have such truncations on both sides, leading to more possible patterns. As an example, we first examine the $D_5$ theory with two and a half M5-branes, with the zero nilpotent orbit on the $\SO(10)$ side and the principal nilpotent orbit on the $\Sp(1)$ side:
\be
\overset{\kso(9)}{4} \ \ \overset{\ksp(1)}{1} \ \ {\overset{\kso(10)}{4}} \ \  \overset{\ksp(1)}{1} \ \ [\SO(10)].
\ee
This theory can be reached by an induced flow from an $E_6$-type theory, with the orbit $2A_1$ on the $E_6$ flavour side and the (principal) orbit $[3]$ on the $\SU(3)$ side:
\begin{align}
    \overset{\kf_4}{5} \ \ 1 \ \ \overset{\ksu(3)}{3} \ \ 1 \ \ \overset{\ke_6}{6} \ \ 1 \ \ \overset{\ksu(2)}{2} \ \ [\SO(7)] \ \ \longrightarrow \ \ \overset{\kso(9)}{4} \ \ \overset{\ksp(1)}{1} \ \ {\overset{\kso(10)}{4}} \ \  \overset{\ksp(1)}{1} \ \ [\SO(10)].
\end{align}

As another example, we consider the $E_7$ theory with $1/3 + 1 + 1/2 = 11/6$ M5-branes, with the $C_3$ flux $-1/3$ on the one end and $1/2$ on the other end. It has the zero orbits on both sides:
\be
    [\SU(2)] \ \ 1 \ \ \overset{\ke_7}{8} \ \ 1 \ \ \overset{\ksu(2)}{2} \ \ \overset{\kso(7)}{3} \ \ \overset{\ksu(2)}{2} \ \ 1 \ \ \overset{\ke_7}{8}  \ \ 1 \ \ \overset{\ksu(2)}{2} \ \ [\SO(7)].
\ee
This theory can be obtained via an induced flow from an $(E_8, E_8)$ conformal matter theory (aka ``the $(G_2, F_4)$ conformal matter theory of rank $11/6$''):
\begin{align}
    & \overset{\ke_8}{9} \ \ 1 \ \ 2 \ \ \overset{\ksu(2)}{2} \ \ \overset{\kg_2}{3} \ \ 1 \ \ \overset{\kf_4}{5} \ \ 1 \ \ \overset{\kg_2}{3} \ \ \overset{\ksu(2)}{2} \ \ 2 \ \ 1 \ \ \overset{\ke_8}{11} \ \ 1 \ \ \overset{\ksu(1)}{2} \ \ \overset{\ksu(2)}{2} \ \ [G_2]. \\
    \longrightarrow & [\SU(2)] \ \ 1 \ \ \overset{\ke_7}{8} \ \ 1 \ \ \overset{\ksu(2)}{2} \ \ \overset{\kso(7)}{3} \ \ \overset{\ksu(2)}{2} \ \ 1 \ \ \overset{\ke_7}{8}  \ \ 1 \ \ \overset{\ksu(2)}{2} \ \ [\SO(7)].
\end{align}
The UV theory is associated with the orbit $G_2(a_1)$ in $G_2$ and the orbit $\widetilde{A}_2$ in $F_4$. To see that the leftmost end is compatible on the tensor branch (where one might naively find a contradiction), note that the $\ke_8$ gauge symmetry is distinct in that such a gauge symmetry on a curve of self-intersection $-n$ has the number of E-strings equal to $N_e = 12-n$. In other words, we should read it as
\be
    \overset{\ke_8}{9} \ \ = \ \ (1)^{\otimes3} \ \ \overset{\ke_8}{(12)},
\ee
in which way it is not hard to see the compatibility on the tensor branch.

By M-/F-theory duality, on the M-theory tensor branch, the reduced flavour symmetry can be described as a ``frozen singularity'', i.e., M-theory on $\bbC^2/\Gamma$ with $\int_{S^3/\Gamma} C_3 = \frac{n}{d}$. See \cite{Tachikawa:2015wka} for a study of frozen singularities across M-/F-theory duality, and \cite{Cvetic:2024mtt} for a more recent study giving an explicit map from an unfrozen singularity to a frozen one.

%\newpage

\textbf{The $\bbZ_3$ centre-flavour symmetry and induced flows} We conclude the discussions on the E-type conformal matter theories with some specific examples in the induced flows from $(E_6, E_6)$-type conformal matter theories to ``$(A_5, A_5)$-type conformal matter theories'', where there is a common $\bbZ_3$ candidate centre-flavour symmetry that one can track along the RG flow.

Following \cite{Heckman:2022suy}, the theories in the $\SU(6)$ nilpotent hierarchy preserving a $\bbZ_3$ centre-flavour symmetry is specified by looking for the partitions of 6 where all parts have multiplicities that is divisible by 3. Therefore, we have two options:
\be
    \calO \in \left\{\left[1^6\right]_{\mathfrak{sl}(6)}, \ \ \left[2^3\right]_{\mathfrak{sl}(6)}\right\}.
\ee
Now, we look for their induced orbits in $E_6$ and their corresponding theories in the IR under a single VEV:
\begin{align}
    &\Ind^{\ke_6}_{\mathfrak{sl}(6)}\left(\left[1^6\right]\right) = A_2: \quad [\SU(3)] \ \ 1 \ \ \underset{\underset{[\SU(3)]}{1}}{\overset{\ke_6}{6}} \ \ 1 \overset{\ksu(3)}{3} \ \ 1 \ \ \overset{\ke_6}{6} \dots\ \ [E_6],\\
    &\Ind^{\ke_6}_{\mathfrak{sl}(6)}\left(\left[2^3\right]\right) = D_4: \quad \overset{\ksu(3)}{3} \ \ 1 \ \ \underset{\underset{[\SU(3)]}{1}}{\overset{\ke_6}{6}} \ \ 1 \overset{\ksu(3)}{3} \ \ 1 \ \ \overset{\ke_6}{6} \dots\ \ [E_6].
\end{align}
Both of these theories admit a flavour symmetry $(\SU(3) \times E_6)/\bbZ_3$ with a $\bbZ_3$ diagonal quotient \cite{Heckman:2022suy}. In addition, there is a $\bbZ_3$ 1-form symmetry coming from the linear combination of the $\bbZ_3$ centre of the individual gauge symmetries on the tensor branch \cite{Apruzzi:2020zot}. Choosing a suitable number of M5-branes and a pair of such orbits, we can obtain an induced flow:
\be
    \overset{\ksu(3)}{3} \ \ 1 \ \ \underset{\underset{[\SU(3)]}{1}}{\overset{\ke_6}{6}} \ \ 1 \overset{\ksu(3)}{3} \ \ 1 \ \ \underset{\underset{[\SU(3)]}{1}}{\overset{\ke_6}{6}} \ \ 1 \ \ [\SU(3)] \ \  \longrightarrow \ \ \overset{\ksu(3)}{2} \ \ \underset{[\SU(3)]}{\overset{\ksu(6)}{2}} \ \ \overset{\ksu(6)}{2} \ \ [\SU(6)], \quad \Delta d_{\bbH} = 1,
\ee
where the Higgs branch quaternionic dimension is reduced from 31 to 30. It would be very interesting to carefully track the behaviour of various $\bbZ_3$ symmetries along the RG flow.

\subsection{DE-Type Dirac Pairings}\label{DEbase}
Up till now, we have been focusing on the cases with A-type Dirac pairings. In the rest of this section, we give examples to illustrate the induced flows when the Dirac pairings are of DE-type. As we explained earlier, such theories admit a Type IIB construction as $k$ NS5-branes probing a Kleinian singularity of DE-type. The RG flows of such theories has been systemically studied in \cite{Lawrie:2024zon}, except for the flows that change the endpoint configurations which we now treat.

To set up the stage, one simplest example would be the flow of a $(2,0)$ theory of DE-type, whose tensor branch description is given by a collection of $-2$ curves forming an intersection matrix of DE-type. This is well-known to come from the resolution of the Kleinian singularity of DE-type. Here, the RG flow can be realized by a complex structure deformation of the Kleinian singularity. Such a flow always corresponds to removing a single node from the Dynkin diagram, and the resulting theory might become reducible. For example, we can have the following flow from the $(2,0)$ theory of type $D_8$ to
\be
\begin{array}{cccccccc}
      & 2 &   &   &   &   &    \\
    2 & 2 & 2 & 2 & 2 & 2 & 2
\end{array}
    \quad \longrightarrow \quad \begin{array}{ccc}
      & 2 &   \\
    2 & 2 & 2
\end{array} \sqcup \begin{array}{ccc}
      &   &   \\
    2 & 2 & 2.
\end{array} \label{eqn:(2,0)_split}
\ee
The resulting $(2,0)$ theory has two irreducible components of types $D_4$ and $A_3$ respectively. This corresponds to the Lie algebra branching $D_8 \rightarrow D_4 \times A_3$ by removing the node at the position shown here.

We now proceed to an atomic flow with the same set of base configuration as above, but with a non-trivial fibre decoration. We start from the M-theory brane picture, where we have the $D_8$-type Dirac pairing coming from 8 M5-branes placed on top of an OM5-plane (that is, the M-theory uplift of the ON-orientifold) and a transverse singularity of type $\bbC^2/\bbZ_k$. The theory is given by
\begin{equation}
\begin{array}{cccccccc}
      & \overset{\ksu(k)}{2} &   &   &   &   &   &   \\
    \overset{\ksu(k)}{2} & \overset{\ksu(2k)}{2} & \overset{\ksu(2k)}{2} & \overset{\ksu(2k)}{2} & \overset{\ksu(2k)}{2} & \underset{[\SU(k)]}{\overset{\ksu(2k)}{2}} & \overset{\ksu(k)}{2}.
\end{array}
\end{equation}
An atomic Higgsing is not possible unless we turn on a VEV inside the $\SU(k)$ flavour symmetry\footnote{Comparing this with \cite{Lawrie:2024zon}, we take the natural UV theory to be the one that comes from the little string theory (with gauge ranks arranged according to the affine Dynkin diagram) and decouple the tensor multiplet associated with the affine node. Such a configuration would have a more natural construction in Type II string theory/M-theory.}.

Now, we look for a VEV in $\SU(k)$ such that the IR theory would admit a further atomic Higgsing to the following pair of components:
\begin{itemize}
    \item a stack of 4 M5 with an OM5 and a transverse $\bbC^2/\bbZ_{k_1}$ singularity;
    \item a stack of 4 M5 with no OM5 on a transverse $\bbC^2/\bbZ_{k_2}$ singularity.
\end{itemize}
For concreteness, let us illustrate this with $k_1 = k_2 = 3$ (so $k = 6$). We get
\be
\begin{array}{ccc}
      & \overset{\ksu(3)}{2} &   \\
    \overset{\ksu(3)}{2} & \underset{[\SU(3)]}{\overset{\ksu(6)}{2}} & \overset{\ksu(3)}{2}
\end{array} \sqcup \begin{array}{ccccc}
    [\SU(3)] & \overset{\ksu(3)}{2} & \overset{\ksu(3)}{2} & \overset{\ksu(3)}{2} & [\SU(3)].
\end{array}
\ee
As we can see, the $D_4$-type theory carries one $\SU(3)$ flavour symmetry with a trivial VEV $[1^3]$. In the second component, the two $\SU(3)$ flavour symmetries should be folded into a single $\SU(3)$ upon fusing them with the OM5, which also carries a trivial VEV $[1^3]$. Therefore, the UV theory right before the atomic flow should carry an induced nilpotent VEV of $[2^3]$. Therefore, the atomic flow is\footnote{To determine the tensor branch description, a technical approach is to rewind this to the UV configuration where the rightmost curve also has $\ksu(2k) = \ksu(12)$ gauge algebra with $\SU(12)$ flavour symmetry, in which the induced nilpotent VEV is further made into $[4^3]$.}
\begin{align}
&\begin{array}{cccccccc}
      & \overset{\ksu(6)}{2} &   &   &   &   &   &   \\
    \overset{\ksu(6)}{2} & \overset{\ksu(12)}{2} & \overset{\ksu(12)}{2} & \underset{[\SU(3)]}{\overset{\ksu(12)}{2}} & \overset{\ksu(9)}{2} & \overset{\ksu(6)}{2} & \overset{\ksu(3)}{2}
\end{array} \\ \vspace{2mm}
\longrightarrow & \begin{array}{cccc}
      & \overset{\ksu(3)}{2} & \\
    \overset{\ksu(3)}{2} & \underset{[\SU(3)]}{\overset{\ksu(6)}{2}} & \overset{\ksu(3)}{2} 
\end{array}  \ \ \sqcup \ \ 
\begin{array}{ccccc}
    [\SU(3)] & \overset{\ksu(3)}{2} & \overset{\ksu(3)}{2} & \overset{\ksu(3)}{2} & [\SU(3)].
\end{array}\label{eqn:(1,0)_split}
\end{align}
The quaternionic dimension of the UV theory is 26, and the quaternionic dimension of the irreducible components in the IR are 13 for the one with the $D_4$ Dirac pairing and 12 for the one with the $A_3$ Dirac pairing\footnote{We remark that the two flows \eqref{eqn:(2,0)_split} and \eqref{eqn:(1,0)_split} both exhibit rich behaviour on the spectrum of the non-invertible duality defects \cite{Lawrie:2023tdz}. Specifically, the automorphism of the quiver is $S_2 = \bbZ_2$ for the UV theory and $S_3 \times S_2$ for the IR theory, where the $S_2$ in the UV embeds into the $S_3$ component in the IR. This means that we actually get an enrichment of the non-invertible duality defects into $S_3$-ality defects \cite{Lu:2024lzf}. It would be interesting to study more implications of such duality-defect-preserving RG flows.}.

Such an atomic flow for a theory with an E-type base is more difficult to identify, especially when all the irreducible components in the IR are $(1,0)$ theories. Without aiming at being exhaustive, we now give a somewhat degenerate example where a $(1,0)$ theory of $E_8$-type Dirac pairing flows to the theory with two components. One is a $(1,0)$ theory with $E_6$-type Dirac pairing, and the other is a $(2,0)$ theory of $A_1$-type:
\begin{align}
&\begin{array}{ccccccccc}
      &  & \overset{\ksu(9)}{2}  &   &   &   &   &   \\
    \overset{\ksu(6)}{2} & \overset{\ksu(12)}{2} & \overset{\ksu(18)}{2} & {\overset{\ksu(15)}{2}} & \overset{\ksu(12)}{2} & \overset{\ksu(9)}{2} & \overset{\ksu(6)}{2} & [\SU(3)] \\
\end{array} \\ \vspace{1cm} \nonumber\\ \vspace{3cm}
\longrightarrow &\begin{array}{ccccc}
      &  & [\SU(3)]  &   &    \\
      &  & \overset{\ksu(6)}{2}  &   &    \\
    \overset{\ksu(3)}{2} & \overset{\ksu(6)}{2} & \overset{\ksu(9)}{2} & {\overset{\ksu(6)}{2}} & \overset{\ksu(3)}{2} \\
\end{array} \ \ \sqcup  \ \ 2.
\end{align}
where the difference of the quaternionic dimension is 1.

\section{Flows via Induced Discrete Homomorphisms into \texorpdfstring{$E_8$}{E8}}\label{indE8}
Let us now handle another type of ``induced'' flows involving orbi-instanton theories. They have a different type of ``boundary condition'' seen from the generalized quiver description of 6d SCFTs, where we place the set of M5-branes inside an end-of-the-world M9-brane. Induced flows among such theories have analogous physics to those discussed in the previous section. However, the mathematical description will not involve the (induced) nilpotent orbits, but the inductions of the homomorphisms from the (direct sums of) discrete subgroups of $\SU(2)$ into $E_8$.

Specifically, inside the M9-brane, we have a Kleinian singularity $\bbC^2/\Gamma$ placed transverse to the M5-brane stack. In particular, in the asymptotic boundary $S^3/\Gamma$, we are allowed to turn on a flat connection of the $E_8$ bundle living on the worldvolume of the M9, which are mathematically known to be classified by the discrete homomorphisms of $\Gamma$ into $E_8$. See \cite{Heckman:2013pva,Heckman:2015bfa,Frey:2018vpw} for how such a discrete homomorphism define an SCFT (as a flow from the SCFT with a trivial $E_8$ connection). The theory is labelled by $(\mu, \alpha)$, where $\mu$ is a nilpotent orbit inside $\kg_{\Gamma}$ (that is, the McKay dual to $\Gamma$) and $\alpha \in \text{Hom}(\Gamma, E_8)$. Therefore, we shall denote an orbi-instanton theory as $\calT(\Gamma, \mu, \alpha)$.

\begin{figure}
    \centering
    \includegraphics[width=0.9\linewidth]{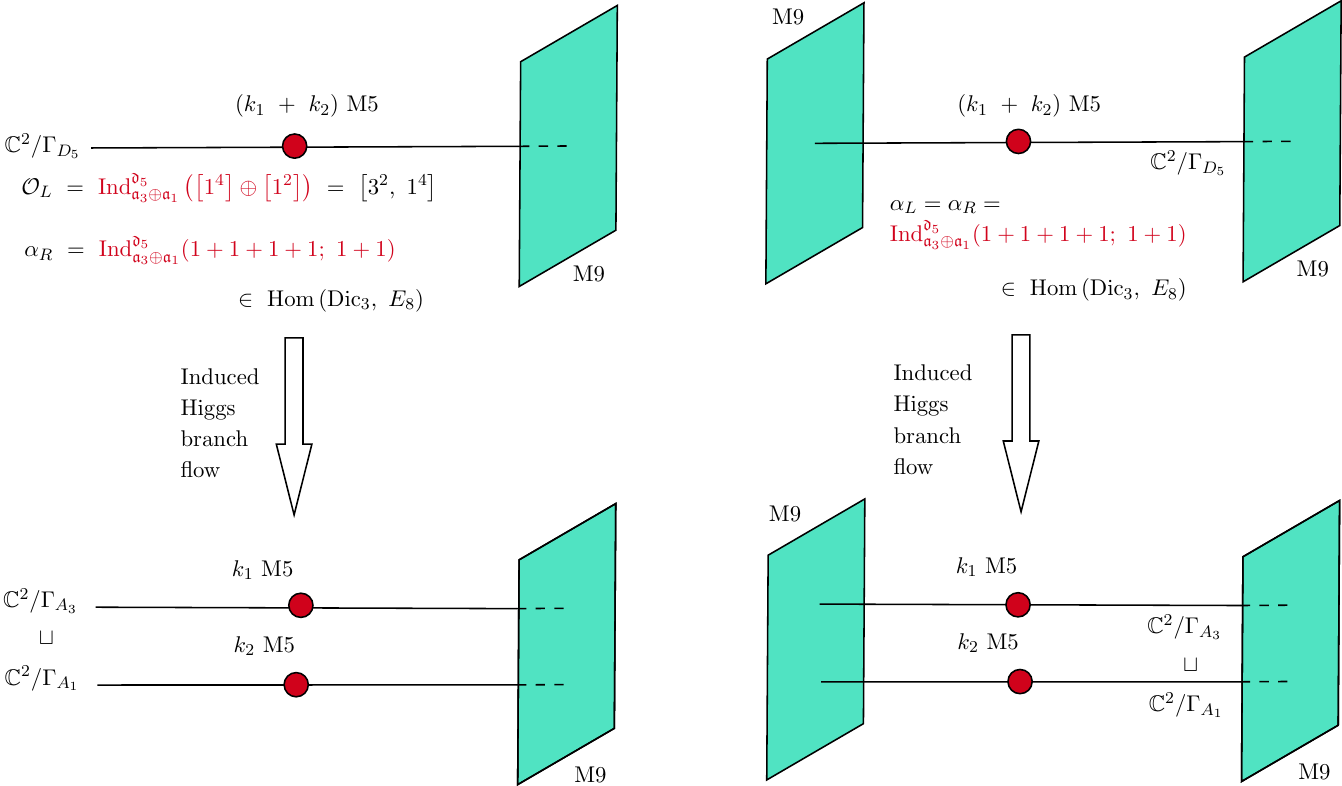}
    \caption{Left: The M-theoretic picture of an induced flow among orbi-instanton theories, which is parallel to that of conformal matter theories except for the extra M9-brane. Here, the $E_8$ holonomy data is specified by an ``induced discrete homomorphism'' from $\text{Dic}_3$ to $E_8$, which will be defined in \S\ref{DEtypeorbins} motivated by the physics of the induced flows. Right: An induced flow among heterotic LSTs, where we put M9-branes on both ends, thus obtaining the Ho\v{r}ava-Witten configuration \cite{Horava:1995qa}.}\label{fig:induced_with_M9}
\end{figure}

Induced flows of the orbi-instanton theories can also be thought of as some analogue of induced flows of the conformal matter theories. See Figure \ref{fig:induced_with_M9} for an illustration of the M-theory understanding of such an induced flow for the orbi-instanton theory with a single M9 brane, or for the $E_8 \times E_8$ heterotic LST with two M9-branes, which will be discussed towards the end of this section. 

Now, for the orbi-instanton theory, if we only focus on the nilpotent orbit side, then $\mu^{\mathrm{UV}}$ in the UV theory is induced from $\mu^{\mathrm{IR}}$ (or $\bigoplus\limits_i \mu_i^{\mathrm{IR}}$) in the IR theory. Even though this situation does not involve the nilpotent orbits, the atomic RG flow of 6d SCFTs naturally gives a notion of ``induction'' on the discrete homomorphism side that is physically expected to have similar structure with the induction of nilpotent orbits. For example, the 6d SCFTs instruct us to define a ``dimension function'' of the discrete homomorphisms (see below), which is expected to be preserved along such inductions. However, there is no immediate (independent) mathematical significance of such induction of discrete homomorphisms. As an example, this general class of RG flows is expected to incorporate a map from $\Hom(\bbZ_2, E_8)\oplus\text{Hom}(\bbZ_3, E_8)$ to $\Hom(\bbZ_5, E_8)$, where the statement that $2 + 3 = 5$ does not give a clear hint on how to relate the two homomorphisms.

A qualitative difference from the nilpotent orbit case is as follows. The trivial homomorphism always induces the trivial homomorphism, but the same is not true for the ``maximal'' homomorphism. In fact, even the existence or the well-definedness of the ``maximal'' homomorphism is not mathematically clear since the meaning of the partial ordering of such discrete homomorphisms can only be physically defined by considering the RG flows among SCFTs.

The discussions on the discrete homomorphisms and the induced flows can be split into two cases:
\begin{itemize}
    \item In the \textbf{A-type} cases, $\text{Hom}(\bbZ_k, E_8)$ is completely classified by Kac \cite{KAC_Lie_MR739850}. Therefore, in the cases when we only concern cyclic groups both in the UV and in the IR, the atomic flows instruct us to write down a clear mathematical notion of the induced discrete homomorphisms. Such meaning can be exactly shown to reproduce the results from the unitary magnetic quivers \cite{Bourget:2023dkj,Bourget:2024mgn}. It would be interesting to understand its mathematical meaning from as many perspectives as possible.
    \item In the \textbf{DE-type} cases where the finite groups are not cyclic groups, an exact analogue of the A-type classfication is not directly available. However, assuming the physical ``definition'' via the Higgsings of the orbi-instanton 6d SCFTs, we may provide a procedure to work out the embedding. When the orthosymplectic magnetic quivers are known, the induced flows would reproduce the decay and fission processes in \cite{Lawrie:2024wan}.
\end{itemize}

The dimension function is a major bonus one can get by considering the RG flows, which was not pursued in \cite{Frey:2018vpw}. The dimension function of a homomorphism $\alpha$ is defined to be the change of the quaternionic dimension of the Higgs branch under the RG flow from the UV orbi-instanton theory with the trivial discrete homomorphism:
\begin{equation}
    d(\mathcal{T}(\Gamma,\mu,\alpha)):=\dim_{\mathbb{H}}(\mathcal{H}(\Gamma,\mu,\alpha_{\text{trivial}}))-\dim_{\mathbb{H}}(\mathcal{H}(\Gamma,\mu,\alpha)),
\end{equation}
where we have used $\mathcal{H}(\Gamma,\mu,\alpha)$ to the denote the Higgs branch of the orbi-instanton theory $\mathcal{T}(\Gamma,\mu,\alpha)$. In particular, this is independent of the nilpotent orbit $\mu$ on the other side of the theory.

It turns out that we would always get the trivial discrete homomorphism in the IR theory under the induced flow from the trivial discrete homomorphisms in the UV theory.
\begin{tcolorbox}[colback=blue!10!white,breakable]
A consequence of this observation is that the dimension function of the discrete homomorphisms into $E_8$ is preserved along inductions.
\end{tcolorbox}
\noindent Specifically, if $\alpha: \Gamma \rightarrow E_8$ is induced from $\alpha': \Gamma' \rightarrow E_8$, then we have $d(\mathcal{T}(\Gamma,\mu,\alpha)) = d(\mathcal{T}(\Gamma,\mu,\alpha'))$. This has been observed in all the examples that we have checked, and it is consistent with the decay and fission algorithm if the magnetic quivers are known\footnote{It would also be interesting to compare this statement with the fact that the induction of the nilpotent orbits preserves the \textbf{codimension} $\mathrm{codim}(\mathcal{O}) \ \equiv\mathfrak{} \dim(\calO_{\mathrm{prin}}) - \dim(\calO)$.}. Therefore, we may simply write $d(\alpha)$ for the dimension function, and this justifies the well-definedness of the dimension function of a homomorphism.

Although the inductions of the discrete homomorphisms are treated in a way that is mathematically independent of the inductions of the nilpotent orbits, we would still mimic the terminologies of the nilpotent orbits for the discrete homomorphisms. It is clear that the zero/trivial homomorphisms coincide with the trivial homomorphisms that send every $\Gamma$ element to $0 \in \ke_8$ in the usual sense. We may also say that a homomorphism is Richardson (resp.~rigid) if it can be induced from some trivial homomorphism (resp.~cannot be induced from any other homomorphisms). Although a mathematically well-defined partial ordering of the discrete homomorphisms is not yet available, the RG flows among the 6d theories defines a physically motivated partial ordering among these homomorphisms. We may say that a homomorphism is principal/regular if the corresponding theory is in the bottom IR. However, notice that it is still not clear whether this is unique or not.

\subsection{A-Type Cases}\label{Atypeorbins}
Let us first review the discrete homomorphisms from $\bbZ_k$ into $E_8$ and their classification. According to \cite[\S8.6]{KAC_Lie_MR739850}, a discrete homomorphism from $\bbZ_k$ into $G$ with rank $l = \mathrm{rk}(G)$, which is in turn given by a $k^{\text{th}}$ order automorphism $\sigma$ of $G$ (i.e., $\sigma^k=1$), is classified by an $(l+1)$-tuple of integers $\bm{s} = (s_0, s_1, \dots, s_l)$ called the Kac label. The order of the automorphism $\sigma$ is related to $\bm{s}$ via the weighted sum by the Coxeter labels $a_0,a_1,\dots,a_l$ (with $a_0=1$):
\be
    k = r \sum_{i=0}^{l} a_i s_i.
\ee
Here, $r$ can be understood as the twist parameter which is the least positive integer such that $\sigma^r$ is an inner automorphism. In other words, there exists a graph automorphism of the Dynkin diagram of $G$ of order $r$. Then $r=1,2,3$. Now, using $e_j$ to denote the standard Chevalley generators associated to the positive simple roots, the relations
\begin{equation}
    \sigma_{s;r}(e_j)=\text{e}^{2\pi is_j/k}e_j\quad(j=0,\dots,l)
\end{equation}
uniquely determine an automorphism $\sigma_{s;r}$ of order $k$. Such automorphisms $\sigma_{s;r}$ exhausts all the $k^{\text{th}}$ order automorphism of $G$ up to the conjugation by an automorphism of $G$.

Following \cite{Frey:2018vpw}, when $s_i\neq0$, we can remove the corresponding node in the $E_8$ Dynkin diagram. Then the residual Dynkin diagram gives the flavour symmetry (on the M9 side) of the generalized quiver.

For $E_8$, we have $a_i = (1, 2, 3, 4, 5, 6, 4', 2', 3')$, and $r$ is always 1. The dimension function reads
\be
d(\alpha) = \sum_{i = 0}^{l} d_i s_i.
\ee
The coefficients $d_i$ are given by
\be
    \begin{array}{|c||c|c|c|c|c|c|c|c|c|} \hline
        a_i & 1 & 2 & 2' & 3  & 3' & 4 & 4' & 5 & 6   \\\hline
        d_i & 0 & 29 & 46 & 57 & 68 & 84 & 91 & 110 & 135 \\ \hline
    \end{array}.\label{eqn:dim_function}
\ee
This dimension formula can be seen from the Coulomb branches of the magnetic quivers, which was explicitly worked out in \cite[\S4.3]{Mekareeya:2017jgc}.

Therefore, as in \cite{Heckman:2015bfa}, we can label the discrete homomorphisms by a (possibly repeated) sum of the Coxeter labels. For example, consider the $A_3$-type orbi-instanton theory:
\be
[E_8] \ \ 1 \ \ 2 \ \ \overset{\ksu(2)}{2} \ \ \overset{\ksu(3)}{2} \ \ \underset{[N_f = 1]}{\overset{\ksu(4)}{2}} \ \ \dots \ \ \overset{\ksu(4)}{2} \ \ [\SU(4)].
\ee
One can turn on the following set of discrete homomorphisms:
\be
    1^4,\ 1^2 + 2,\ 1^2 + 2',\ 1 + 3,\ 1 + 3',\ 2^2,\ 2 + 2',\ (2')^2,\ 4,\ 4'.
\ee
This set of discrete homomorphisms produces a set of IR theories \cite{Heckman:2013pva} with the Hasse diagram given in \cite[Figure 1]{Frey:2018vpw}.

Let us now state the general pattern of the atomic induced flows for the A-type cases. We say that an induction from $\bm{s}=\bm{s}^1\oplus\bm{s}^2 \in \Hom(\bbZ_{k_1}, E_8)\oplus\Hom(\bbZ_{k_2}, E_8)$ to $\bm{s}' \in \Hom(\bbZ_{k_1 + k_2}, E_8)$ if
\be
    \bm{s}' = \bm{s}^1 + \bm{s}^2 = \left(s^1_0 + s^2_0, \dots, s^1_8 + s^2_8\right).
\ee
Analogous to the notation of the induced orbits, we may write
\begin{equation}
    \bm{s}'=\text{Ind}^{\mathfrak{l}'}_{\mathfrak{l}}(\bm{s}).
\end{equation}
For the A-type cases as above, we write $\mathfrak{l}=(\mathbb{Z}_{k_1},\mathbb{Z}_{k_2})$ and $\mathfrak{l}'=\mathbb{Z}_k$. When it would not cause any confusions, we shall omit the superscript $\mathfrak{l}'$ and the subscript $\mathfrak{l}$. Notice that we can also have $\Hom(\bbZ_{k_2}, E_8)$ to be the trivial $\Hom(\bbZ_{1}, E_8)$, which only contains $\bm{s}^2=(1,0,\dots,0)$.
\begin{tcolorbox}[colback=blue!10!white,breakable]
Then for an induced flow of an orbi-instanton theory, the discrete homomorphism in the UV theory is induced from the discrete homomorphism in the IR theory.
\end{tcolorbox}

Moreover, an induced flow is atomic if the corresponding induction is atomic, i.e., if there does not exist any $\bm{s}''$ such that
\begin{equation}
    \bm{s}'=\text{Ind}^{\mathfrak{l}'}_{\mathfrak{l}''}(\bm{s}'')=\text{Ind}^{\mathfrak{l}'}_{\mathfrak{l}''}\left(\text{Ind}^{\mathfrak{l}''}_{\mathfrak{l}}(\bm{s})\right).
\end{equation}
This is the case if the rank of the Lie group $G$ McKay dual to $\Gamma_G$ is changed by 1 along the induction. It also indicates the transitivity of the inductions of the discrete homomorphisms. The induced flows can be decomposed into multiple steps of atomic flows. For instance, the flow from the theory with a discrete homomorphism in $\text{Hom}(\mathbb{Z}_6,E_8)$ to the one with a discrete homomorphism in $\text{Hom}(\mathbb{Z}_3,E_8)^{\oplus2}$ is atomic. However, the flow from this UV theory to the one with a discrete homomorphism in $\text{Hom}(\mathbb{Z}_2,E_8)^{\oplus3}$ is not atomic since we can have $\text{Hom}(\mathbb{Z}_2,E_8)\oplus
\text{Hom}(\mathbb{Z}_4,E_8)$ as an intermediate step.

In fact, the magnetic quivers are also known from \cite{Mekareeya:2017jgc} for the A-type orbi-instanton theories. The induced flows should agree with certain decay and fission processes as in \cite{Bourget:2023dkj,Bourget:2024mgn}.

Here, we have only given the meaning of the inductions for $\Hom(\bbZ_k, E_8)$. In the next subsection, we will explain and illustrate the physical definition of such inductions for the homomorphisms from the non-cyclic finite subgroups from $\SU(2)$ into $E_8$.

\paragraph{Examples} Let us illustrate our discussions with some examples. Take the trivial homomorphism $[1^5]\in\text{Hom}(\mathbb{Z}_5,E_8)$ and the nilpotent orbit $[2,1^3]_{\mathfrak{sl}(5)}$:
\be
\mathcal{T}_{\text{UV}}=[E_8] \ \ 1 \ \ 2 \ \ \overset{\ksu(2)}{2} \ \ \overset{\ksu(3)}{2} \ \ \overset{\ksu(4)}{2} \ \ \underset{[N_f = 1]}{\overset{\ksu(5)}{2}} \ \ \underset{[N_f = 1]}{\overset{\ksu(5)}{2}} \ \ \overset{\ksu(4)}{2} \ \ [\SU(3)].
\ee
This admits an atomic induced flow due to the induction from the trivial homomorphism $[1^4]\in\text{Hom}(\mathbb{Z}_4,E_8)$ (where we have again omitted the trivial $\mathbb{Z}_1$ part). For the nilpotent orbit, there is an induction from $[1^4]_{\mathfrak{sl}(4)}$ to $[2,1^3]_{\mathfrak{sl}(5)}$ as discussed in our previous section. Therefore, we have
\be
\mathcal{T}_{\text{IR}}=[E_8] \ \ 1 \ \ 2 \ \ \overset{\ksu(2)}{2} \ \ \overset{\ksu(3)}{2} \ \ \underset{[N_f = 1]}{\overset{\ksu(4)}{2}} \ \ \overset{\ksu(4)}{2} \ \ \overset{\ksu(4)}{2} \ \ \overset{\ksu(4)}{2} \ \ [\SU(4)],
\ee
where the $E_8$ flavour symmetry is left intact. By the tensor branch understanding from \cite{Bao:2024eoq}, it is not hard to identify the atomic flow as a plateau Higgsing. This flow has the M-theory interpretation of performing a complex structure deformation of the $A_4$-type Kleinian singularity transverse to the stack of M5-branes into the $A_3$-type.

We proceed to the second example with a non-trivial discrete homomorphism. Let us take the homomorphism $[2+2'+1]$ in $\text{Hom}(\mathbb{Z}_5,E_8)$:
\be
    \mathcal{T}_{\text{UV}}=[\SO(12)] \ \ \overset{\ksp(1)}{1} \ \ \overset{\ksu(2)}{2} \ \ \underset{[N_f = 1]}{\overset{\ksu(4)}{2}} \ \ \underset{[N_f = 1]}{\overset{\ksu(5)}{2}} \ \ \underset{[N_f = 1]}{\overset{\ksu(5)}{2}} \ \ \overset{\ksu(4)}{2} \ \ [\SU(3)].
\ee
There is an atomic induced flow giving the IR theory:
\be
    \mathcal{T}_{\text{IR}}=[\SO(12)] \ \ \overset{\ksp(1)}{1} \ \ \overset{\ksu(2)}{2} \ \ \underset{[SU(2)]}{\overset{\ksu(4)}{2}} \ \ \overset{\ksu(4)}{2} \ \ \overset{\ksu(4)}{2} \ \ \overset{\ksu(4)}{2} \ \ [\SU(4)].
\ee
The homomorphism is now given by $[2+2']$ in $\text{Hom}(\mathbb{Z}_4,E_8)$. This is again a plateau Higgsing.

This flow has the M-theory interpretation of performing a complex structure deformation of the $A_4$-type Kleinian singularity transverse to the stack of M5-branes into the $A_3$-type. It would be very interesting to understand the precise change of the $E_8$ holonomy dictated by the induction in a more geometric way, and we hope to come back to this question in the future.

As a final example, we give an atomic induced flow to an reducible theory with multiple orbi-instanton components. The UV theory is given by an $A_5$-type orbi-instanton theory with the orbit $\mu = [3, 2, 1]$ and the discrete homomorphism $\alpha = 3 + 2' + 1$:
\begin{equation}
    \mathcal{T}_{\text{UV}}=[\SO(10)] \ \ \overset{\ksp(1)}{1} \ \ \underset{[\SU(2)]}{\overset{\ksu(5)}{2}} \ \ \underset{[N_f = 1]}{\overset{\ksu(6)}{2}} \ \ \underset{[N_f = 1]}{\overset{\ksu(6)}{2}} \ \ \underset{[N_f = 1]}{\overset{\ksu(5)}{2}} \ \ \underset{[N_f = 1]}{\overset{\ksu(3)}{2}}.
\end{equation}
The IR theory has two components, one with $\alpha_1 = 3$ and $\mu_1 = [2, 1]$ and the other with $\alpha_2 = 2' + 1$ and $\mu_2 = [1^3]$:
\begin{equation}
    \calT_{\text{IR}} = [E_6] \ \ 1 \ \ \underset{[\SU(3)]]}{\overset{\ksu(3)}{2}} \ \  \underset{[N_f = 1]}{\overset{\ksu(3)}{2}} \ \ \underset{[N_f = 1]}{\overset{\ksu(2)}{2}} \ \ \sqcup  \ \ [\SO(14)] \ \ \overset{\ksp(1)}{1} \ \ \underset{[N_f = 1]}{\overset{\ksu(3)}{2}} \ \ {\overset{\ksu(3)}{2}} \ \ [\SU(3)].
\end{equation}

\subsection{DE-Type Cases}\label{DEtypeorbins}
We now proceed to discussing the orbi-instanton theories of DE-types. The primary difference in these cases is that there are no general mathematical classifications for such homomorphisms as simple as in the A-type cases. In fact, the only available characterizations of such cases are the classifications of $\Hom(G, E_8)$ for $G = \Gamma_{D_5}, \Gamma_{D_7}$ and $\Gamma_{E_8}$ \cite{frey2001conjugacy}.

Nevertheless, it was conjectured in \cite{Heckman:2015bfa,Frey:2018vpw} that the one-to-one correspondence between the 6d SCFTs of orbi-instanton types and $\Hom(\Gamma, E_8)$ for the A-type discrete subgroups of $\SU(2)$ continues to hold for the DE-types. In this case, our perspective of atomic Higgsings and induced homomorphisms provides a non-trivial way to connect $\Hom(\Gamma, E_8)$ for different $\Gamma$. Notice that such correspondence is for 6d SCFTs with a fixed period. In other words, starting from a UV theory, we might reach an SCFT with a shorter quiver that only differs by a number of periods after several steps of atomic Higgsings. We shall then only assign those with the longest quivers with the homomorphisms to ensure the well-definedness of the orderings and inductions. This means that, for the shortest possible orbi-instantons, one can still turn on all possible flat connections.

Based on the conjectured one-to-one correspondence above between the 6d orbi-instanton SCFTs and $\Hom(\Gamma, E_8)$, we now proceed to our main statement:
\begin{tcolorbox}[colback=blue!10!white,breakable]
For a Levi subalgebra $\kl = \bigoplus\limits_i \kl_i$ of $\mathfrak{g}$, we may consider an induced flow from a $\kg$-type orbi-instanton theory to a reducible theory with the components being the orbi-instanton theories of $\kl_i$-types. The induced flow from an orbi-instanton theory of type $\kg$ to an orbi-instanton theory of type $\kl$ gives a physical definition of the induction from $\bigoplus\limits_i\Hom(\Gamma_{\kl_i}, E_8)$ to $\Hom(\Gamma_{\kg}, E_8)$.
\end{tcolorbox}

As an induction from $\bigoplus\limits_i\Hom(\Gamma_{\kl_i}, E_8)$ to $\Hom(\Gamma_{\kg}, E_8)$ can only be physically defined when $\kl \subset \kg$, for non-trivial inclusions, we always have
\be
    \mathrm{rank}(\kg) - \mathrm{rank}(\mathfrak{s}) \geq 1,
\ee
where we have decomposed $\mathfrak{l}=\mathfrak{s}\oplus\mathfrak{a}$ into a semisimple part $\mathfrak{s}$ and an abelian part $\mathfrak{a}$. The equal sign would be called an atomic induction. For such $\kg, \kl$ and a homomorphism $\alpha \in \bigoplus\limits_i\Hom(\Gamma_{\kl_i}, E_8)$, we have
\be
\Ind_{\kl}^{\kg}(\alpha) \in \Hom(\Gamma_{\kg}, E_8)
\ee
such that the orbi-instanton SCFT of type $\kg$ with discrete homomorphism $\Ind_{\kl}^{\kg}(\alpha) \in \Hom(\Gamma_{\kg}, E_8)$ and nilpotent orbit $\nu = \Ind_{\kl}^{\kg}(\calO_{\mathrm{trivial}})$ admits an induced flow (of quaternionic dimension $(\mathrm{rank}(\kg) - \mathrm{rank}(\mathfrak{s}))$ to the orbi-instanton SCFT of type $\kl$ with discrete homomorphism $\alpha$ and trivial nilpotent VEV $\calO_{\mathrm{trivial}}$. 

We learn from the physics of RG flows that such inductions of discrete homomorphisms also admit \textbf{transitivity}. An induction of rank difference $\mathrm{rank}(\kg) - \mathrm{rank}(\mathfrak{s})$ can always be broken down to $\mathrm{rank}(\kg) - \mathrm{rank}(\mathfrak{s})$ atomic inductions (each of rank difference 1 by definition). From the physics of RG flows, we also know that the induction of discrete homomorphisms should preserve the dimension function, i.e.,
\be
    d(\alpha)=d\left(\Ind_{\kl}^{\kg}(\alpha)\right).
\ee

We remark that, looking at the RG flow between orbi-instanton theories, the change in the quaternionic dimension can be solely traced back to the rank difference between $\kg$ and $\mathfrak{s}$. One should not ``add up the contributions for the induced orbit and the induced discrete homomorphisms''.

There is a lack of a mathematical algorithm (or even definition) to obtain the induction data of the discrete homomorphisms. Nevertheless, we can still obtain a general statement of the trivial homormophisms. The trivial flat $E_8$ connection would always induce the trivial discrete homomorphism in $\Gamma_{\kg}$. This can be seen by noticing that performing a complex structure deformation of the Kleinian singularity in the presence of a trivial $E_8$ bundle would not introduce any non-trivial $E_8$ bundle.

\begin{figure}
    \centering
    \includegraphics[width=0.9\linewidth]{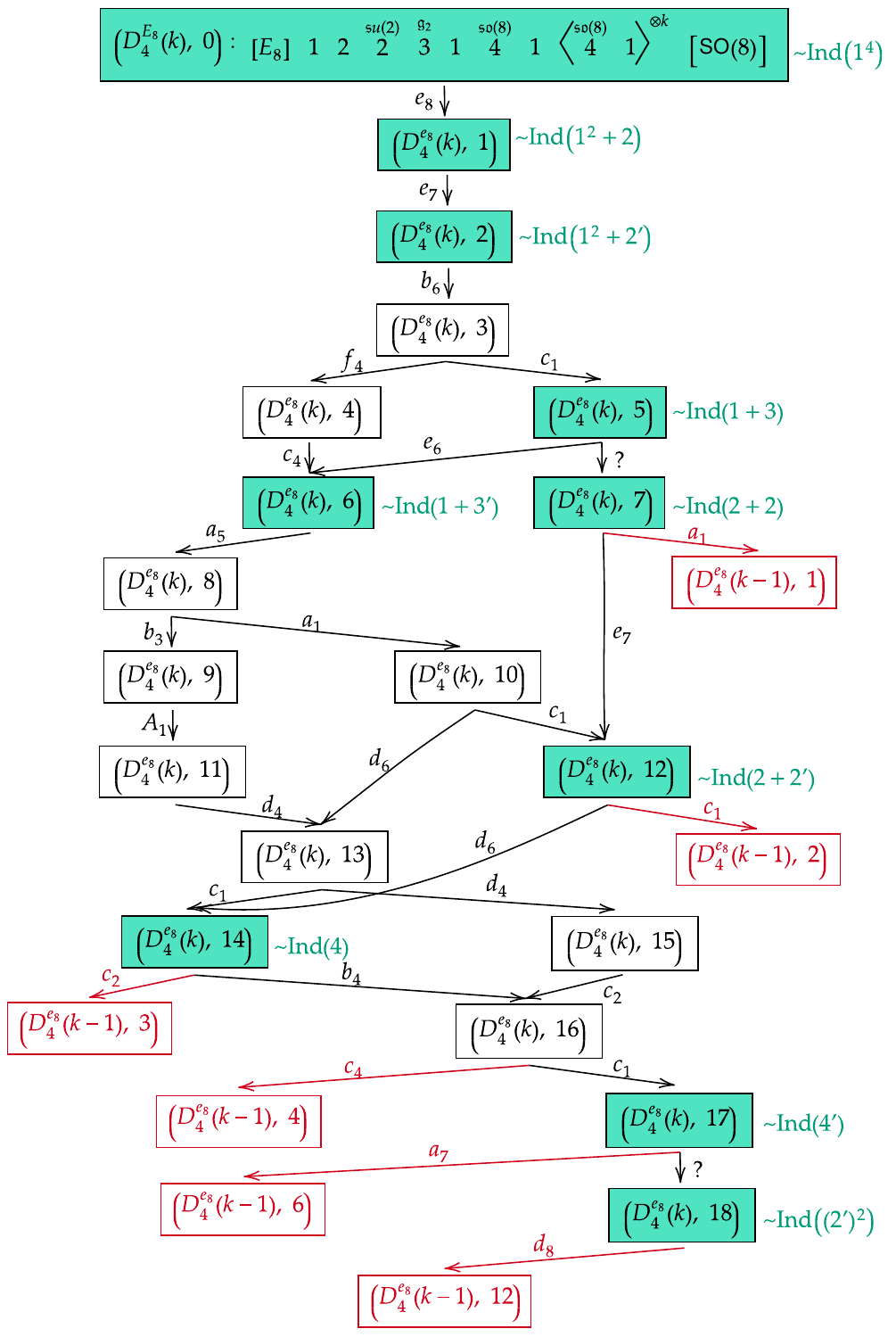}
    \caption{The Hasse diagram of the $D_4$ orbi-instanton hierarchy for the discrete homomorphisms with $k$ extra M5-branes. Notice that we are not showing the full Hasse diagram of all the Higgsings from the UV theory. The discrete homomorphisms for the shaded nodes can be induced from those in $\text{Hom}(\mathbb{Z}_4,E_8)$ whose induction data are listed next to the nodes. This diagram is supposed to be iterated to generate an infinite diagram when $k$ goes to infinity. The places of the iterations are indicated by the red nodes. See Table \ref{D4_orbi-instanton_table} for the F-theory description of each node.}
    \label{fig:D4_orbi_instanton_Hasse}
\end{figure}

\paragraph{A D-type example} To illustrate our physically motivated definition of the inductions for $\Hom(\Gamma_{\kg}, E_8)$, we give an example on the inductions from $\Hom(\bbZ_4, E_8)$ to $\Hom(\Gamma_{D_4}\cong\mathrm{Dic}_2, E_8)$. All the orbi-instanton theories of type $D_4$ were worked out in \cite[Appendix B.1]{Frey:2018vpw}, but the possible RG flows among them was not explicitly written down there. We present such a Hasse diagram in Figure \ref{fig:D4_orbi_instanton_Hasse}, where the 6d SCFTs represented by each node (together with their quaternionic dimensions and all atomic RG flows) are given separately in Table \ref{D4_orbi-instanton_table}.

In Figure \ref{fig:D4_orbi_instanton_Hasse}, we put special emphasis on the discrete homomorphisms that lie in the image of
\be
    \Ind_{\mathfrak{a}_3}^{\mathfrak{d}_4}: \Hom(\bbZ_4, E_8) \rightarrow \Hom(\text{Dic}_2, E_8).
\ee
Such theories are labelled by a shaded background, and we also give the unique homomorphism $\alpha \in \Hom(\bbZ_4, E_8)$, specified by their Kac label, such that $\Ind_{\mathfrak{a}_3}^{\mathfrak{d}_4}(\alpha)$ labels the current theory of type $D_4$. Notice that in Table \ref{D4_orbi-instanton_table}, the nilpotent orbit on the right side for each theory is the zero orbit of $\mathfrak{so}(8)$. If we change the zero nilpotent orbit to the nilpotent VEV\footnote{Again, we have omitted the labels I, II for the very even orbits. The inductions distinguishing the two very even orbits for the same partition can be determined as reviewed in Appendix \ref{indorb}. Recall that throughout the paper, we would always omit the abelian parts of the Levi subalgebras for brevity.} $\left[2^4\right] = \Ind_{\mathfrak{sl}(4)}^{\kso(8)}\left(\left[1^4\right]\right)$, each such shaded node gives an atomic induced flow from a $D_4$-type orbi-instanton theory with the homomorphism $\Ind_{\mathfrak{a}_3}^{\mathfrak{d}_4}(\alpha)$ to an $A_3$-type orbi-instanton theory with the homomorphism $\alpha$ and the trivial nilpotent VEV $\left[1^4\right]$. A catalogue of all such flows is given in Table \ref{D4_A3_orbi-instanton_table} (where we have used the same $D^{e_8}_4$ labels for simplicity).

\begin{longtable}{|c|c|c|c|}
\hline
$D_4^{e_8}$ label &  Tensor branch description & $\dim_{\mathbb{H}}$ & (descendant \#; flow type) \\ \hline \hline
$(D_4^{e_8}(k), 0)$ & $[E_8] \ 1 \ 2 \ \overset{\ksu(2)}{2}\ \overset{\kg_2}{3} \ 1 \ \left\langle\overset{\kso(8)}{4} \ 1\right\rangle^{\otimes (k+1)} \ [\SO(8)]$ & $166+30k$ & $(1; e_8)$ \\ \hline
$(D_4^{e_8}(k), 1)$ & $[E_7] \ 1 \ \overset{\ksu(2)}{2}\ \overset{\kg_2}{3} \ 1 \ \left\langle\overset{\kso(8)}{4} \ 1\right\rangle^{\otimes (k+1)} \ [\SO(8)]$ & $137+30k$ & $(2; e_7)$ \\ \hline
$(D_4^{e_8}(k), 2)$ & $[\SO(13)] \ \overset{\ksp(1)}{1}\ \overset{\kg_2}{3} \ 1 \ \left\langle\overset{\kso(8)}{4} \ 1\right\rangle^{\otimes (k+1)} \ [\SO(8)]$ & $120+30k$ & $(3; b_6)$ \\ \hline
$(D_4^{e_8}(k), 3)$ & $[F_4] \ 1 \ \underset{[\Sp(1)]}{\overset{\kg_2}{3}} \ 1 \ \left\langle\overset{\kso(8)}{4} \ 1\right\rangle^{\otimes (k+1)} \ [\SO(8)]$ & $110+30k$ & $(4; f_4),\ (5; c_1)$ \\ \hline
$(D_4^{e_8}(k), 4)$ & $[\Sp(4)] \ \overset{\kg_2}{2} \ 1 \ \left\langle\overset{\kso(8)}{4} \ 1\right\rangle^{\otimes (k+1)} \ [\SO(8)]$ & $102+30k$ & $(6; c_4)$ \\ \hline
$(D_4^{e_8}(k), 5)$ & $[E_6] \ 1 \ {\overset{\ksu(3)}{3}} \ 1 \ \left\langle\overset{\kso(8)}{4} \ 1\right\rangle^{\otimes (k+1)} \ [\SO(8)]$ & $109+30k$ & $(6; e_6),\ (7; ?)$ \\ \hline
$(D_4^{e_8}(k), 6)$ & $[\SU(6)] \ {\overset{\ksu(3)}{2}} \ 1 \ \left\langle\overset{\kso(8)}{4} \ 1\right\rangle^{\otimes (k+1)} \ [\SO(8)]$ & $98+30k$ & $(8; a_5)$ \\ \hline
$(D_4^{e_8}(k), 7)$ & $[E_7] \ 1 \ {\overset{\ksu(2)}{2}} \ \underset{[\Sp(1)]}{\overset{\kso(7)}{3}}\ 1 \ \left\langle\overset{\kso(8)}{4} \ 1\right\rangle^{\otimes k} \ [\SO(8)]$ & $108+30k$ & $(12; e_7),\ {\color{red} ((D_4^{e_8}(k-1), 1); a_1) }$ \\ \hline
$(D_4^{e_8}(k), 8)$ & $[\SO(7)] \ {\overset{\ksu(2)}{2}} \ 1 \ \left\langle\overset{\kso(8)}{4} \ 1\right\rangle^{\otimes (k+1)} \ [\SO(8)]$  & $93+30k$ & $(9; b_3),\ (10; a_1)$ \\ \hline
$(D_4^{e_8}(k), 9)$ & $[\SU(2)] \ 2 \ 1 \ \left\langle\overset{\kso(8)}{4} \ 1\right\rangle^{\otimes (k+1)} \ [\SO(8)]$ & $89+30k$ & $(11; A_1)$ \\ \hline
$(D_4^{e_8}(k), 10)$ & $[\SO(12)] \ {\overset{\ksp(1)}{1}} \ \overset{\kso(8)}{3}\ 1 \ \left\langle\overset{\kso(8)}{4} \ 1\right\rangle^{\otimes k} \ [\SO(8)]$ & $92+30k$ & $(12; c_1),\ (13; d_6)$ \\ \hline
$(D_4^{e_8}(k), 11)$ & $([\SO(8)] \ 1)^{\otimes 2} \ \overset{\kso(8)}{4} \ 1 \ \left\langle\overset{\kso(8)}{4} \ 1\right\rangle^{\otimes k} \ [\SO(8)]$ & $88+30k$ & $(13; d_4)$ \\ \hline
$(D_4^{e_8}(k), 12)$ & $[\SO(12)] \ {\overset{\ksp(1)}{1}} \ \underset{[\Sp(1)]}{\overset{\kso(7)}{3}}\ 1 \ \left\langle\overset{\kso(8)}{4} \ 1\right\rangle^{\otimes k} \ [\SO(8)]$ & $91+30k$ & $(14; d_6),\ {\color{red} ((D_4^{e_8}(k-1), 2); c_1) }$ \\ \hline
$(D_4^{e_8}(k), 13)$ & $[\SO(8)] \ 1 \ \underset{[\Sp(1)^3]}{\overset{\kso(8)}{3}} \ 1 \ \left\langle\overset{\kso(8)}{4} \ 1\right\rangle^{\otimes k} \ [\SO(8)]$ & $83+30k$ & $(14; c_1),\ (15; d_4)$ \\ \hline
$(D_4^{e_8}(k), 14)$ & $[\SO(9)] \ 1 \ \underset{[\Sp(2)]}{\overset{\kso(7)}{3}} \ 1 \ \left\langle\overset{\kso(8)}{4} \ 1\right\rangle^{\otimes k} \ [\SO(8)]$ & $82+30k$ & $(16; b_4),\ {\color{red} ((D_4^{e_8}(k-1), 3); c_2) }$ \\ \hline
$(D_4^{e_8}(k), 15)$ & $[\Sp(2)^3] \ {\overset{\kso(8)}{2}} \ 1 \ \left\langle\overset{\kso(8)}{4} \ 1\right\rangle^{\otimes k} \ [\SO(8)]$ & $78+30k$ & $(16; c_2)$ \\ \hline
$(D_4^{e_8}(k), 16)$ & $[\Sp(4) \times \Sp(1)] \ {\overset{\kso(7)}{2}} \ 1 \ \left\langle\overset{\kso(8)}{4} \ 1\right\rangle^{\otimes k} \ [\SO(8)]$ & $76+30k$ & $(17; c_1),\ {\color{red} ((D_4^{e_8}(k-1), 4); c_4) }$ \\ \hline
$(D_4^{e_8}(k), 17)$ & $[\SU(8)] \ {\overset{\ksu(4)}{2}} \ 1 \ \left\langle\overset{\kso(8)}{4} \ 1\right\rangle^{\otimes k} \ [\SO(8)]$ & $75+30k$ & $(18; ?),\ {\color{red} ((D_4^{e_8}(k-1), 6); a_7) }$ \\ \hline
$(D_4^{e_8}(k), 18)$ & $[\SO(16)] \ {\overset{\ksp(2)}{1}} \ \overset{\kso(7)}{3}\ 1 \ \left\langle\overset{\kso(8)}{4} \ 1\right\rangle^{\otimes (k-1)} \ [\SO(8)]$ & $74+30k$ & ${\color{red} ((D_4^{e_8}(k-1), 12); d_8) }$ \\ \hline
\caption{The theories that are associated to the discrete homomorphisms from $\Gamma_{D_4}$ into $E_8$. The red labels indicate the ones whose generalized quivers have shorter periods, and the same types of Higgsings can be iterated.}\label{D4_orbi-instanton_table}
\end{longtable}

{\footnotesize
\begin{longtable}{|c|c|c||c|c|c|}%{p{0.28\textwidth-2\tabcolsep}*{6}{p{0.12\textwidth-2\tabcolsep}}}
\hline
$D_4^{e_8}$ label &  Tensor branch description\ ($\calO_R = \left[2^4\right]$) & $\dim_{\mathbb{H}}$ & $[\bbZ_4, E_8]$ & IR $\Hom(\bbZ_4, E_8)$ theory & $\dim_{\mathbb{H}}$ \\ \hline \hline
$(D_4^{e_8}(1), 0)$ & $[E_8] \ 1 \ 2 \ \overset{\ksu(2)}{2}\ \overset{\kg_2}{3} \ 1 \ \overset{\kso(8)}{4} \ 1 \ \ \overset{\kso(7)}{3} \ \ [\Sp(2)]$ & $190$ & $1^4$  &  $[E_8]\ 1 \ 2 \ \overset{\ksu(2)}{2} \ \overset{\ksu(3)}{2} \ \underset{[N_f = 1]}{\overset{\ksu(4)}{2}} \ \overset{\ksu(4)}{2} \ [\SU(4)]$ & $189$ \\ \hline
$(D_4^{e_8}(1), 1)$ & $[E_7] \ 1 \ \overset{\ksu(2)}{2}\ \overset{\kg_2}{3} \ 1 \ \overset{\kso(8)}{4} \ 1 \ \ \overset{\kso(7)}{3} \ \ [\Sp(2)]$ & $161$ & $1^2+2$  &  $[E_7]\ 1 \ \overset{\ksu(2)}{2} \ \overset{\ksu(3)}{2} \ \underset{[N_f = 1]}{\overset{\ksu(4)}{2}} \ \overset{\ksu(4)}{2} \ [\SU(4)]$ & $160$ \\ \hline
$(D_4^{e_8}(1), 2)$ & $[\SO(13)] \ \overset{\ksp(1)}{1}\ \overset{\kg_2}{3} \ 1 \ \overset{\kso(8)}{4} \ 1 \ \ \overset{\kso(7)}{3} \ \ [\Sp(2)]$ & $144$ & $1^2+2'$  &  $[\SO(14)]\ \overset{\ksp(1)}{1} \ \overset{\ksu(3)}{2} \ \underset{[N_f = 1]}{\overset{\ksu(4)}{2}} \ \overset{\ksu(4)}{2} \ [\SU(4)]$ & $143$ \\ \hline
$(D_4^{e_8}(1), 5)$ & $[E_6] \ 1 \ {\overset{\ksu(3)}{3}} \ 1 \ \overset{\kso(8)}{4} \ 1 \ \ \overset{\kso(7)}{3} \ \ [\Sp(2)]$ & $133$ & $1+3$  &  $[E_6]\ 1 \ \underset{[SU(2)]}{\overset{\ksu(3)}{2}} \ \underset{[N_f = 1]}{\overset{\ksu(4)}{2}} \ \overset{\ksu(4)}{2} \ [\SU(4)]$ & $132$ \\ \hline
$(D_4^{e_8}(1), 6)$ & $[\SU(6)] \ {\overset{\ksu(3)}{2}} \ 1 \ \overset{\kso(8)}{4} \ 1 \ \ \overset{\kso(7)}{3} \ \ [\Sp(2)]$ & $122$ & $1+3'$  &  $[\SU(8)]\ {\overset{\ksu(3)}{1}} \ \underset{[N_f = 1]}{\overset{\ksu(4)}{2}} \ \overset{\ksu(4)}{2} \ [\SU(4)]$ & $121$  \\ \hline
$(D_4^{e_8}(1), 7)$ & $[E_7] \ 1 \ {\overset{\ksu(2)}{2}} \ \underset{[\Sp(1)]}{\overset{\kso(7)}{3}}\ 1 \  \overset{\kso(7)}{3} \ \ [\Sp(2)]$ & $132$ & $2^2$  &  $[E_7]\ 1 \ \overset{\ksu(2)}{2} \ \underset{[\SU(2)]}{\overset{\ksu(4)}{2}} \ \overset{\ksu(4)}{2} \ [\SU(4)]$ & $131$  \\ \hline
$(D_4^{e_8}(1), 12)$ & $[\SO(12)] \ {\overset{\ksp(1)}{1}} \ \underset{[\Sp(1)]}{\overset{\kso(7)}{3}}\ \overset{\kso(7)}{3} \ \ [\Sp(2)]$ & $115$ & $2+2'$  &  $[\SO(12)]\ \overset{\ksp(1)}{1} \ \underset{[\SU(2)]}{\overset{\ksu(4)}{2}} \ \overset{\ksu(4)}{2} \ [\SU(4)]$ & $114$  \\ \hline
$(D_4^{e_8}(1), 14)$ & $[\SO(9)] \ 1 \ \underset{[\Sp(2)]}{\overset{\kso(7)}{3}} \ 1 \ \ \overset{\kso(7)}{3} \ \ [\Sp(2)]$ & $106$ & $4$  &  $[\SO(10)]\ 1 \ \underset{[\SU(4)]}{\overset{\ksu(4)}{2}} \ \overset{\ksu(4)}{2} \ [\SU(4)]$ & $105$ \\ \hline
$(D_4^{e_8}(1), 17)$ & $[\SU(8)] \ {\overset{\ksu(4)}{2}} \ 1 \ \ \overset{\kso(7)}{3} \ \ [\Sp(2)]$ & $99$ & $4'$  &  $[\SU(8) \times \SU(2)]\ \ {\overset{\ksu(4)}{1}} \ \overset{\ksu(4)}{2} \ [\SU(4)]$ & $98$ \\ \hline
$(D_4^{e_8}(1), 18)$ & $[\SO(16)] \ {\overset{\ksp(2)}{1}} \ \overset{\kso(7)}{2}\ [\Sp(2) \times \Sp(1)]$ & $98$ & $(2')^2$  &  $[\SO(16)]\ \ {\overset{\ksp(2)}{1}} \ \overset{\ksu(4)}{2} \ [\SU(4)]$ & $97$  \\ \hline
\caption{Inductions from $\Hom(\bbZ_4, E_8)$ to $\Hom(\text{Dic}_2, E_8)$, realized by 6d SCFTs. We remark that the last theory exhibits the behaviour of the \textit{short quiver}, but we still get an atomic induced flow.}\label{D4_A3_orbi-instanton_table}
\end{longtable}
}

Let us highlight a few examples from the above discussions as some further illustrations. First of all, we give an example of the induced flow of the endpoint-changing type. We can always perform the ``trivial induction'' from any orbi-instanton theory by removing $k$ M5-branes, which produces a separate $(2,0)$ theory of $k-1$ type. For instance,
\begin{align}
&([\SO(8)] \ 1)^{\otimes 2} \ \overset{\kso(8)}{4} \ 1 \ \left\langle\overset{\kso(8)}{4} \ 1\right\rangle^{\otimes k} \ [\SO(8)] \\
\longrightarrow & ([\SO(8)] \ 1)^{\otimes 2} \ \overset{\kso(8)}{4} \ 1 \ \left\langle\overset{\kso(8)}{4} \ 1\right\rangle^{\otimes k-m} \ [\SO(8)] \ \ \sqcup \ \ [E_8] \ \ 1 \ \ \underbrace{2 \ \ 2 \ \ \dots \ \ 2}_{m-1}.
\end{align}
The magnetic quivers are known in this case \cite{Bao:2024eoq,Lawrie:2024wan}, which would give the same result. We shall not repeat the magnetic quivers here.

We can equally treat examples without known magentic quivers. For example, the top UV theory has the following induced flow:
\begin{align}
&[E_8] \ 1 \ 2 \ \overset{\ksu(2)}{2}\ \overset{\kg_2}{3} \ 1 \ \left\langle\overset{\kso(8)}{4} \ 1\right\rangle^{\otimes k} \ [\SO(8)] \\
\longrightarrow & [E_8] \ 1 \ 2 \ \overset{\ksu(2)}{2}\ \overset{\kg_2}{3} \ 1 \ \left\langle\overset{\kso(8)}{4} \ 1\right\rangle^{\otimes k-m} \ [\SO(8)] \ \ \sqcup \ \ [E_8] \ \ 1 \ \ \underbrace{2 \ \ 2 \ \ \dots \ \ 2}_{m-1}.
\end{align}
If this UV theory has an orthosymplectic magnetic quiver, then this flow together with the one in Table \ref{D4_A3_orbi-instanton_table} would give a natural guess of the possible magnetic quiver as in Appendix \ref{guessMQ}.

Now, we consider an endpoint-changing flow governed by a non-trivial induction. To see how the induced homomorphism is obtained, let us start with the IR theory with two components with the homomorphism in $\text{Hom}(\mathbb{Z}_2,E_8)\oplus\text{Hom}(\mathbb{Z}_4,E_8)$. We pick the homomorphism 2 in the $\text{Hom}(\mathbb{Z}_2,E_8)$ part:
\be
    [E_7]\ 1\ \underset{[\SU(2)]}{\overset{\ksu(2)}{2}}\  \overset{\ksu(2)}{2}\ [\SU(2)].
\ee
In the $\text{Hom}(\mathbb{Z}_4,E_8)$ part, we choose the homomorphism $1+1+2'$:
\be
[\SO(14)]\ \overset{\ksp(1)}{1}\ {\overset{\ksu(3)}{2}} \ \underset{[N_f = 1]}{\overset{\ksu(4)}{2}} \ \overset{\ksu(4)}{2} \ [\SU(4)].
\ee

The UV theory of the induced flow should be the one that has the induced homomorphism of $(2,1+1+2')$ and the induced orbit $[3^2, 1^4]$ of $D_5$. If both the induced homomorphism and the induced orbit were trivial, the theory would have quaternionic dimension 235:
\be
[E_8] \ 1 \ 2 \ \overset{\ksu(2)}{2}\ \overset{\kg_2}{3} \ 1 \  \overset{\kso(9)}{4} \ \underset{[N_f = 1/2]}{\overset{\ksp(1)}{1}} \ \overset{\kso(10)} {4} \ \ \overset{\ksp(1)}{1} \ \overset{\kso(10)} {4} \ \ \overset{\ksp(1)}{1} \ \overset{\kso(10)}{4} \ \overset{\ksp(1)}{1} \ \overset{\kso(10)} {4} \ \overset{\ksp(1)}{1} \ \overset{\kso(10)}{4} \ \overset{\ksp(1)}{1} \ [\SO(10)].
\ee
Turning on the induced orbit $[3^2, 1^2]$, we get the theory with quaternionic dimension 312:
\be
[E_8] \ 1 \ 2 \ \overset{\ksu(2)}{2}\ \overset{\kg_2}{3} \ 1 \  \overset{\kso(9)}{4} \ \underset{[N_f = 1/2]}{\overset{\ksp(1)}{1}} \ \overset{\kso(10)} {4} \ \ \overset{\ksp(1)}{1} \ \overset{\kso(10)} {4} \ \ \overset{\ksp(1)}{1} \ \overset{\kso(10)}{4} \ \overset{\ksp(1)}{1} \ \overset{\kso(10)} {4} \ \underset{[N_f = 1]}{\overset{\ksp(1)}{1}} \ \overset{\kso(8)}{3} \ \ \left[\SU(2)^2\right].
\ee
Now, the induced homomorphism is the one of quaternionic dimension $29+46=75$. By tracking through the atomic Hasse diagram of $D_5$ (which we do not include for brevity), we get a theory of quaternionic dimension 237, which admits the desired induced flow to the pair of irreducible theories with quaternionic dimensions $(143,93)$:
\begin{align}
&[\SO(12)]\ \overset{\ksp(1)}{1}\ \overset{\kso(7)}{3} \ \underset{[N_f = 1]}{\overset{\ksp(1)}{1}} \ \overset{\kso(10)} {4} \ \ \overset{\ksp(1)}{1} \ \overset{\kso(10)} {4} \ \ \overset{\ksp(1)}{1} \ \overset{\kso(10)}{4} \ \overset{\ksp(1)}{1} \ \overset{\kso(10)} {4} \ \underset{[N_f = 1]}{\overset{\ksp(1)}{1}} \ \overset{\kso(8)}{3} \ \ \left[\SU(2)^2\right] \\
\longrightarrow & [E_7] \ 1 \ \underset{[\SU(2)]}{\overset{\ksu(2)}{2}}\  \overset{\ksu(2)}{2}\ [\SU(2)]\ \ \sqcup\ \  [\SO(14)]\ \overset{\ksp(1)}{1}\ {\overset{\ksu(3)}{2}} \ \underset{[N_f = 1]}{\overset{\ksu(4)}{2}} \ \overset{\ksu(4)}{2} \ [\SU(4)].
\label{eq:orbi_instanton_dtype_endpoint_changing}
\end{align}

\paragraph{An E-type example} Let us now consider some induced flows from the exceptional type orbi-instanton theories. We take the IR theory to be the theory $(D_4^{e_8}(0), 3)$ from our catalogue:
\be
    [F_4] \ 1 \ \underset{[\Sp(1)]}{\overset{\kg_2}{3}} \ 1 \ \overset{\kso(8)}{4} \ 1 \ \overset{\kso(8)}{4} \ 1 \ [\SO(8)].
\ee
The discrete homomorphism for this theory is sufficiently simple to reach from the trivial one by a chain of atomic Higgsings: $e_8,e_7,b_6$. In fact, we shall realize that if we consider the induced flows from the orbi-instanton theories of type $D_5, E_6, E_7, E_8$, then the induced discrete homomorphisms will follow similar patterns and the quaternionic dimension can be found to be $29+17+10 = 56$. The atomic Hasse diagram for the orbi-instanton theories of type $D_5, E_6, E_7, E_8$ also admits an identical Higgsing chain that is unique (if we hold the other end fixed) from the UV theory. Therefore, combining with the induced orbits:
\be
    \Ind_{\mathfrak{so}(8)}^{\mathfrak{so}(10)}\left(\left[1^8\right]\right) = \left[3, 1^7\right], \ \ \Ind_{\mathfrak{so}(8)}^{\mathfrak{e}_6}\left(\left[1^8\right]\right) = 2A_2, \ \ 
    \Ind_{\mathfrak{so}(8)}^{\mathfrak{e}_7}\left(\left[1^8\right]\right) = (A_5)'', \ \ 
    \Ind_{\mathfrak{so}(8)}^{\mathfrak{e}_8}\left(\left[1^8\right]\right) = E_6,
\ee
the chain of the atomic flows can be identified as
\begin{align}
    & \ \ [F_4] \ 1 \ \underset{[\Sp(1)]}{\overset{\kg_2}{3}} \ 1 \ {\overset{\kf_4}{5}} \ 1 \ \overset{\kg_2}{3}\ \overset{\ksu(2)}{2} \ 2 \ 1  \ \overset{\ke_8}{(10)} \ 1 \ 2\ \overset{\ksu(2)}{2} \ \overset{\kg_2}{3} \ 1 \ \overset{\kf_4}{5} \ 1 \ [G_2] \\
    \longrightarrow & \ \ [F_4] \ 1 \ \underset{[\Sp(1)]}{\overset{\kg_2}{3}} \ 1 \ {\overset{\kf_4}{5}} \ 1 \ \overset{\kg_2}{3}\ \overset{\ksu(2)}{2} \ 1  \ \overset{\ke_7}{8} \ 1 \ \overset{\ksu(2)}{2} \ \overset{\kg_2}{3} \ 1 \ \overset{\kf_4}{5} \ 1 \ [G_2] \\
    \longrightarrow & \ \ [F_4] \ 1 \ \underset{[\Sp(1)]}{\overset{\kg_2}{3}} \ 1 \ {\overset{\kf_4}{5}} \ 1 \ \overset{\ksu(3)}{3} \ 1 \ \overset{\ke_6}{6} \ 1 \ \overset{\ksu(3)}{3} \ 1 \ \overset{\kf_4}{5} \ 1 \ [G_2] \\
    \longrightarrow & \ \ [F_4] \ 1 \ \underset{[\Sp(1)]}{\overset{\kg_2}{3}} \ 1 \ \overset{\kso(9)}{4} \ \underset{[N_f = 1/2]}{\overset{\ksp(1)}{1}} \ \overset{\kso(10)}{4} \ \underset{[N_f = 1/2]}{\overset{\ksp(1)}{1}} \ \overset{\kso(9)}{4} \ 1 \ [SO(7)] \\
    \longrightarrow & \ \ [F_4] \ 1 \ \underset{[\Sp(1)]}{\overset{\kg_2}{3}} \ 1 \ \overset{\kso(8)}{4} \ 1 \ \overset{\kso(8)}{4} \ 1 \ \overset{\kso(8)}{4} \ 1 \ [\SO(8)].
    \label{eq:orbi_instanton_etype_induced}
\end{align}

\subsection{Little String Theories}\label{LSTs}
To wrap up our section of the induced flows involving M9-branes, we make a very straightforward generalization by including another M9-brane. This means that we are switching to the Ho\v{r}ava-Witten setup, which shrinks to the $E_8 \times E_8$ heterotic string theory when the two M9-branes get close to each other. In this setup, we get little string theories as opposed to SCFTs, marked by the unique finite scale $M_{\text{LST}}$ that is the string scale \cite{Witten:1995zh,Aspinwall:1996vc,Seiberg:1997zk,Losev:1997hx,Intriligator:1997dh,Hanany:1997gh,Brunner:1997gf,Bhardwaj:2015oru,DelZotto:2022ohj}.

Now, our 11d spacetime is $\bbR_\perp \times \bbC^2/\Gamma \times M_6$. We have a common $\bbC^2/\Gamma$ singularity stretching along $\bbR_\perp$ between the two M9-branes. Therefore, on each M9-brane, we have the option of turning on a flat $E_8$ connection parametrized by a discrete homomorphism $\Hom(\Gamma, E_8)$. In the F-theory description, we will get a linear quiver for the UV theory, and each homomorphism will affect one side of the UV theory.

Let us consider an example that takes some cases in \eqref{eq:orbi_instanton_dtype_endpoint_changing} and \eqref{eq:orbi_instanton_etype_induced} and includes another M9 wall on the other end with another (generically different) discrete homomorphism. The induced flows between the LSTs are
\begin{align}
    & [F_4] \ 1 \ \underset{[\Sp(1)]}{\overset{\kg_2}{3}} \ 1 \ {\overset{\kf_4}{5}} \ 1 \ \overset{\ksu(3)}{3} \ 1 \ \overset{\ke_6}{6} \ 1 \ \overset{\ksu(3)}{3} \ 1 \ \overset{\kf_4}{5} \ 1 \ \overset{\kg_2}{3} \ \overset{\ksu(2)}{2} \ 1 \ [E_7]\\
    \longrightarrow & [F_4] \ 1 \ \underset{[\Sp(1)]}{\overset{\kg_2}{3}} \ 1 \ \overset{\kso(9)}{4} \ \underset{[N_f = 1/2]}{\overset{\ksp(1)}{1}} \ \overset{\kso(10)}{4} \ \underset{[N_f = 1/2]}{\overset{\ksp(1)}{1}} \ \overset{\kso(9)}{4} \ 1 \ {\overset{\kg_2}{3}} \ {\overset{\ksu(2)}{2}}\ 1 \ [E_7].
\end{align}
By comparing this induced flow with the induced flow \eqref{eq:orbi_instanton_dtype_endpoint_changing} between the orbi-instanton theories, we see that the tensor branch descriptions for the relevant combination of the curves in the middle exhibit the identical pattern: ``513161315 to 41414'' with suitable fibres. From this, we learn that the F-theory description is capable of capturing the universal behaviours of induced flows even if we change the boundary condition.

It was proposed in \cite{DelZotto:2022ohj} that the 2-group structure constant
\begin{equation}
    \widehat{\kappa}_R=\sum_I\ell_Ih^{\vee}_{\kg_I}
\end{equation}
of a given LST would decrease when flowing to the IR\footnote{We will check the monotonicity of the Weyl anomaly for the SCFTs in \S\ref{athm}.}. For an LST of rank $n_T$, the subscript $I=1,\dots,n_T+1$. Let $\eta$ be the matrix encoding the intersections of the curves such that $\eta^{II}$ are the positive integers between 0 and 12 and the off-diagonal $\eta^{IJ}$ are the non-positive integers indicating the adjacency of the curves. The little string charges $\ell_I$ form the (unique) null vector of $\eta$ such that
\begin{equation}
    \eta^{IJ}l_J=0,\quad\text{gcd}(\ell_I)=1,\quad\ell_I>0.
\end{equation}
The dual Coxeter numbers of the gauge algebras $\mathfrak{g}_I$ are denoted as $h^{\vee}_{\mathfrak{g}_I}$.

We shall now check that $\widehat{\kappa}_R^{\text{UV}} > \widehat{\kappa}_R^{\text{IR}}$ for the above example. The little string charges of the two theories can be computed as
\be
    \ell^{\text{UV}} = (1, 1, 2, 1, 3, 2, 3, 1, 3, 2, 3, 1, 2, 1, 1, 1),\quad\ell^{\text{IR}} = (1, 1, 2, 1, 2, 1, 2, 1, 2, 1, 1, 1).
\ee
After contracting with the suitable $\kg_I$, we get $\widehat{\kappa}_R^{\text{UV}} = 52$, and $\widehat{\kappa}_R^{\text{IR}} = 40$, indeed satisfying $\widehat{\kappa}_R^{\text{UV}} > \widehat{\kappa}_R^{\text{IR}}$.

Let us consider another example which flows into a pair of irreducible LSTs. For each theory, we take the pair of discrete homomorphisms to be the same for simplicity. The flow goes as
\begin{align}
&[\SO(12)]\ \overset{\ksp(1)}{1}\ \overset{\kso(7)}{3} \ \underset{[N_f = 1]}{\overset{\ksp(1)}{1}} \ \left\langle \overset{\kso(10)}{4} \ \ \overset{\ksp(1)}{1} \right\rangle^{\otimes 6} \ \ \overset{\kso(10)}{4} \ \ \underset{[N_f = 1]}{\overset{\ksp(1)}{1}} \overset{\kso(7)}{3} \ \overset{\ksp(1)}{1}\ [\SO(12)] 
\\
\longrightarrow & [E_7] \ 1 \ \underset{[\SU(2)]}{\overset{\ksu(2)}{2}}\  \underset{[\SU(2) \times \SU(2)_{\text{deloc}}]}{\overset{\ksu(2)}{2}}\ 1 \ [E_7]\ \ \sqcup\ \  [\SO(14)]\ \overset{\ksp(1)}{1}\ {\overset{\ksu(3)}{2}} \ \underset{[N_f = 1]}{\overset{\ksu(4)}{2}} \ \underset{[N_f = 1]}{\overset{\ksu(4)}{2}}\ \overset{\ksu(3)}{2}\ \overset{\ksp(1)}{1} \ [\SO(14)].
\label{eq:LST_dtype_endpoint_changing}
\end{align}
The change of the quaternionic dimension is $275-93-181=1$ as expected. To reproduce this answer, we recall that the LST $0$ has the Higgs branch of quaternionic dimension 0 while the LSTs $[E_8]\ 1 \ 1\ [E_8]$ and $\overset{\ksp(1)}{0} [\SO(32) \times \SU(2)]$ both have the Higgs branches of quaternionic dimension 29. In this case, we have $\widehat{\kappa}_R^{\text{UV}} = 102$ while $\widehat{\kappa}_R^{\text{IR}} = \widehat{\kappa}_R^{\text{IR}, 1} + \widehat{\kappa}_R^{\text{IR}, 2} = 4 + 18 = 22$, again satisfying $\widehat{\kappa}_R^{\text{UV}} > \widehat{\kappa}_R^{\text{IR}}$.

\section{The \texorpdfstring{$a$}{a}-Monotonicity for Induced Flows}\label{athm}
In this section, we shall check the monotonicity of the Weyl anomaly for the induced flows of the SCFTs. Unlike in 2d and 4d, the monotonicity for the $a$-anomaly in 6d (i.e. the coefficient of Euler density in $\langle T^\mu_{\ \ \mu} \rangle$) has not yet been established in general (however, see \cite{Cordova:2015vwa} for the statement on all tensor branch flows). Indeed, there are important evidences \cite{Elvang:2012st,Heckman:2021nwg} emphasizing the qualitative difference of 6d from lower dimensional cases, due to the presence of string excitations. Hence, there are non-trivial conceptual motivations behind examining the $a$-monotonicity for as many cases as possible. In fact, this also serves as one of the important motivations behind our programme of identifying all atomic Higgsings in 6d.

One optimistic goal is to express the relevant quantities
\begin{equation}
    \Delta X:=\Delta X_{\text{IR}}-\Delta X_{\text{UV}},
\end{equation}
where $X=\alpha,\beta,\gamma,\delta,a$, using the data of ($\kg$, $\kl$, $\calO_\kl$, $\calO_\kg = \Ind^{\kg}_{\kl}(\calO_\kl)$; $k_1$, $k_2$, $k = k_1 + k_2$). Here, $k_1$ and $k_2$ are the numbers of M5-branes in the two stacks after splitting. In the followings, we shall mainly focus on some specific families of theories. It would be nice to have some general expressions out of the classical examples and systematically check the $a$-monotonicity explicitly, generalizing \cite{Mekareeya:2016yal}.

\paragraph{The A-type theories} Let us first argue that the $a$-anomaly does decrease under the induced flows for the A-type long quivers (although such a flow should be guaranteed to be valid from the perspective of the fission of the magnetic quiver). We recall that the anomaly coefficients in the 8-form anomaly polynomial (where $p_i(T)$ is the $i^{\text{th}}$ Pontryagin class of the 6d tangent bundle, and $c_2(R)$ is the second Chern class of the $\SU(2)_R$ bundle)
\be
    I_8 = \alpha c_2(R)^2 + \beta c_2(R) p_1(T) + \gamma p_1(T)^2 + \delta p_2(T)
\ee
are given by \cite{Mekareeya:2016yal}
\begin{align}
    &\alpha=\frac{1}{24}\left(12\sum_{i,j=1}^{N-1}\left(C^{-1}\right)_{ij}r_ir_j+2(N-1)-\sum_{i=1}^{N-1}r_i^2\right),\\
    &\beta=\frac{1}{24}\left(N-1-\frac{1}{2}\sum_{i=1}^{N-1}r_i^2\right),\\
    &\gamma=\frac{1}{5760}\left(\frac{7}{2}\sum_{i=1}^{N-1}r_if_i+30(N-1)\right),\\
    &\delta=-\frac{1}{2880}\left(\sum_{i=1}^{N-1}r_if_i+60(N-1)\right),
\end{align}
where $N-1$ is the number of tensor multiplets (which is equal to the number of $-2$ curves) and $r_i$ is the rank of the gauge algebra on the $i^{\text{th}}$ curve. The balance is given by $-f_i=r_{i+1}-r_{i-1}-2r_i$. The Cartan matrix of $A_{N-1}$ is denoted as $C$.

Given a long quiver, let us take the nilpotent orbit $\mathcal{O}_L$ (as $\mathcal{O}_R$ follows the same argument). Since $\Delta\gamma=\Delta\delta=0$, it suffices to consider
\begin{equation}
    24(\alpha-\beta)=12\sum_{i,j=1}^{N-1}\left(C^{-1}\right)_{ij}r_ir_j+N-1-\frac{1}{2}\sum_{i=1}^{N-1}r_i^2.
\end{equation}
Suppose that the $A_{N-1}$ orbit labelled by the partition $\bm{r}$ is induced from an $A_{K_s-1}$ orbit with partition $\bm{s}$ and an $A_{K_t-1}$ orbit with partition $\bm{t}$. We have\footnote{Notice that this also includes the situation where $K_t=1$.}
\begin{equation}
    \bm{r}^\text{T}=\bm{s}^\text{T}\sqcup\bm{t}^\text{T},\label{rsplitst}
\end{equation}
where the reshuffling of the columns is implicit on the right hand side to guarantee that we still have a Young diagram after taking the union. Moreover,
\begin{equation}
    r_i=\sum_{j=1}^i\left(\bm{r}^\text{T}\right)_j,\quad s_i=\sum_{j=1}^i\left(\bm{s}^\text{T}\right)_j,\quad t_i=\sum_{j=1}^i\left(\bm{t}^\text{T}\right)_j,\label{risumrT}
\end{equation}
for the gauge algebras $\ksu(r_i)$ governed by $\mathcal{O}_L$ (and likewise for $\ksu(s_i)$, $\ksu(t_i)$)\footnote{We notice that the ranks of the gauge groups may be related to the closures of the nilpotent orbits and their symplectic resolutions as follows. Given a nilpotent orbit $\mathcal{O}$ with partition $\bm{r}$ in $\mathfrak{sl}(n)$, retain the notation in \eqref{risumrT}. Then its closure is given by $\overline{\mathcal{O}}=\{A\in\mathfrak{sl}(n)|\dim_{\mathbb{C}}(\ker A^j)\geq r_j\}$. It admits a symplectic resolution $\pi:T^*F\rightarrow\overline{\mathcal{O}}$, where $F$ is the flag variety defined as $F=\{(V_1,\dots,V_l)|\dim_{\mathbb{C}}(V_j)=r_j,~V_j\subset V_{j+1},~\forall j\}$. Its cotangent bundle $T^*F$ is isomorphic to the coincidence variety $\{(A,V_{\cdot})\in\mathfrak{sl}(n)\times F|AV_j\subseteq V_{j-1},~\forall j\}$. Of course, for closures of nilpotent orbits of other types, not all of them admit symplectic resolutions.}. This can be illustrated as
\begin{equation}
    \includegraphics[width=10cm]{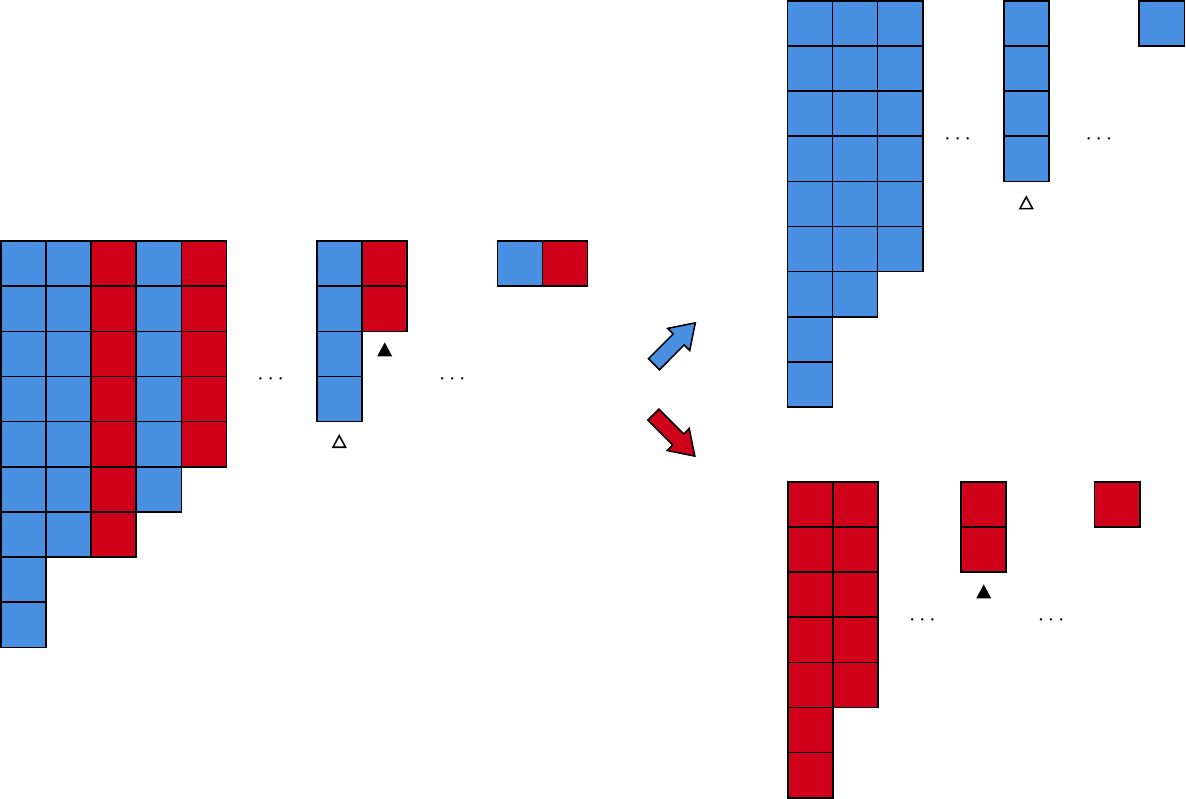}.
\end{equation}
Suppose that the blue (resp.~red) column labelled by the triangle is the $i^\text{th}$ column in the blue (resp.~red) Young diagram\footnote{Recall that in our convention, we read the partition in terms of the rows and its transpose in terms of the columns.}. Then $s_i$ (resp.~$t_i$) counts the number of boxes of all the blue (resp.~red) columns left to this labelled column (including the labelled column itself).

It is clear that $(N-1)>(K_s-1)+(K_t-1)$ as $N=K_s+K_t$. For the remaining factors, let us first consider the off-diagonal part, namely the $r_ir_j$ terms with $i\neq j$. As the inverse Cartan matrix for $A_{N-1}$ is given by \cite{wei2017inverses}
\begin{equation}
    \left(C^{-1}\right)_{ij}=\min\{i,j\}-\frac{ij}{N},
\end{equation}
we have
\begin{align}
    \sum_{i\neq j}^{N-1}\left(C_{ij}\right)^{-1}r_ir_j&=\sum_{i\neq j}^{N-1}\left(\min\{i,j\}-\frac{ij}{N}\right)\left(\sum_{k=1}^i\left(\bm{r}^\text{T}\right)_k\right)\left(\sum_{k=1}^j\left(\bm{r}^\text{T}\right)_k\right)\nonumber\\
    &>\sum_{i\neq j}^{K_s-1}\left(\min\{i,j\}-\frac{ij}{K_s}\right)\left(\sum_{k=1}^i\left(\bm{s}^\text{T}\right)_k\right)\left(\sum_{k=1}^j\left(\bm{s}^\text{T}\right)_k\right)+[s\rightarrow t]\nonumber\\
    &=\sum_{i\neq j}^{K_s-1}\left(C_{ij}\right)^{-1}s_is_j+\sum_{i\neq j}^{K_t-1}\left(C_{ij}\right)^{-1}t_it_j.
\end{align}
The inequality comes from the following two facts:
\begin{itemize}
    \item $\min\{i,j\}-\frac{ij}{N}>\min\{i,j\}-\frac{ij}{K_s}$ as $N>K_s$ (and likewise for $K_t$);
    \item For each $s_i$ or $t_i$, one can find an associated $r_k$ such that its last column/summand in \eqref{risumrT} corresponds to the last column/summand of $s_i$ or $t_i$. In particular, each $s_i$ or $t_i$ has a distinct corresponding $r_k$ due to \eqref{rsplitst}. Moreover, $r_k\geq s_i$ since
    \begin{equation}
        \left\{\left(\bm{s}^\text{T}\right)_j\bigg|1\leq j\leq i\right\}\subseteq\left\{\left(\bm{r}^\text{T}\right)_j\bigg|1\leq j\leq k\right\},
    \end{equation}
    and likewise for $t_i$. Pictorially, this can be illustrated as
    \begin{equation}
        \includegraphics[width=8cm]{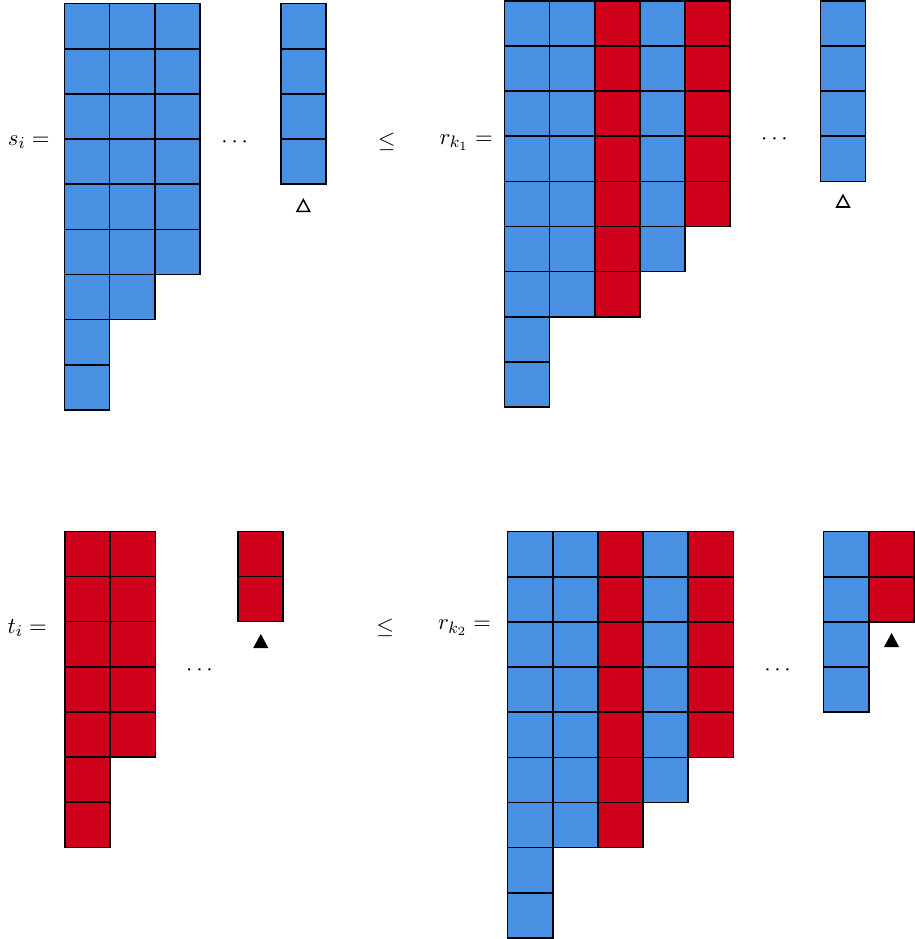}.
    \end{equation}
\end{itemize}

For the diagonal ones, namely the $r_ir_j$ terms with $i=j$, the coefficients in $24(\alpha-\beta)$ read
\begin{equation}
    12\left(C^{-1}\right)_{ii}-\frac{1}{2}=\min\{i,i\}-\frac{i^2}{N}-\frac{1}{2}=\frac{i(N-i)}{N}-\frac{1}{2}\geq\frac{N-1}{N}-\frac{1}{2}\geq0,
\end{equation}
where the equalities are saturated when $N=2$ (which is the possible minimal $N$). Therefore, the coefficients of $r_ir_i$ are no less than those of $s_is_i$ and $t_it_i$, similar to the argument for the off-diagonal terms. As a result, $\Delta\alpha-\Delta\beta<0$, and hence, $\Delta a<0$.

\paragraph{The D-type conformal matters} For the D-type conformal matter theories, the monotonicity of the $a$-anomaly follows a similar argument. Again, as $\Delta\gamma$ and $\Delta\delta$ are zero, it suffices to consider the coefficients $\alpha$ and $\beta$ in the anomaly polynomial. Now, we have \cite{Mekareeya:2016yal}
\begin{align}
    &\alpha=\frac{1}{24}\left(6\sum_{i,j=1}^{N-1}\left(C^{-1}\right)_{ij}r_ir_j+12\sum_iq_i+7N-1-n_V\right),\\
    &\beta=\frac{1}{24}\left(-3\sum_i(2+q_i)+\frac{1}{2}(N-1)-\frac{1}{2}n_V\right),\\
    &\gamma=\frac{1}{24}\left(\frac{7}{240}\left(-n_V+\frac{1}{2}\sum_i(p_iq_i+q_{i+1}p_i+f_ip_i+g_iq_i)\right)+\frac{23}{240}(N-1)+\frac{3}{8}N\right),\\
    &\delta=-\frac{1}{1440}\left(-n_V+\frac{1}{2}\sum_i(p_iq_i+q_{i+1}p_i+f_ip_i+g_iq_i)+29(N-1)\right),
\end{align}
where $N-1$ is the total number of $-1$- and $-4$ curves and $r_i$ is twice the rank of the gauge algebra on the $i^{\text{th}}$ curve. We have also introduced $p_i=r_{2i}$ and $q_i=r_{2i-1}$ so that the gauge algebras are $\mathfrak{so}(p_i)$ and $\mathfrak{sp}(q_i/2)$. The number of vector multiplets is given by
\begin{equation}
    n_V=\sum_i\left(\frac{1}{2}p_i(p_i-1)+\frac{1}{2}q_i(q_i+1)\right).
\end{equation}
Moreover,
\begin{equation}
    f_i=2p_i-16-q_i-q_{i+1},\quad g_i=2q_i+16-p_i-p_{i+1}.
\end{equation}
Now, the ``Cartan'' matrix $C_{ij}$ has alternating 1 and 4 as diagonal terms instead of 2 (where the first and the last diagonal terms are both 1). Therefore, its inverse reads
\begin{equation}
    \left(C^{-1}\right)_{ij}=\frac{1}{2^{\mathfrak{p}(i)+\mathtt{p}(j)-1}}\left(\min\{i,j\}-\frac{ij}{2K}\right),
\end{equation}
where $\mathfrak{p}(n)=\frac{1+(-1)^n}{2}$ which outputs 0 (resp.~1) for an odd (resp.~even) number.

Suppose a D-type orbit $\bm{r}$ is induced from $\bm{s}\oplus\bm{t}$ where $\bm{s}$ and $\bm{t}$ are A-type and D-type orbits respectively\footnote{For brevity, we simply use the corresponding partition to denote the nilpotent orbit. For the Levi subalgebras $A_{n-1}\oplus A_0$ and $A_{n-3}\oplus2A_1$, the arguments are similar and can be treated as $A_{n-1}\oplus A_0\oplus D_0$ and $A_{n-3}\oplus D_2$.}. If there are $(\bm{r})_j$ boxes in the $j^\text{th}$ row of $\bm{r}$ (and likewise for $\bm{s}$ and $\bm{t}$), then
\begin{equation}
    \bm{r}=\left(\widehat{\bm{r}}\right)_D,\quad(\widehat{\bm{r}})_j:=2(\bm{s})_j+(\bm{t})_j
\end{equation}
following \eqref{soind}. As $\bm{r}$ is the D-collapse of $\widehat{\bm{r}}$, we always have
\begin{equation}
    \left(\bm{r}^\text{T}\right)_j\geq\left(\widehat{\bm{r}}^\text{T}\right)_j.
\end{equation}
In other words, the $j^\text{th}$ column in $\bm{r}$ would always have more boxes than the $j^\text{th}$ column in $\widehat{\bm{r}}$ has. Since we still have
\begin{equation}
    r_k=\sum_{j=1}^k\left(\bm{r}^\text{T}\right)_j,
\end{equation}
where $r_{2i}=p_i$, $r_{2i-1}=q_i$ (and likewise for $\widehat{r}_k$, $s_k$ and $t_k$)\footnote{Here, we have similarly defined $\widehat{r}_k=\sum\limits_{j=1}^k\left(\widehat{\bm{r}}^\text{T}\right)_j$.}. Following the similar argument as in the A-type case, we have
\begin{equation}
    \mathcal{C}(r_i)\geq\mathcal{C}\left(\widehat{r}_i\right)\geq\mathcal{C}(s_i\text{ or }t_i),\quad\mathcal{C}(r_ir_j)\geq\mathcal{C}\left(\widehat{r}_i\widehat{r}_j\right)\geq\mathcal{C}(s_is_j\text{ or }t_it_j),\label{Dcoeffcompare}
\end{equation}
where $\mathcal{C}(x)$ denotes the coefficient of the term $x$ in $24(\alpha-\beta)$. To apply \eqref{Dcoeffcompare}, we have also used the fact that there is an associated $r_i$ for each $s_i$ or $t_i$. This can be seen as follows. Suppose we have a configuration $1414\dots141$ with $K$ $(-1)$ curves and $(K-1)$ $(-4)$ curves in the UV. After an induced flow, we have two pieces. One configuration is $1414\dots141$ with $k$ $(-1)$ curves and $(k-1)$ $(-4)$ curves. The other has a string of $(K-k-1)$ $(-2)$ curves. In particular, the number of $-1$ curves (resp.~$-4$ curves) is greater than the number of $-1$- and $-2$ curves (resp.~$-4$ curves) in the IR. We can then realize \eqref{Dcoeffcompare} by assigning the $-m$ curves (with the corresponding $t_i$) in the IR to the $-m$ curves (with the corresponding $r_i$) in the UV, where $m=1,4$. Moreover, we can assign the $-2$ curves (with the corresponding $s_i$) in the IR to the (remaining) $-1$ curves (with the corresponding $r_i$) in the UV. As a result, $24(\Delta\alpha-\Delta\beta)<0$ from comparing the coefficients, and we still have $\Delta a<0$.

In this paper, we have only shown the $a$-monotonicity for the A-type and D-type $(2,0)$ SCFTs on the A-type bases. For the E-type gauge theories, this can be checked case by case. Besides these theories, one may also consider those on the D-type and E-type bases, as well as the orbi-instanton theories. We expect that the $a$-monotonicity would still hold following the results of the anomaly polynomials in \cite{Mekareeya:2017jgc,Baume:2023onr} although the computations would be more involved.

\section*{Acknowledgement}
We would like to thank Darrin D.~Frey, Amihay Hanany, Deshuo Liu, Dmytro Matvieievskyi, Hiraku Nakajima, Yuji Tachikawa, Gabi Zafrir, and Zhenghao Zhong for enjoyable discussions. JB is supported by a JSPS fellowship. HYZ is supported by WPI Initiative, MEXT, Japan at Kavli IPMU, the University
of Tokyo.

\appendix

\section{Mathematical Preliminaries to Nilpotent Orbits}\label{preliminaries}
In this appendix, we briefly some review relevant Lie algebra data that are useful for us. The standard textbook is \cite{collingwood1993nilpotent}. See also \cite{Chacaltana:2012zy} for a comprehensive study on nilpotent orbits in physics.

\paragraph{Nilpotent elements} In a Lie algebra $\kg$, any element $X$ naturally gives an adjoint action
\be
    \mathrm{ad}_X: \kg \rightarrow \mathrm{end}(\kg), \ \ Y \mapsto \mathrm{ad}_X(Y)=[X, Y], \ \ \forall \ Y \in \kg.
\ee
We say that an element $X \in \kg$ is \textit{nilpotent} if its adjoint action is a nilpotent action, i.e., if $\adx$ is a nilpotent endomorphism of $\kg$ as a vector space. Similarly, we can define $X \in \kg$ to be \textit{semisimple} if $\adx$ is a semisimple endomorphism of $\kg$, that is, every $\adx$-invariant subspace has an $\adx$-invariant complement. The nilpotent and semisimple elements are important due to the Jordan decomposition (see \cite[\S1.1]{collingwood1993nilpotent}).

\paragraph{Adjoint orbits of nilpotent elements} To classify the nilpotent and semisimple elements of $\kg$, we would like to mod out some equivalence relations. For instance, in the case of $\mathfrak{sl}(n)$, the sizes of the Jordan blocks are the data that we want to extract while their arrangements are less important. In general, the equivalence classes are determined by the \textit{adjoint} action, which can be written as
\be
    \mathrm{Aut}(\kg)^\text{o} \cdot X = \{\phi(X) | \phi \in \mathrm{Aut}(g)^\text{o}\},
\ee
where $\mathrm{Aut}(\kg)^\text{o}$ is the identity component of $\mathrm{Aut}(\kg)$. Notice that we have $\phi\cdot\adx\cdot\phi^{-1}=\text{ad}_{\phi(X)}$. Now, the \textit{nilpotent orbit} in $\kg$ is the conjugacy class of the nilpotent element $X$ under the adjoint actions, which we denote as $\calO_X$. The semisimple orbit of a semisimple element is defined in a similar manner. In fact, given a semisimple Lie algebra with Cartan subalgebra $\mathfrak{h}$ and Weyl group $W$, the semisimple orbits are in bijective correspondence with $\mathfrak{h}/W$, and there are hence infinitely many distinct semisimple orbits. Topologically, any semisimple orbit in a reductive Lie algebra is simply connected. These facts can be found for example in \cite{collingwood1993nilpotent}. In contrast, the topology of nilpotent orbits is more involved, and the nilpotent orbits are also our main focus in the paper.

The dimension of a nilpotent orbit can be computed using the \textit{centralizer/commutant subalgebra} of $X$ in $\kg$:
\be
    \kg^{X} = \{Y \in \kg \ | \ [X, Y] = 0\}.
\ee
The complex dimension of a nilpotent orbit is then $\dim_{\mathbb{C}}(\calO_X)=\dim_{\mathbb{C}}(\kg) - \dim_{\mathbb{C}}(\kg^X)$. Therefore, within a fixed Lie algebra $\kg$, an element $X$ whose nilpotent orbit $\calO_X$ of larger dimension is the one whose commutant subalgebra $\kg^X$ is of smaller dimension.

\paragraph{Closure inclusions of nilpotent orbits} The nilpotent orbits, albeit mutually disjoint from each other as equivalence classes, it is possible to define a partial ordering on them via the Zariski \textit{closure inclusion}:
\be
    \calO_X \preceq \calO_Y \quad \text{if} \quad \calO_X \subseteq \overline{\calO}_Y.
\ee
Under this partial ordering, all nilpotent orbits of a Lie algebra form a \textit{Hasse diagram}\footnote{We remark that Hasse diagrams arise in multiple contexts in our paper. Here, we are discussing in a purely mathematical context about Hasse diagrams of the nilpotent orbits under the closure inclusions. In the main body of the paper, the Hasse diagrams are concerned with the SCFTs under the Higgs branch RG flows. We need to carefully distinguish these two contexts. They are not the same in general although they could coincide in some specific contexts such as the nilpotent Higgsings of class $\mathcal{S}$ theories or 6d SCFTs.}.

\paragraph{Nilpotent orbits in classical Lie algebras} Nilpotents orbits in the simple Lie algebras of classical types can be completely classified and labelled by partitions of integers:
\begin{itemize}
    \item The $A_{n-1}=\mathfrak{sl}(n)$ nilpotent orbits are in one-to-one correspondence with the partitions of $n$.
    \item The $B_n=\kso(2n+1)$ nilpotent orbits are in one-to-one correspondence with the partitions of $2n+1$ where the even parts occur with even multiplicities.
    \item The $C_n=\ksp(n)$ nilpotent orbits are in one-to-one correspondence with the partitions of $2n$ where the odd parts occur with even multiplicities.
    \item The $D_n=\kso(2n)$ nilpotent orbits are labelled by the partitions of $2n$ where the even part occur with even multiplicities. In the case of \textit{very even} partitions with only even parts (and with even multiplicities), each partition labels two nilpotent orbits, which are distinguished by the superscript I, II. (Otherwise, each partition that is not very even labels one nilpotent orbit.) More concretely, the weighted Dynkin diagrams of the two very even orbits would have the weights on the bifurcating ends exchanged\footnote{According to the Jacobson-Morozov theorem, one can associate a standard $\mathfrak{sl}(2)$-triple to a nilpotent orbit, and the weighted Dynkin diagram is determined by the (diagonal) matrices in the triple.}.
\end{itemize}
As the partitions for the BCD types are not arbitrary, we shall refer to the partitions that correspond to the orbits in these cases as the X-partitions (where X is B, C or D). We say that a partition $\bm{p}=[p_1,\dots,p_n]$ dominates $\bm{p}'=[p'_1,\dots,p'_{n'}]$ and write $\bm{p}'\preceq\bm{p}$ if
\begin{equation}
    \sum_{i=1}^kp_i\geq\sum_{i=1}^kp'_i
\end{equation}
for all $k$ (where 0 can always be added to the one with fewer parts). Then $\mathcal{O}_{\bm{p}}\preceq\mathcal{O}_{\bm{q}}$ is equivalent to $\bm{p}\preceq\bm{q}$.

\paragraph{Nilpotent orbits in exceptional Lie algebras} For exceptional Lie algebras, their nilpotent orbits are often denoted by the \emph{Bala-Carter labels}. To understand the BC labels, we first recall that a \emph{parabolic subalgebra} $\mathfrak{p}$ of a semisimple Lie algebra $\mathfrak{g}$ is a subalgebra containing a Borel subalgebra. It admits a Levi decomposition $\mathfrak{p}=\mathfrak{l}\oplus\mathfrak{n}$, where $\mathfrak{l}$ is reductive and $\mathfrak{n}$ is the nilradical of $\mathfrak{p}$. In particular, $\mathfrak{l}$ is called the \emph{Levi subalgebra} of $\mathfrak{g}$. If $\dim(\mathfrak{l})=\dim(\mathfrak{n}/[\mathfrak{n},\mathfrak{n}])$, then we say that the parabolic subalgebra $\mathfrak{p}$ is distinguished. It turns out that there is a one-to-one correspondence between the nilpotent orbits of $\mathfrak{g}$ and the $\text{Aut}(\mathfrak{g})^\text{o}$-conjugacy classes of pairs $(\mathfrak{l},\mathfrak{p})$ where $\mathfrak{l}$ is a Levi subalgebra of $\mathfrak{g}$ and $\mathfrak{p}$ is a distinguished parabolic subalgebra of the semisimple algebra $[\mathfrak{l},\mathfrak{l}]$.

Now, the BC label of a nilpotent orbit reads $X_N(a_i)$, where $X_N$ is the Cartan type of the semisimple part of $\mathfrak{l}$. It further contains $i$, the number of simple roots in any Levi subalgebra of $\mathfrak{p}$. If $i=0$, it is often denoted as $X_N$ with $a_0$ omitted. When there are two orbits having the same $X_N$ and the same $i$, one is chosen arbitrarily to have the label $X_N(a_i)$ and the other $X_N(b_i)$. There could also be non-conjugate isomorphic Levi subalgebras. In such cases, one often puts tildes or primes on the labels to distinguish them. The lists of the nilpotent orbits and their BC labels for the exceptional Lie algebras can be found for example in \cite{collingwood1993nilpotent,spaltenstein2006classes}.

We notice that some of the BC labels used in \cite{spaltenstein2006classes} are different from those in other literature such as \cite{collingwood1993nilpotent,Chacaltana:2012zy}. For reference, let us list the different ones as in Table \ref{differentBClabels}.
\begin{longtable}{|c|ccc||c|ccc|}
\hline
\begin{tabular}{c}Lie\\type\end{tabular} & BC1 & BC2 & BC3 & \begin{tabular}{c}Lie\\type\end{tabular} & BC1 & BC2 & BC3 \\ \hline\hline
$E_6$ & $E_6(a_3)$ & $A_5+A_1$ & - & $E_8$ & $E_8(a_5)$ & $D_8(a_1)$ & - \\ \hline
$E_7,E_8$ & $E_7(a_3)$ & $D_6+A_1$ & - & $E_8$ & $E_8(b_5)$ & $E_7(a_2)+A_1$ & $E_6+A_2$ \\ \hline
$E_7,E_8$ & $E_7(a_4)$ & $D_6(a_1)+A_1$ & - & $E_8$ & $E_8(a_6)$ & $A_8$ & $D_8(a_2)$ \\ \hline
$E_7$ & $E_7(a_5)$ & $D_6(a_2)+A_1$ & $A_5+A_2$ & $E_8$ & $E_8(b_6)$ & $D_8(a_3)$ & $E_6(a_1)+A_2$ \\ \hline
$E_7$ & $E_6(a_3)$ & $(A_5+A_1)'$ & - & $E_8$ & $E_8(a_7)$ & $2A_4$ & \begin{tabular}{c}$A_5+A_2+A_1$\\$A_6(a_2)+2A_1$\end{tabular} \\ \hline
$E_7$ & $A_5+A_1$ & $(A_5+A_1)''$ & - & $E_8$ & $E_7(a_5)$ & $A_5+A_2$ & $D_6(a_2)+A_1$ \\ \hline
$E_8$ & $E_8(a_3)$ & $E_7+A_1$ & - & $E_8$ & $E_6(a_3)+A_1$ & $A_5+2A_1$ & - \\ \hline
$E_8$ & $E_8(a_4)$ & $D_8$ & - & $E_8$ & $E_6(a_3)$ & $(A_5+A_1)''$ & - \\ \hline
$E_8$ & $E_8(b_4)$ & $E_7(a_1)+A_1$ & - & $E_8$ & $A_5+A_1$ & $(A_5+A_1)'$ & - \\
\hline
\caption{The different BC labels in the literature. For the three columns of the BC labels, the left column lists the commonly used labels such as in \cite{collingwood1993nilpotent,Chacaltana:2012zy}, and the middle column contains the labels in \cite{spaltenstein2006classes}. In the right column, we also write other alternative BC labels when possible. These different labels can be obtained by choosing different representatives of the nilpotent orbits, as can be found in \cite{de2009induced}.}\label{differentBClabels}
\end{longtable}

\paragraph{Special nilpotent orbits} There is an order-reversing map $\mathtt{d}$, the \emph{Lusztig-Spaltenstein map}, on the set of nilpotent orbits in a given Lie algebra. For the A-type case, this simply sends an orbit to the one with the transposed partition, namely $\mathtt{d}(\mathcal{O}_{\bm{p}})=\mathcal{O}_{\bm{p}^{\text{T}}}$, and it is involutive. For the other classical types, we first need to introduce the concept of the \emph{X-collapse} (where X is B, C or D). Given any partition $\bm{p}$, the X-collapse $\bm{p}_{\text{X}}$ is the unique largest X-partition dominated by $\bm{p}$. Now, the LS dual is given by $\mathtt{d}:\bm{p}\mapsto(\bm{p}^{\text{T}})_{\text{X}}$. For the two very even orbits associated to the same partition, the labels I and II would also get exchanged under $\mathtt{d}$. As a result, the map is not involutive, but rather $\mathtt{d}^2(\bm{p})\succeq\bm{p}$. Nevertheless, we still have $\mathtt{d}^3(\bm{p})=\mathtt{d}(\bm{p})$. Then an orbit (resp.~a partition) is said to be \emph{special} if $\mathtt{d}^2(\mathcal{O}_{\bm{p}})=\mathcal{O}_{\bm{p}}$ (resp.~$\mathtt{d}^2(\bm{p})=\bm{p}$). Otherwise, it is non-special. There is a criterion for the specialness by inspecting the partitions and their transposes directly, and readers are referred to \cite[Proposition 6.3.7]{collingwood1993nilpotent}. For the exceptional cases, since there are finitely many of the orbits, it is possible to have exhaustive lists, which can be found for example in \cite{collingwood1993nilpotent,spaltenstein2006classes,carter1985finite}.

In the classical cases, a \emph{minimal degeneration} is a manipulation on a given partition that takes an orbit $\mathcal{O}$ to another orbit $\mathcal{O}'$ such that there exists no other orbit $\mathcal{O}''$ such that $\mathcal{O}\succeq\mathcal{O}''\succeq\mathcal{O}'$ \cite{kraft1989special,kraft1982geometry} (see also \cite{Balasubramanian:2023iyx}). It could sometimes be useful to think of these Kraft-Procesi moves when considering the nilpotent orbits in the same special piece, namely those having the same image under the LS map. Overall, they were classified and labelled as (a)$\sim$(g) in \cite{kraft1989special,kraft1982geometry}. For the (a) and (g) moves, only one box would be moved (while the others involve moving two boxes). These two moves are called the \emph{small degenerations}, and let us spell out them here. An (a) move acts on two adjacent rows in a partition where one row has $2n+1$ boxes and the other has $2n-1$ boxes. After this move, both rows would have $2n$ boxes. A (g) move involves $2m+2$ rows in a partition where the first (resp.~last) row has $2n+1$ (resp.~$2n-1$) boxes and each row in the middle $2m$ rows has $2n$ boxes. After this move, one box in the first row would be moved to the last row, and all the $2m+2$ rows would be of the same length $2n$.

\subsection{Induced Orbits}\label{indorb}
The main dramatis personae in this paper is the concept of the induced orbits. This enables us to construct/induce new orbits in a larger algebra from orbits in a smaller one. Let $\kp$ be a parabolic sublgebra of $\kg$ with Levi decomposition $\kl \oplus \kn$. Suppose that $\calO_\kl$ is a nilpotent orbit in $\kl$. Then there is a unique nilpotent orbit $\mathcal{O}_{\mathfrak{g}}$ in $\mathfrak{g}$ meeting $\mathcal{O}_{\mathfrak{l}}+\mathfrak{n}$ in an open dense set. The nilpotent orbit $\mathcal{O}_{\mathfrak{g}}$ is said to be \emph{induced} from $\mathcal{O}_{\mathfrak{l}}$ and is denoted by \cite{lusztig1979induced}
\begin{equation}
    \calO_\kg = \Ind^\kg_{\kp} (\calO_\kl).
\end{equation}
The dimension of the induced orbit is given by $\dim(\mathcal{O}_{\mathfrak{g}})=\dim(\mathcal{O}_{\mathfrak{l}})+2\dim(\mathfrak{n})$. The intersection $\mathcal{O}_{\mathfrak{g}}\cap(\mathcal{O}_{\mathfrak{l}}+\mathfrak{n})$ consists of a single $P_{\text{ad}}$-orbit, where $P_{\text{ad}}$ is the connected Lie group of $\text{Aut}(\mathfrak{g})^{\text{o}}$ with Lie algebra $\mathfrak{p}$. It is the unique orbit in $\mathfrak{g}$ of this dimension that meets $\mathcal{O}_{\mathfrak{l}}+\mathfrak{n}$.

In fact, the induced orbit depends only on the Levi subalgebra $\mathfrak{l}$, but not on the choice of a parabolic subalgebra $\mathfrak{p}$. In other words, $\text{Ind}^{\mathfrak{g}}_{\mathfrak{p}}(\mathcal{O}_{\mathfrak{l}})=\text{Ind}^{\mathfrak{g}}_{\mathfrak{p}'}(\mathcal{O}_{\mathfrak{l}})$ for $\mathfrak{p}=\mathfrak{l}\oplus\mathfrak{n}$ and $\mathfrak{p}'=\mathfrak{l}\oplus\mathfrak{n}'$. Hence, we shall often use the following notations interchangeably:
\begin{equation}
    \mathcal{O}_{\mathfrak{g}}=\text{Ind}^{\mathfrak{g}}_{\mathfrak{p}}(\mathcal{O}_\mathfrak{l})=\text{Ind}^{\mathfrak{g}}_{\mathfrak{l}}(\mathcal{O}_\mathfrak{l})=\text{Ind}(\mathcal{O}_\mathfrak{l}).
\end{equation}

Let us list some results of the induced orbits as they can be physically interpreted as a particular type of Higgsings which we refer to as induced Higgsings/flows. For $\mathfrak{g}=\mathfrak{sl}(n)$, consider the nilpotent orbit $\mathcal{O}_{\bm{p}^1}\oplus\dots\oplus\mathcal{O}_{\bm{p}^r}$ labelled by partitions $\bm{p}^k=[p^k_1,\dots,p^k_n]$ in a Levi subalgebra (where without loss of generality, we may assume that $\bm{p}^k$ has $n$ parts by padding it with 0 when necessary). This would induce the orbit $\text{Ind}(\mathcal{O})=\mathcal{O}_{\Sigma\bm{p}}$ in $\mathfrak{sl}(n)$, where $\Sigma\bm{p}$ denotes the partition with the $i^{\text{th}}$ part $p^1_i+\dots+p^r_i$. In such cases, an atomic Higgsing always has $r=2$.

For $\mathfrak{g}$ of other classical types, we can write the orbit in the Levi subalgebra $\mathfrak{l}$ as\footnote{The Levi subalgebra $\mathfrak{l}$ could be found by removing a node in the Dynkin diagram of $\mathfrak{g}$ (for the atomic steps). A node on the end of a leg could be removed, and either $\bm{d}$ or $\bm{f}$ could be thought of as the trivial partition in such cases. For the D-type Dynkin diagrams, the node at the intersection of the three legs could also be removed. Then $\bm{f}$ could be understood as the partition for $\mathfrak{so}(4)\cong\mathfrak{sl}(2)\times\mathfrak{sl}(2)$.}
\begin{equation}
    \mathcal{O}_{\mathfrak{l}}=\mathcal{O}_{\bm{d}}\oplus\mathcal{O}_{\bm{f}},
\end{equation}
where $\mathcal{O}_{\bm{d}}$ (resp.~$\mathcal{O}_{\bm{f}}$) is a nilpotent orbit in $\mathfrak{sl}(l)$ (resp.~some subalgebra $\mathfrak{g}'$ of the same type as $\mathfrak{g}$). Now, define a new partition $\bm{p}=[p_1,\dots,p_N]$ such that
\begin{equation}
    p_i=2d_i+f_i.\label{soind}
\end{equation}
Then the partition of $\text{Ind}(\mathcal{O}_{\mathfrak{l}})$ is given by the partition $\bm{p}_{\text{X}}$ (for X being B, C or D). For very even orbits that appear in $\mathfrak{g}=\mathfrak{so}(4n)$, if the dimension of the standard representation of $\mathfrak{g}'$ (i.e., $2k$ for $\mathfrak{sp}(k)$ and $k$ for $\mathfrak{so}(k)$) is non-zero, then the label I or II is the same as that of $\mathcal{O}_{\bm{f}}$. If this dimension is zero, then the label I or II is the same as (resp.~different from) that of $\mathfrak{l}$ if $n$ is even (resp.~odd). Here, we assign a label (either I or II) to the Levi subalgebra $\mathfrak{l}$ as this corresponds to removing one of the two nodes on the bifurcating ends in the Dynkin diagram of $\mathfrak{g}=\mathfrak{so}(4n)$. Moreover, a special (resp.~non-special) orbit in $\mathfrak{g}$ can only be induced from an orbit whose partition for the $\mathfrak{g}'$ part is special (resp.~non-special).

\subsection{Kempken-Spaltenstein Algorithm of Induced Orbits}\label{KSalgorithm}
The \emph{Kempken-Spaltenstein (KS) algorithm} is a useful tool to obtain the rigid orbits that induce the nilpotent orbits in the BCD types \cite{premet2014derived}. One can find the rigid induction data of a given nilpotent orbit after a sequence of steps using the KS algorithm. Although we are mostly interested in the atomic Higgsings that are not necessarily given by inductions of the rigid orbits, it would also be helpful to find the atomic induced flows with the knowledge of the KS algorithm.

Given a partition $\bm{p}=[p_1,\dots,p_r]$ of an integer $n$, there could be two possible types of \emph{reductions}. A type 1 reduction can be made at the position $i$ if $p_i\geq p_{i+1}+2$. This would result in the partition $\bm{p}'$ of $2n-2i$ satisfying
\begin{equation}
    p'_j=\begin{cases}
        p_j-2,&j\leq i,\\
        p_j,&j>i.
    \end{cases}
\end{equation}
A type 2 reduction can be made at the position $i$ if $p_i$ is even (resp.~odd) for $\mathfrak{sp}(n/2)$ (resp.~$\mathfrak{so}(n)$) and $p_{i-1}>p_i=p_{i+1}>p_{i+2}$ (where we have the convention that $p_0>p_1$). This would result in the partition $\bm{p}'$ of $2n-2i$ satisfying
\begin{equation}
    p'_j=\begin{cases}
        p_j-2,&j<i,\\
        p_j-1,&i\leq j\leq i+1,\\
        p_j,&j>i+1.
    \end{cases}
\end{equation}
A position $i$ is called \emph{adimissible} if either a type 1 or a type 2 reduction can be performed at this position.

For the purpose of atomic Higgsings, it suffices to consider one step of such reductions of a given partition (although it may not give all the possible atomic induced flows as we always get the zero orbit for the $\mathfrak{sl}(l)$ part). Nevertheless, for completeness, let us state the full algorithm here. To get the rigid induction data, we simply need to apply the two types of the reductions recursively\footnote{This recursive process is due to the proposition that $\text{Ind}^{\mathfrak{g}}_{\mathfrak{l}_2}\left(\text{Ind}^{\mathfrak{l}_2}_{\mathfrak{l}_1}\left(\mathcal{O}_{\mathfrak{l}_2}\right)\right)=\text{Ind}^{\mathfrak{g}}_{\mathfrak{l}_1}\left(\mathcal{O}_{\mathfrak{l}_1}\right)$, where $\mathfrak{l}_{1,2}$ are Levi subalgebras of $\mathfrak{g}$ satisfying $\mathfrak{l}_1\subset\mathfrak{l}_2$.} until there are no admissible positions in the resulting partition. The sequence of the admissible positions where we take the reductions is called a maximal admissible sequence.

\section{Atomic Inductions for \texorpdfstring{$E_6$}{E6} Nilpotent Orbits}\label{atomicInduction_e6}
In this appendix, we list all the induced orbits $\calO_{E_6}$ in $\mathfrak{e}_6$ that comes from a maximal proper Levi subalgebra $\mathfrak{l}$:
\be
    \calO_{E_6} = \mathrm{Ind}^{\mathfrak{e}_6}_{\kl}(\calO_{\kl}).
\ee
By ``maximal'', we mean that $\kl=\mathfrak{s}\oplus\mathfrak{gl}(1)$ with $\mathfrak{s}$ of rank 5, that is, $\mathfrak{s} \in \{A_5, D_5, A_4 + A_1, 2A_2 + A_1\}$. The results are given in Tables \ref{table:e6_ind_a5}, \ref{table:e6_ind_d5}, \ref{table:e6_ind_a4_a1}, and \ref{table:e6_ind_2a2_a1} respectively. For each choice of $\kl$, we list $\calO_{E_6}$ and all the orbits giving the inductions in $\kl$, along with the complex dimensions of these orbits\footnote{As the $\mathfrak{gl}(1)$ part is trivial and smooth, we can just write $\mathcal{O}_{\mathfrak{s}}$. It should be clear that when $\mathfrak{s}$ is not simple, the orbit $\mathcal{O}_{\mathfrak{s}}$ is the direct sum of the components.}.

\begin{longtable}{|c|c|c|c||c|c|c|c|} \hline
$\calO_{A_5}$ & $\dim_\bbC (\calO_{A_5})$ & induced $\calO_{E_6} $ & $\dim_\bbC (\calO_{E_6})$ & $\calO_{A_5}$ & $\dim_\bbC (\calO_{A_5})$ & induced $\calO_{E_6} $ & $\dim_\bbC (\calO_{E_6})$ \\ \hline\hline
$[1^6]$      &  0 & $A_2$      & 42 & $[3^2]$      & 24 & $E_6(a_3)$ & 66 \\ \hline
$[2, 1^4]$   & 10 & $A_3$      & 52 & $[4, 1^2]$   & 24 & $E_6(a_3)$ & 66 \\ \hline
$[2^2, 1^2]$ & 16 & $D_4(a_1)$ & 58 & $[4, 2]$     & 26 & $D_5$      & 68 \\ \hline
$[2^3]$      & 18 & $D_4$      & 60 & $[5, 1]$     & 28 & $E_6(a_1)$ & 70 \\ \hline
$[3, 1^3]$   & 18 & $A_4$      & 60 & $[6]$        & 30 & $E_6$      & 72 \\ \hline
$[3, 2, 1]$  & 22 & $D_5(a_1)$ & 64 & & & &  \\ \hline
\caption{Induced orbits $\mathcal{O}_{E_6}$ from $(\mathfrak{s} = A_5, \calO_{\kl})$.}
\label{table:e6_ind_a5}
\end{longtable}

\begin{longtable}{|c|c|c|c||c|c|c|c|} \hline
$\calO_{D_5}$ & $\dim_\bbC (\calO_{D_5})$ & induced $\calO_{E_6}$ & $\dim_\bbC (\calO_{E_6})$ & $\calO_{D_5}$ & $\dim_\bbC (\calO_{D_5})$ & induced $\calO_{E_6}$ & $\dim_\bbC (\calO_{E_6})$ \\ \hline\hline
$[1^{10}]$      &  0 & $2A_1$      & 32 & $[3^3, 1]$      & 30 & $A_4+A_1$      & 62  \\ \hline
$[2^2, 1^6]$    & 14 & $A_2+A_1$   & 46 & $[4^2, 1^2]$    & 32 & $D_5(a_1)$      & 64  \\ \hline
$[3, 1^7]$      & 16 & $2A_2$      & 48 & $[5, 2^2, 1]$   & 32 & $A_5$ & 64  \\ \hline
$[2^4, 1^2]$    & 20 & $A_3$       & 52 & $[5, 3, 1^2]$   & 34 & $E_6(a_3)$ & 66  \\ \hline
$[3, 2^2, 1^3]$ & 24 & $A_3+A_1$   & 56 & $[5^2]$         & 36 & $D_5$ & 68  \\ \hline
$[3^2, 1^4]$    & 26 & $D_4(a_1)$  & 58 & $[7, 1^3]$      & 36 & $D_5$      & 68  \\ \hline
$[3^2, 2^2]$    & 28 & $A_4$       & 60 & $[7, 3]$        & 38 & $E_6(a_1)$ & 70  \\ \hline
$[5, 1^5]$      & 28 & $A_4$       & 60 & $[9, 1]$        & 40 & $E_6$      & 72  \\ \hline
\caption{Induced orbits in $\mathcal{O}_{E_6}$ from $(\mathfrak{s} = D_5, \calO_{\kl})$.}
\label{table:e6_ind_d5}
\end{longtable}

\begin{longtable}{|c|c|c|c|c|c|} \hline
$\calO_{A_1}$ & $\dim_\bbC (\calO_{A_1})$ & $\calO_{A_4}$ & $\dim_\bbC (\calO_{A_4})$ & induced $\calO_{E_6}$ & $\dim_\bbC (\calO_{E_6})$ \\ \hline\hline
$[1^{2}]$      &  0 & $[1^5]$      &  0  & $A_2+ 2A_1$      & 50  \\ \hline
$[1^{2}]$      &  0 & $[2, 1^3]$   &  8  & $D_4(a_1)$      & 58  \\ \hline
$[1^{2}]$      &  0 & $[2^2, 1]$   & 12  & $A_4 + A_1$      & 62  \\ \hline
$[1^{2}]$      &  0 & $[3, 1^2]$   & 14  & $d_5(a_1)$      & 64  \\ \hline
$[1^{2}]$      &  0 & $[3, 2]$     & 16  & $E_6(a_3)$      & 66  \\ \hline
$[1^{2}]$      &  0 & $[4, 1]$     & 18  & $D_5$      & 68  \\ \hline
$[1^{2}]$      &  0 & $[5]$        & 20  & $E_6(a_1)$      & 70  \\ \hline
$[2]$          &  2 & $[1^5]$      &  0  & $A_3$           & 52  \\ \hline
$[2]$          &  2 & $[2, 1^3]$   &  8  & $A_4$           & 60  \\ \hline
$[2]$          &  2 & $[2^2, 1]$   & 12  & $D_5(a_1)$      & 64  \\ \hline
$[2]$          &  2 & $[3, 1^2]$   & 14  & $E_6(a_3)$      & 66  \\ \hline
$[2]$          &  2 & $[3, 2]$     & 16  & $D_5$           & 68  \\ \hline
$[2]$          &  2 & $[4, 1]$     & 18  & $E_6(a_1)$      & 70  \\ \hline
$[2]$          &  2 & $[5]$        & 20  & $E_6$           & 72  \\ \hline
\caption{Induced orbits in $\mathcal{O}_{E_6}$ from $(\mathfrak{s} = A_4+A_1, \calO_{\kl})$.}
\label{table:e6_ind_a4_a1}
\end{longtable}

\newpage

\begin{longtable}[H]{|c|c|c|c|c|c|c|c|} \hline
$\calO_{A_1}$ & $\dim_\bbC (\calO_{A_1})$ & $\calO_{A_2}^a$ & $\dim_\bbC \left(\calO_{A_2}^a\right)$ & $\calO_{A_2}^b$ & $\dim_\bbC \left(\calO_{A_2}^b\right)$ & induced $\calO_{E_6}$ & $\dim_\bbC (\calO_{E_6})$ \\ \hline\hline
$[1^{2}]$      &  0 & $[1^{3}]$      &  0 & $[1^3]$      &  0  & $D_4(a_1)$      & 58  \\ \hline
$[1^{2}]$      &  0 & $[1^{3}]$      &  0 & $[2, 1]$     &  4  & $A_4 + A_1$     & 62  \\ \hline
$[1^{2}]$      &  0 & $[1^{3}]$      &  0 & $[3]$        &  6  & $D_5(a_1)$      & 64  \\ \hline
$[1^{2}]$      &  0 & $[2,  1]$      &  4 & $[2, 1]$     &  4  & $E_6(a_3)$      & 66  \\ \hline
$[1^{2}]$      &  0 & $[2,  1]$      &  4 & $[3]$        &  6  & $D_5$           & 68  \\ \hline
$[1^{2}]$      &  0 & $[3]$          &  6 & $[3]$        &  6  & $E_6(a_1)$      & 70  \\ \hline
$[2]$          &  2 & $[1^{3}]$      &  0 & $[1^3]$      &  0  & $D_5$           & 60  \\ \hline
$[2]$          &  2 & $[1^{3}]$      &  0 & $[2, 1]$     &  4  & $D_5(a_1)$      & 64  \\ \hline
$[2]$          &  2 & $[1^{3}]$      &  0 & $[3]$        &  6  & $E_6(a_3)$      & 66  \\ \hline
$[2]$          &  2 & $[2,  1]$      &  4 & $[2, 1]$     &  4  & $D_5$           & 68  \\ \hline
$[2]$          &  2 & $[2,  1]$      &  4 & $[3]$        &  6  & $E_6(a_1)$      & 70  \\ \hline
$[2]$          &  2 & $[3]$          &  6 & $[3]$        &  6  & $E_6$           & 72  \\ \hline
\caption{Induced orbits in $\mathcal{O}_{E_6}$ from $(\mathfrak{s} = 2A_2+A_1, \calO_{\kl})$.}
\label{table:e6_ind_2a2_a1}
\end{longtable}

The nilpotent orbits that do not appear in any of the tables are exactly the rigid orbits in $\mathfrak{e}_6$: $0$, $A_1$, $3A_1$, $2A_2 + A_1$. In other words, any induced orbits in $\mathfrak{e}_6$ can be induced from at least one of the maximal Levi subalgebras $\mathfrak{l}$ listed above.

\section{Comments on Magnetic Quivers and Induced Flows}\label{guessMQ}
In \S\ref{DEtypeorbins}, we discussed the atomic induced Higgsings for the DE-type orbi-instanton theories. In general, the magnetic quivers for these cases are not known except
\begin{equation}
    \left(D^{e_8}_4,11\right):\quad([\SO(8)] \ 1)^{\otimes 2} \ \overset{\kso(8)}{4} \ 1 \ \left\langle\overset{\kso(8)}{4} \ 1\right\rangle^{\otimes k} \ [\SO(8)].
\end{equation}
The corresponding magnetic quiver reads
\begin{equation}
    \includegraphics[width=10cm]{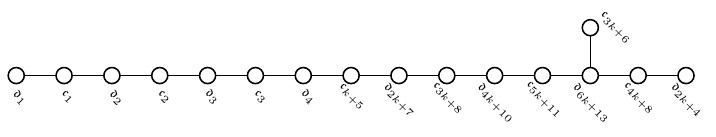},\label{D4e811MQ}
\end{equation}
which can be obtained from the Type IIA brane construction \cite{Bao:2024eoq,Lawrie:2024wan}. In particular, the nilpotent VEV $\left[1^8\right]_{\mathfrak{so}(8)}$ corresponds to the tail
\begin{equation}
    \includegraphics[width=5cm]{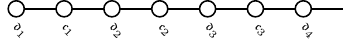}
\end{equation}
in the magnetic quiver. When the nilpotent VEV is $\left[2^4\right]^{\text{I,II}}_{\mathfrak{so}(8)}$, we can simply replace the above tail with
\begin{equation}
    \includegraphics[width=3cm]{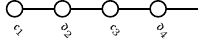}.
\end{equation}

Suppose that an orthosymplectic magnetic quiver is Higgsed to a unitary magnetic quiver (with a single component) after an atomic induced flow. Then it is not hard to see that the resulting unitary magnetic quiver has the same gauge ranks as the orthosymplectic one, and we only need to change the group types to the unitary groups. Based on this observation, it is tempting to guess some of the D-type orbi-instanton magnetic quivers as the A-type ones are known \cite{Mekareeya:2017jgc}. Notice that the dimension change of the moduli space is automatically satisfied. In terms of the Type IIA brane systems, the dynamical branes are moved far away from the orientifolds as if the orientifolds are removed from the original systems. Although there are no more evidences for such extrapolations, we may guess the magnetic quivers for the theories in Table \ref{D4_A3_orbi-instanton_table}. For instance, the $\left(D^{e_8}_4,0\right)$ theory has the following tail for the discrete homomorphism:
\begin{equation}
    \includegraphics[width=7cm]{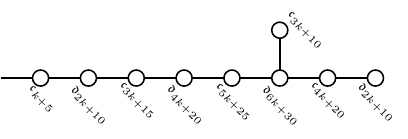}.
\end{equation}
Given the magnetic quivers for the A-types as in \cite{Mekareeya:2017jgc}, the other magnetic quivers for the theories in Table \ref{D4_A3_orbi-instanton_table} can be guessed in the same manner. In the orthosymplectic quivers, it seems that the parts with alternating balances $\pm1$ ($+1$ for $\mathfrak{d}$ and $-1$ for $\mathfrak{c}$) give the global symmetries (possibly with further enhancements). However, this is simply because the A-type unitary magnetic quivers have the balanced nodes encoding the global symmetries. When the gauge nodes are changed to the orthosymplectic ones, the balances would become $\pm1$. Therefore, there is still a lack of physical evidences to see whether the overbalanced and underbalanced nodes could give any information of the global symmetries.

For the remaining theories in Table \ref{D4_orbi-instanton_table}, it could also be possible to guess their magnetic quivers based on the atomic Higgsings among the theories. For instance, $\left(D^{e_8}_4,13\right)$ lives between $\left(D^{e_8}_4,11\right)$ and $\left(D^{e_8},14\right)$. From the transverse slices given in \cite{Bao:2024eoq,Lawrie:2024wan}, a natural guess of the magnetic quiver for $\left(D^{e_8}_4,13\right)$ would be replacing the $-\mathfrak{d}_{2k+7}-\mathfrak{c}_{3k+8}-\mathfrak{d}_{4k+10}-\mathfrak{c}_{5k+11}-$ part in \eqref{D4e811MQ} with $-\mathfrak{d}_{2k+6}-\mathfrak{c}_{3k+7}-\mathfrak{d}_{4k+9}-\mathfrak{c}_{5k+10}-$.

We should emphasize again that the above tails in the magnetic quivers are very naive guesses. Although they are consistent with the atomic induced Higgsings, such extrapolations would still need more physical and/or mathematical reasonings to verify or refute the guesses. It could be possible that the magnetic quivers would look very differently and may not even have any orthosymplectic magnetic quiver descriptions.

% \section{Relation to generalized symmetries}

% We could identify flows that preserve / break a certain symmetry, or have a certain symmetries enhanced.

% But it would be much more interesting if we can really learn some knowledge that is very hard to see (or at least very difficult to realize) if we do not think in terms of generalized symmetries.

% Could say something on the 2-Group structure constant of the LSTs! But still unclear of how surprising it can get. Maybe the study of how these 2-Group structure constant can be a bit fun, if not super exciting.

\addcontentsline{toc}{section}{References}
\bibliographystyle{utphys.bst}
\bibliography{ref}

\end{document}